\documentclass[11pt]{article}
\usepackage{amsfonts,amsmath, amssymb, bm}
\usepackage{times}
\usepackage{graphicx,afterpage} 
\usepackage{epstopdf}
\usepackage{float,multicol,multirow}
\usepackage[caption=false]{subfig}
\usepackage{color}
\usepackage[table]{xcolor}
\usepackage{hyperref}
\usepackage{fullpage}

\def\v#1{{\mathbf #1}}

\def\x{{\mathbf x}}

\def\dotfil{\leaders\hbox to 1.5mm{.}\hfill}
\newcounter{rmnum}
\def\RN#1{\setcounter{rmnum}{#1}\uppercase\expandafter{\romannumeral\value{rmnum}}}
\def\rn#1{\setcounter{rmnum}{#1}\expandafter{\romannumeral\value{rmnum}}}

\newcommand{\TNorm}[1]{\mbox{}\left\|#1\right\|_2}

\newcommand{\TNormS}[1]{\mbox{}\left\|#1\right\|_2^2}

\newcommand{\VZeroNorm}[1]{\mbox{}\left\|#1\right\|_0  }
\newcommand{\VOneNorm}[1]{\mbox{}\left\|#1\right\|_1  }
\newcommand{\VOneNormS}[1]{\mbox{}\left\|#1\right\|_1^2  }
\newcommand{\VOneNormQ}[1]{\mbox{}\left\|#1\right\|_1^4  }
\newcommand{\VThreeNorm}[1]{\mbox{}\left\|#1\right\|_3  }
\newcommand{\VPNorm}[1]{\mbox{}\left\|#1\right\|_p}
\newcommand{\VQNorm}[1]{\mbox{}\left\|#1\right\|_q}

\newcommand{\E}{\mathbb{E}}
\newcommand{\sgn}[1]{{\bf sgn}{\left(#1\right)}}

\newcommand{\setlinespacing}[1]%
           {\setlength{\baselineskip}{#1 \defbaselineskip}}

\newcommand{\abs }[1]{\left|#1\right|}

\newtheorem{lemma}{Lemma}
\newtheorem{theorem}{Theorem}

\newenvironment{proof}{\noindent {\em Proof:}}{\hspace*{\fill}\mbox{$\diamond$}}

\newcommand{\mat}[1]{{\ensuremath{\bm{\mathrm{#1}}}}}

\def\x{{\mathbf x}}

\def\u{{\mathbf u}}
\def\v{{\mathbf v}}

\def\matA{\mat{A}}

\def\matD{\mat{D}}
\def\matE{\mat{E}}
\def\matG{\mat{G}}

\def\matU{\mat{U}}
\def\matV{\mat{V}}

\def\matX{\mat{X}}
\def\matY{\mat{Y}}

\usepackage[ruled]{algorithm2e}

\SetAlFnt{\small}
\SetAlCapFnt{\small}
\SetAlCapNameFnt{\small}
\SetAlCapHSkip{0pt}
\IncMargin{-\parindent}

\title{A Randomized Rounding Algorithm for Sparse PCA}

\author{
	Kimon Fountoulakis\thanks{International Computer Science Institute, Department of Statistics, University of California Berkeley, Berkeley, CA, USA, kfount@berkeley.edu}
	\and
	Abhisek Kundu\thanks{Speech and Biomedical Analytics Group, Xerox Research Centre India, Bangalore, India, abhisekkundu@gmail.com}
	\and
	Eugenia-Maria Kontopoulou\thanks{Department of Computer Science, Purdue University, West Lafayette IN, USA, ekontopo@purdue.edu}
	\and
	Petros Drineas\thanks{Department of Computer Science, Purdue University, West Lafayette IN, USA, pdrineas@purdue.edu}
}

\date{}

\begin{document}

\maketitle

\begin{abstract}
We present and analyze a simple, two-step algorithm to approximate the optimal solution of the sparse PCA problem. Our approach first solves an $\ell_{1}$-penalized version of the NP-hard sparse PCA optimization problem and then uses a randomized rounding strategy to sparsify the resulting dense solution. Our main theoretical result guarantees an additive error approximation and provides a tradeoff between sparsity and accuracy. Our experimental evaluation indicates that our approach is competitive in practice, even compared to state-of-the-art toolboxes such as \texttt{Spasm}.
\end{abstract}

\paragraph{Keywords.} Sparce pca, rounding randomized algorithm.

\section{Introduction}
Large matrices are a common way of representing modern, massive datasets, since an $m \times n$ real-valued matrix $\matX$ provides a natural structure for encoding information about $m$ objects, each of which is described by $n$ features. Principal Components Analysis (PCA) and the Singular Value Decomposition (SVD) are fundamental data analysis tools, expressing a data matrix in terms of a sequence of orthogonal vectors of decreasing importance. While these vectors enjoy strong optimality properties and are often interpreted as fundamental latent factors that underlie the observed data, they are linear combinations of up to all the data points and features. As a result, they are notoriously difficult to interpret in terms of the underlying processes generating the data~\cite{MD09}.

The seminal work of~\cite{AGJ2007} introduced the concept of Sparse PCA, where sparsity constraints are enforced on the singular vectors in order to improve interpretability. As noted in~\cite{AGJ2007,MD09,PDK2013}, an example where sparsity implies interpretability is document analysis, where sparse principal components can be mapped to specific topics by inspecting the (few) keywords in their support. Formally, Sparse PCA can be defined as\footnote{Recall that the $\ell_p$ norm of a vector $\x \in \mathbb{R}^n$ is defined as $\VPNorm{\x}^p = \sum_{i=1}^n \abs{\x_i}^p$ when $0<p<\infty$; we will only use $p=1$, $2$ and $3$. Furthermore, note that $\|x\|_0$ is the $\ell_0$ ``norm", which is the number of non-zero entries of the input vector $x$.} (see eqn.~(1) in~\cite{PDK2013}):
\begin{equation}\label{eqn:sparsePCA}
{\cal Z}_{opt} = \max_{\x \in \mathbb{R}^n,\ \TNorm{\x}\le 1} \x^T\matA\x, \quad
\mbox{s.t.}\quad \VZeroNorm{\x} \leq k.
\end{equation}
In the above formulation, the parameter $k$ controls the sparsity of the resulting vector and is part of the input; $\matA = \matX^T \matX \in \mathbb{R}^{n \times n}$ is the symmetric positive semidefinite (PSD) covariance matrix that represents all pairwise feature similarities for the data matrix $\matX$ and $\x_{opt}$ denotes a vector that achieves the optimal value ${\cal Z}_{opt}$ in the above formulation. In words, the above optimization problem seeks a \textit{sparse}, unit norm vector $\x_{opt}$ that maximizes the data variance $\x_{opt}^T \matA \x_{opt}$. The following two facts are well-known: first, solving the above optimization problem is NP-hard~\cite{MWA2006} and, second, this hardness is due to the sparsity constraint. Indeed, if the sparsity constraint was removed, then the resulting optimization problem can be easily solved: the optimal vector is the top (left or right) singular vector of $\matA$ and the related maximal value is equal to the top singular value of $\matA$. Finally, we note that more than one sparse singular vectors can be constructed using a simple deflation procedure; see Section~\ref{sec:setup} for details. Following the lines of~\cite{PDK2013}, we will only focus on the formulation of eqn.~(\ref{eqn:sparsePCA}).


\vspace{0.02in}\noindent\textbf{Related work.} The simplest sparse PCA approaches are to either rotate~\cite{J95} or threshold~\cite{CJ95} the top singular vector of the matrix $\matA$. Such simple methods are computationally efficient and tend to perform very well in practice (see Section~\ref{sec:experiments}). However, there exist cases where they fail (see~\cite{CJ95} and Section~\ref{sec:experiments} here). An alternative line of research focused on solving relaxations of eqn.~(\ref{eqn:sparsePCA}). For example, an $\ell_1$ relaxation of eqn.~(\ref{eqn:sparsePCA}) was first used in SCoTLASS~\cite{JTU2003}. Another possible relaxation is a regression-type approximation~\cite{ZHT12}, which was implemented
in~\cite{KCLE12}. (We will compare our approach to this method.) Finally, efficient optimization methods have been developed for the sparse PCA problem. For example, the generalized power method was proposed in~\cite{JNRS2010}: this method calculates stationary points for penalized versions of eqn.~(\ref{eqn:sparsePCA}).

Despite the many approaches that were developed for sparse PCA, only a handful of them provide any type of theoretical guarantees regarding the quality of the obtained (approximate) solution. For example, the semidefinite relaxation of~\cite{AGJ2007} was analyzed in~\cite{AW2008}, albeit for the special case where $\matA$ is a spiked covariance matrix with a sparse maximal singular vector. Briefly,~\cite{AW2008} studies conditions for the dimensions $m$ and $n$ of the data matrix $\matX$, and the sparsity parameter $k$, so that the semidefinite relaxation of~\cite{AGJ2007} recovers the sparsity pattern of the optimal solution of eqn.~\eqref{eqn:sparsePCA}.
Other attempts for provable results include the work of~\cite{ABG2008}, which was later analyzed in~\cite{ABG2014}. In the latter paper, the authors show bounds for the semidefinite relaxation of~\cite{ABG2008}, in the special case that the data points are sampled using Gaussian models with a single sparse leading singular vector. Strong compressed-sensing-type conditions were used in~\cite{YZ2013} to guarantee recovery of the optimal solution of eqn.~\eqref{eqn:sparsePCA} using a  truncated power method. However,~\cite{YZ2013} requires that the optimal solution is approximately sparse and also that the noise matrix has sparse submatrices with small spectral norm.
Finally,~\cite{PDK2013} describes a greedy combinatorial approach for sparse PCA and provides relative-error bounds for the resulting solution under the assumption that the covariance matrix $\matA$ has a decaying spectrum. It is important to note that in all the above papers special assumptions are necessary regarding the input data in order to guarantee the theoretical bounds. Our work here does not make any assumptions on the input data, but we do pay the cost of increased sparsity in the final solution (see Theorem~\ref{thm:main} for details).
There are also connections between sparse approximations and subspace learning methods, which are widely used in machine learning and data mining. Recently, methods that enforce sparsity have been developed for Human Activity Recognition \cite{plaplacian} and Image Classification \cite{Imageclass}. Moreover, some of these methods have been applied to multidimensional data \cite{Tao:2007:GTD:1313053.1313256}.

\vspace{0.02in}\noindent\textbf{Our algorithm.} We present and analyze a simple, two-step algorithm to approximate the optimal solution of the problem of eqn.~(\ref{eqn:sparsePCA}). Our approach first finds a stationary point of an $\ell_1$-penalized version of problem \eqref{eqn:sparsePCA}. Then, a randomized rounding strategy is employed to sparsify the resulting dense solution of the $\ell_1$-penalized problem. More precisely, we first solve:
\begin{equation}\label{eqn:sparsePCArelaxed}
\tilde{\cal Z}_{opt} = \max_{\x \in \mathbb{R}^n,\ \TNorm{\x}\le1} \x^T\matA\x, \quad
\mbox{s.t.}\quad \VOneNorm{\x} \leq \sqrt{k}.
\end{equation}
Notice that we replaced the constraint on the $\ell_0$ norm of the vector $\x$ by a (tighter) constraint on the $\ell_1$ norm of $\x$.
It is important to mention that problem \eqref{eqn:sparsePCArelaxed} is difficult and all we can
hope in practice is to calculate a stationary point. However, one should not discount the quality of stationary points. In Section \ref{sec:experiments} we
show that by calculating stationary points we capture as much of the variance as computationally expensive convex relaxations. Having said that, the theoretical analysis
which follows assumes that we work with the globally optimal solution of problem \eqref{eqn:sparsePCArelaxed}.

Let $\tilde{\x}_{opt}$ be a vector that achieves the optimal value $\tilde{\cal Z}_{opt}$ for problem \eqref{eqn:sparsePCArelaxed}.
Clearly, $\tilde{\x}_{opt}$ is not necessarily sparse. Therefore, we employ a randomized rounding strategy to sparsify it by keeping larger entries of $\tilde{\x}_{opt}$ with higher probability. Specifically, we employ Algorithm~\ref{alg:sparsify} on the vector $\tilde{\x}_{opt}$ to get a sparse vector $\hat{\x}_{opt}$ that is our approximate solution to the sparse PCA formulation of eqn.~{(\ref{eqn:sparsePCA})}.
\begin{algorithm}[t]
\SetAlgoNoLine
\KwIn{ $\x \in \mathbb{R}^{n}$, integer $s$.}
\KwOut{$\hat{\x} \in \mathbb{R}^{n}$, $\E\left(\VZeroNorm{\hat{\x}}\right) \leq s$.}
\For{$i=1,\ldots, n$}{
Set: $$\hat{\x}_i = \left\{ \begin{array}{ll}
          \frac{1}{p_i}\x_i, & \text{with probability}\ {p_i = \min\left\{\frac{s\abs{\x_i}}{\VOneNorm{\x}},1\right\} }\\
          0, & \text{otherwise}
        \end{array}\right.$$}
\caption{Vector sparsification}
\label{alg:sparsify}
\end{algorithm}


It is obvious that, in expectation, the vector $\hat{\x}_{opt}$ has at most $s$ non-zero entries. We will discuss the appropriate choice for $s$ in Theorem~\ref{thm:main} below.

\vspace{0.02in}\noindent\textbf{Our results: theory.} Surprisingly, this simple randomized rounding approach has not been analyzed in prior work on Sparse PCA. Theorem~\ref{thm:main} is our main theoretical result and guarantees an additive error approximation to the NP-hard problem of eqn.~(\ref{eqn:sparsePCA}). For simplicity of presentation, we will assume that the rows (and therefore columns) of the matrix $\matA$ have at most unit norms\footnote{We can relax this assumption by increasing $s$ -- our sampling factor -- by a factor that depends on the upper bound of the row and column norms of $\matA$.}.
\begin{theorem}\label{thm:main}
Let $\x_{opt}$ be the optimal solution of the Sparse PCA problem of eqn.~(\ref{eqn:sparsePCA})
satisfying $\TNorm{\x_{opt}}=1$ and $\VZeroNorm{\x_{opt}}\leq k$. Let $\hat{\x}_{opt}$ be the
vector returned when Algorithm~\ref{alg:sparsify} is applied on the optimal solution
$\tilde{\x}_{opt}$ of the optimization problem of eqn.~(\ref{eqn:sparsePCArelaxed}), with
$s=200k/\epsilon^2$, where $\epsilon \in (0,1]$ is an accuracy parameter. Then, $\hat{\x}_{opt}$ has the following properties:
\begin{enumerate}
\item $\E \VZeroNorm{\hat{\x}_{opt}} \leq s.$
\item With probability at least 3/4,
$$\TNorm{\hat{\x}_{opt}} \leq 1+0.15\epsilon.$$
\item With probability at least 3/4,
\begin{equation}\label{eqn:thm:main}
\hat{\x}_{opt}^T \matA \hat{\x}_{opt} \geq \x_{opt}^T \matA \x_{opt} - \epsilon.\end{equation}
\end{enumerate}
\end{theorem}
In words, the above theorem states that our sparse vector $\hat{\x}_{opt}$ is almost as good as the optimal vector $\x_{opt}$ in terms of capturing (with constant probability) almost as much of the spectrum of $\matA$ as $\x_{opt}$ does. This comes at a penalty: the sparsity of $\x_{opt}$, which is equal to $k$, has to be relaxed to $O\left(k/\epsilon^2\right)$. This provides an elegant trade-off between sparsity and accuracy\footnote{A less important artifact of our approach is the fact that the Euclidean norm of the vector $\hat{\x}_{opt}$ is slightly larger than one.}. It is worth emphasizing that one should not worry about the small success probability of our approach: by repeating the rounding $t$ times and keeping the vector $\hat{\x}_{opt}$ that satisfies the second bound and maximizes $\hat{\x}_{opt}^T \matA \hat{\x}_{opt}$, we can immediately guarantee that we will achieve both bounds with probability at least $1-2^{-t}$.

\vspace{0.02in}\noindent\textbf{Our results:experiments.} We empirically evaluate our approach on real and synthetic data. We chose to compare our algorithm with the state-of-the-art \texttt{Spasm} toolbox of~\cite{KCLE12,ZHT12}. We also compare our solution with the simple \texttt{MaxComp} heuristic, which  computes the top singular vector of matrix $\matA$ and returns a sparse vector by greedily keeping the top $k$ largest components (in absolute value) and setting the remaining ones to zero. Our empirical evaluation indicates that our simple, provably accurate approach is competitive in practice. 
\section{Proof of Theorem 1.1}
\subsection{Preliminaries}\label{sxn:proof:prelim}

Consider the indicator random variables $\delta_i$ for all $i =1\ldots n$ which take the following values:
$$\delta_i = \left\{ \begin{array}{ll}
          \frac{1}{p_i} & ,\text{with probability}\ p_i \\
          0 & ,\text{otherwise}
        \end{array}\right.$$
It is easy to see that $\hat{\x}_{i} = \delta_i \tilde{\x}_i$ for all $i =1 \ldots n$. The following trivial properties hold for all $i$ and will be used repeatedly in the proofs: $\E \delta_i = 1$, $\E \left(1-\delta_i\right) = 0$, and $\E \left(1-\delta_i\right)^2 = p_i^{-1}-1$.
For simplicity of notation we will drop the the subscript \textit{opt} from $\x_{opt}$, $\tilde{\x}_{opt}$, and $\hat{\x}_{opt}$ in all proofs in this section.

\subsection{A bound for $\VZeroNorm{\hat{\x}_{opt}}$}\label{sxn:appendix:bound1}

Recall that we will use the sampling probabilities $p_i$ as defined in Algorithm~\ref{alg:sparsify}.
By definition, $p_i \leq s\abs{\tilde{\x}_i}/\VOneNorm{\tilde{\x}}$, therefore
\begin{equation}
\E\left(\VZeroNorm{\hat{\x}}\right) = \sum_{i=1}^n p_i \leq \sum_{i=1}^n \frac{s\abs{\tilde{\x}_i}}{\VOneNorm{\tilde{\x}}} = s.
\end{equation}
which proves the first bound in Theorem~\ref{thm:main}.

\subsection{A bound for  $\TNorm{\hat{\x}_{opt}}$}\label{sxn:appendix:bound2}

The following lemma immediately implies the second bound of Theorem~\ref{thm:main} by setting $s = 200k/\epsilon^2$.

\begin{lemma}\label{lem:twonormbound}
	Given our notation, with probability at least 3/4,
	$$\TNorm{\hat{\x}_{opt}} \leq 1+2\sqrt\frac{k}{s}.$$
\end{lemma}

\begin{proof}
	%
	It is more intuitive to provide a bound on the expectation of $\TNormS{\hat{\x}-\tilde{\x}}$ and then leverage the triangle inequality in order to bound $\TNorm{\hat{\x}}$. Using the indicator variables $\delta_i$ and linearity of expectation, we get
	\begin{eqnarray*}
		\E \TNormS{\hat{\x}-\tilde{\x}} &=& \E \sum_{i=1}^n \left(1-\delta_i\right)^2 \tilde{\x}_i^2\\
		                                &=& \sum_{i=1}^n \tilde{\x}_i^2 \E\left(1-\delta_i\right)^2\\
		                                &=& \sum_{i=1}^n \left(\frac{1}{p_i}-1\right) \tilde{\x}_i^2.
	\end{eqnarray*}
	We will now prove the following inequality, which will be quite useful in later proofs:
	\begin{equation}\label{eqn:pd11}
	\sum_{i=1}^n \left(\frac{1}{p_i}-1\right) \tilde{\x}_i^2 \leq \frac{k}{s}.
	\end{equation}
	Towards that end, we will split the set of indices $\{1\ldots n\}$ in two subsets: the set $I^{=1}$ corresponding to indices $i$ such that $p_i = 1$ and the set $I^{<1}$ corresponding to indices $i$ such that $p_i < 1$. Note that for all $i \in I^{<1}$ it must be the case that
	$$p_i = \frac{s\abs{\tilde{\x}_i}}{\VOneNorm{\tilde{\x}}}.$$
	We now proceed as follows:
	\begin{eqnarray*}
		\E \TNormS{\hat{\x}-\tilde{\x}}
		&=& \sum_{i \in I^{=1}} \left(\frac{1}{p_i}-1\right) \tilde{\x}_i^2 + \sum_{i \in I^{<1}} \left(\frac{1}{p_i}-1\right) \tilde{\x}_i^2\\
		&=&\sum_{i \in I^{<1}} \left(\frac{1}{p_i}-1\right) \tilde{\x}_i^2\\
		&\leq& \sum_{i \in I^{<1}} \frac{1}{p_i} \tilde{\x}_i^2\\
		&=& \sum_{i \in I^{<1}} \frac{\VOneNorm{\tilde{\x}}}{s\abs{\tilde{\x}_i}} \tilde{\x}_i^2\\
		&\leq& \frac{\VOneNorm{\tilde{\x}}}{s} \sum_{i=1}^n \abs{\tilde{\x}_i}\\
		&=& \frac{\VOneNormS{\tilde{\x}}}{s} \leq \frac{k}{s}.
	\end{eqnarray*}
	For the last inequality we used the fact that $\VOneNorm{\tilde{\x}} \leq \sqrt{k}$. We now use Markov's inequality to conclude that, with probability at least 3/4,
	\begin{equation}\label{eqn:res1}
	\TNormS{\hat{\x}-\tilde{\x}} \leq \frac{4k}{s}.
	\end{equation}
	To conclude the proof note that, from the triangle inequality,
	$$\TNorm{\hat{\x} - \tilde{\x}} \geq \abs{\TNorm{\hat{\x}} - \TNorm{\tilde{\x}}}$$
	and thus
	$$\TNorm{\hat{\x}}\leq \TNorm{\tilde{\x}} + \TNorm{\hat{\x} - \tilde{\x}} \leq 1 + \TNorm{\hat{\x} - \tilde{\x}}.$$
	Combining with eqn.~(\ref{eqn:res1}), after taking the square root of both sides, concludes the proof of the lemma.
\end{proof}

\subsection{Proving eqn.~(\ref{eqn:thm:main})}
The following lemma states that the solution of the convex relaxation of the Sparse PCA problem is at least as good as the solution of the Sparse PCA problem.
\begin{lemma}\label{lem:relaxation}
	Given our notation, $\x_{opt}$ is a feasible solution of the relaxed Sparse PCA formulation of eqn.~(\ref{eqn:sparsePCArelaxed}). Thus,
	$\tilde{\cal Z}_{opt} = \tilde{\x}_{opt}^T \matA \tilde{\x}_{opt}^T \geq \x_{opt}^T \matA \x_{opt}.$
\end{lemma}

\begin{proof}
	Recall that $\x_{opt}$ is a unit norm vector whose zero norm is at most $k$. Then, if we let $\sgn{\x}$ denote the vector of signs for $\x$ (with the additional convention that if $\x_i$ is equal to zero then the $i$-th entry of $\sgn{\x}$ is also set to zero), we get
	$$
	\VOneNorm{\x_{opt}} = \abs{\sgn{\x_{{opt}} }^T\x_{opt}}
	\leq \TNorm{\sgn{\x_{{opt}} }} \TNorm{\x_{opt}}
	\leq \sqrt{k}.
	$$
	The second inequality follows since $\sgn{\x_{opt}}$ has at most $k$ non-zero entries. Thus, $\x_{opt}$ is feasible for the optimization problem of eqn.~(\ref{eqn:sparsePCArelaxed}) and the conclusion of the lemma follows immediately.
\end{proof}

Getting a lower bound for $\hat{\x}^T \matA\hat{\x}$ is the toughest part of Theorem~\ref{thm:main}. Towards that end, the next lemma bounds the error $\abs{\tilde{\x}_{opt}^T\matA\tilde{\x}_{opt}-\hat{\x}_{opt}^T \matA\hat{\x}_{opt}}$ as a function of two other quantities which will be easier to bound.
\begin{lemma}\label{lem:mainbound}
Given our notation,
$$
\abs{\tilde{\x}_{opt}^T\matA\tilde{\x}_{opt}-\hat{\x}_{opt} \matA\hat{\x}_{opt}} \leq \;
2\abs{\tilde{\x}_{opt}^T\matA\left(\tilde{\x}_{opt}-\hat{\x}_{opt}\right)} +
|\left(\hat{\x}_{opt}-\tilde{\x}_{opt}\right)^T \matA\left(\tilde{\x}_{opt}-\hat{\x}_{opt}\right)|.
$$
\end{lemma}

\begin{proof}
We start with
\begin{eqnarray}
\nonumber \abs{\tilde{\x}^T\matA\tilde{\x}-\hat{\x}^T \matA\hat{\x}} &=&  \abs{\tilde{\x}^T\matA\tilde{\x}-\hat{\x}^T\matA\tilde{\x}+\hat{\x}^T\matA\tilde{\x}-\hat{\x}^T \matA\hat{\x}}\\ & \leq&
\label{eqn:pd2}  \abs{\left(\tilde{\x}-\hat{\x}\right)^T \matA\tilde{\x}} + \abs{\hat{\x}^T\matA\left(\tilde{\x}-\hat{\x}\right)}.
\end{eqnarray}
Next,
\begin{eqnarray}
\nonumber \abs{\left(\hat{\x}-\tilde{\x}\right)^T \matA\left(\tilde{\x}-\hat{\x}\right)} & =&
\nonumber\abs{\hat{\x}^T \matA\left(\tilde{\x}-\hat{\x}\right)-\tilde{\x}^T\matA\left(\tilde{\x}-\hat{\x}\right)}\\ &\geq&\label{eqn:pd1}
\abs{\abs{\hat{\x}^T \matA\left(\tilde{\x}-\hat{\x}\right)}-\abs{\tilde{\x}^T\matA\left(\tilde{\x}-\hat{\x}\right)}},
\end{eqnarray}
which implies
\begin{eqnarray}
\abs{\hat{\x}^T \matA\left(\tilde{\x}-\hat{\x}\right)} & \leq & \abs{\tilde{\x}^T\matA\left(\tilde{\x}-\hat{\x}\right)} +
\abs{\left(\hat{\x}-\tilde{\x}\right)^T \matA\left(\tilde{\x}-\hat{\x}\right)}.
\end{eqnarray}
We now combine eqns.~(\ref{eqn:pd2}) and~(\ref{eqn:pd1}) to get

\begin{eqnarray*}
	\abs{\tilde{\x}^T\matA\tilde{\x}-\hat{\x}^T \matA\hat{\x}} &\leq& \abs{\left(\tilde{\x}-\hat{\x}\right)^T\matA\tilde{\x}} +
	\abs{\tilde{\x}^T\matA\left(\tilde{\x}-\hat{\x}\right)} +
	\abs{\left(\hat{\x}-\tilde{\x}\right)^T \matA\left(\tilde{\x}-\hat{\x}\right)}\\
	&=& 2\abs{\tilde{\x}^T\matA\left(\tilde{\x}-\hat{\x}\right)} +
	\abs{\left(\hat{\x}-\tilde{\x}\right)^T \matA\left(\tilde{\x}-\hat{\x}\right)}.
\end{eqnarray*}
\end{proof}

Our next lemma will provide a bound for the first of the two quantities of interest in Lemma~\ref{lem:mainbound}.
\begin{lemma}\label{lem:firstbound}
Given our notation, with probability at least 7/8,
$$\abs{\tilde{\x}_{opt}^T\matA \left(\tilde{\x}_{opt} - \hat{\x}_{opt}\right)} \leq \sqrt{8k/s}.$$
\end{lemma}

\begin{proof}
	Let $\matD\in\mathbb{R}^{n\times n}$ be a diagonal matrix with entries $\matD_{ii} = \delta_i$ for all $i=1\ldots n$.
	Hence, we can write $\hat{\x}=\matD\tilde{\x}$. We have that
	\begin{equation*}
	\left(\tilde{\x}-\hat{\x}\right)^T\matA\tilde{\x} = \tilde{\x}^T\left(I_n - \matD\right)\matA\tilde{\x} = \tilde{\x}^T \left[
	\begin{array}{c}
	(1-\delta_1)\matA_{1*}\tilde{\x} \\
	(1-\delta_2)\matA_{2*}\tilde{\x} \\
	\vdots \\
	(1-\delta_n)\matA_{n*}\tilde{\x}
	\end{array}
	\right] = \sum_{i=1}^n\left(1-\delta_i\right)\tilde{\x}_i \matA_{i*}\tilde{\x},
	\end{equation*}
	where $\matA_{i*}$ is the $i$-th row of the matrix $A$ as a row vector. Squaring the above expression, we get
	\begin{equation} \left((\tilde{\x}-\hat{\x})^T\matA\tilde{\x}\right)^2
	\sum_{i=1}^n \sum_{j=1}^n\left(\tilde{\x}_i \matA_{i*}^T\tilde{\x}\right)\left(\tilde{\x}_j \matA_{j*}\tilde{\x}\right)\left(1-\delta_i\right)\left(1-\delta_j\right).\end{equation}
	Recall that $\E\left(1-\delta_{i}\right)=0$ for all $i$; thus, for all $i\neq j$, $1-\delta_i$ and $1-\delta_j$ are independent random variables and therefore the expectation of their product is equal to zero. Thus, we can simplify the above expression as follows:
	\begin{eqnarray*}
		\E \left(\left(\tilde{\x}-\hat{\x}\right)^T\matA\tilde{\x}\right)^2 &=& \sum_{i=1}^n \E \left(1-\delta_{i}\right)^2
		\left(\tilde{\x}_i\matA_{i*}\tilde{\x}\right)^2\\
		&=&\sum_{i=1}^n  \left(\frac{1}{p_i}-1\right) \tilde{\x}_i^2 \left(\matA_{i*}\tilde{\x}\right)^2 \\
		&\le& \sum_{i=1}^n \left(\frac{1}{p_i}-1\right) \tilde{\x}_i^2.
	\end{eqnarray*}
	In the last inequality we used $\abs{\matA_{i*} \tilde{\x}} \leq \TNorm{\matA_{i*}}\TNorm{\tilde{\x}} \leq 1$. The last term in the above derivation can be bounded as shown in eqn.~(\ref{eqn:pd11}), and thus we conclude
	$$ \E \left(\left(\tilde{\x}-\hat{\x}\right)^T\matA\tilde{\x}\right)^2 \leq k/s.$$
	Markov's inequality now implies that with probability at least 7/8
	$$ \left(\left(\tilde{\x}-\hat{\x}\right)^T\matA\tilde{\x}\right)^2 \leq 8k/s.$$
	\end{proof}

Our next lemma will provide a bound for the second of the two quantities of interest in Lemma~\ref{lem:mainbound}. The proof of the lemma is tedious and a number of cases need to be considered.
\begin{lemma}\label{lem:secondbound}
Given our notation, with probability at least 7/8,
$$\abs{\left(\tilde{\x}_{opt}-\hat{\x}_{opt}\right)^T\matA\left(\tilde{\x}_{opt}-\hat{\x}_{opt}\right)} \leq \left(24k^2/s^2+6k^2/s^3+54\sqrt{k}/s\right)^{1/2}.$$
\end{lemma}

\begin{proof}
%
Using the indicator variables $\delta_i$ and linearity of expectation, we get
\begingroup
\small
\begin{eqnarray}
\nonumber \E \left(\left(\tilde{\x}-\hat{\x}\right)^T\matA\left(\tilde{\x}-\hat{\x}\right)\right)^2
= \E \left(\sum_{i,j=1}^n \matA_{ij} \left(1-\delta_i\right)\tilde{\x}_i \left(1-\delta_j\right)\tilde{\x}_j\right)^2= &\\
\label{eqn:basic} \sum_{i_1,j_1,i_2,j_2=1}^n \matA_{i_1j_1} \matA_{i_2j_2}  \tilde{\x}_{i_1}\tilde{\x}_{i_2}\tilde{\x}_{j_1}\tilde{\x}_{j_2} \times
 \E\left(\left(1-\delta_{i_1}\right)\left(1-\delta_{i_2}\right)\left(1-\delta_{j_1}\right)\left(1-\delta_{j_2}\right)\right).&
\end{eqnarray}
\endgroup\\
Recall that $\E\left(1-\delta_{i}\right)=0$ for all $i$. Notice that if any of the four indices $i_1,i_2,j_1,j_2$ appears only once,
then the expectation $\E\left(\left(1-\delta_{i_1}\right)\left(1-\delta_{i_2}\right)\left(1-\delta_{j_1}\right)\left(1-\delta_{j_2}\right)\right)$
corresponding to those indices equals zero. This expectation is non-zero if the four indices are paired in couples or if all four are equal.
That is, non-zero expectation happens if
\begin{eqnarray*}
\mbox{(A)} &:& i_1=i_2 \neq j_1=j_2 \hspace{1.0cm} \mbox{ ($n^2-n$ terms)}   \\
\mbox{(B)} &:& i_1=j_1 \neq i_2=j_2 \hspace{1.0cm} \mbox{ ($n^2-n$ terms)}   \\
\mbox{(C)} &:& i_1=j_2 \neq i_2=j_1 \hspace{1.0cm} \mbox{ ($n^2-n$ terms)}   \\
\mbox{(D)} &:& i_1=i_2 =    j_1=j_2 \hspace{1.0cm} \mbox{ ($n$ terms)}   .
\end{eqnarray*}
For case~(A), let $i_1=i_2=k$ and let $j_1=j_2=\ell$, in which case the corresponding terms in eqn.~(\ref{eqn:basic}) become (notice that $\delta_k$ and $\delta_{\ell}$ are independent random variables since $k \neq \ell$):
%
\begin{eqnarray}
\nonumber \sum_{k,\ell=1,\ k \neq \ell}^n \matA_{k\ell}^2 \tilde{\x}_{k}^2\tilde{\x}_{\ell}^2
\E\left(\left(1-\delta_{k}\right)^2\left(1-\delta_{\ell}\right)^2\right)
&=\\ \nonumber \sum_{k,\ell=1,\ k \neq \ell}^n \matA_{k\ell}^2 \tilde{\x}_{k}^2\tilde{\x}_{\ell}^2
\left(\frac{1}{p_k}-1\right)\left(\frac{1}{p_\ell}-1\right)
&\leq\\ \nonumber \sum_{k=1}^n \tilde{\x}_{k}^2\left(\frac{1}{p_k}-1\right)\sum_{\ell=1}^n \tilde{\x}_{\ell}^2\left(\frac{1}{p_\ell}-1\right)
\label{eqn:caseA}&=\\ \left(\sum_{k=1}^n \tilde{\x}_{k}^2\left(\frac{1}{p_k}-1\right)\right)^2 \leq \frac{k^2}{s^2}.
\end{eqnarray}
%
In the first inequality we used $\abs{\matA_{k\ell}}\leq 1$ for all $k$ and $\ell$ and added extra positive terms (corresponding to $k = \ell$), which reinforce the inequality. The last inequality follows from eqn.~(\ref{eqn:pd11}).

For case~(B), let $i_1=j_1=k$ and let $i_2=j_2=\ell$, in which case the corresponding terms in eqn.~(\ref{eqn:basic}) become (notice that $\delta_k$ and $\delta_{\ell}$ are independent random variables since $k \neq \ell$):
\begin{eqnarray}
\nonumber \sum_{k,\ell=1,\ k \neq \ell}^n \matA_{kk} \matA_{\ell\ell} \tilde{\x}_{k}^2\tilde{\x}_{\ell}^2
\E\left(\left(1-\delta_{k}\right)^2\left(1-\delta_{\ell}\right)^2\right) &=\\ \nonumber
\sum_{k,\ell=1,\ k \neq \ell}^n \matA_{kk}\matA_{\ell\ell} \tilde{\x}_{k}^2\tilde{\x}_{\ell}^2
\left(\frac{1}{p_k}-1\right)\left(\frac{1}{p_\ell}-1\right)
\nonumber & \leq \\\nonumber \sum_{k=1}^n \tilde{\x}_{k}^2\left(\frac{1}{p_k}-1\right)\sum_{\ell=1}^n \tilde{\x}_{\ell}^2\left(\frac{1}{p_\ell}-1\right)
& =\\ \label{eqn:caseB}\left(\sum_{k=1}^n \tilde{\x}_{k}^2\left(\frac{1}{p_k}-1\right)\right)^2 \leq \frac{k^2}{s^2}.
\end{eqnarray}
In the first inequality we used $\matA_{kk}\leq 1$ for all $k$ and the fact that the diagonal entries of a symmetric positive definite matrix are non-negative, which allows us to add extra positive terms (corresponding to $k = \ell$) to reinforce the inequality. The remainder of the derivation is identical to case~(A).

For case~(C), let $i_1=j_2=k$ and let $i_2=j_1=\ell$, in which case the corresponding terms in eqn.~(\ref{eqn:basic}) become (notice that $\delta_k$ and $\delta_{\ell}$ are independent random variables since $k \neq \ell$):
%
%
$$\sum_{k,\ell=1,\ k \neq \ell}^n \matA_{k\ell} \matA_{\ell k} \tilde{\x}_{k}^2\tilde{\x}_{\ell}^2\E\left(\left(1-\delta_{k}\right)^2\left(1-\delta_{\ell}\right)^2\right)=$$
$$\sum_{k,\ell=1,\ k \neq \ell}^n \matA_{k\ell}^2 \tilde{\x}_{k}^2\tilde{\x}_{\ell}^2
\left(\frac{1}{p_k}-1\right)\left(\frac{1}{p_\ell}-1\right)\leq\label{eqn:caseC} \frac{k^2}{s^2}.$$
%
%
In the first equality we used the fact that $\matA$ is symmetric; the remainder of the derivation is identical to case~(A).

Finally, for case~(D), let $i_1=i_2=j_1=j_2=k$, in which case the corresponding terms in eqn.~(\ref{eqn:basic}) become:

\begin{eqnarray}
\nonumber \sum_{k=1}^n \matA_{kk}^2 \tilde{\x}_{k}^4
\E\left(1-\delta_{k}\right)^4 &=& \sum_{k=1}^n \matA_{kk}^2 \tilde{\x}_{k}^4
\left(\frac{6}{p_k}-\frac{4}{p_k^2}+\frac{1}{p_k^3}-3\right)\\
\nonumber &\leq& \sum_{k=1}^n \tilde{\x}_{k}^4
\left(\frac{6}{p_k}-\frac{4}{p_k^2}+\frac{1}{p_k^3}-3\right)\\
\nonumber &=& \sum_{k \in I^{<1}} \tilde{\x}_{k}^4
\left(\frac{6}{p_k}-\frac{4}{p_k^2}+\frac{1}{p_k^3}-3\right)\\
\nonumber &\leq& \sum_{k \in I^{<1}} \tilde{\x}_{k}^4
\left(\frac{6}{p_k}+\frac{1}{p_k^3}\right).
\end{eqnarray}
In the above derivation, we used $\abs{\matA_{kk}} \leq 1$. We also split the set of indices $\{1\ldots n\}$ in two subsets: the set $I^{=1}$ corresponding to indices $k$ such that $p_k = 1$ and the set $I^{<1}$ corresponding to indices $k$ such that $p_k < 1$. Note that for all $k \in I^{<1}$ it must be the case that
$p_k = s\abs{\tilde{\x}_i}/\VOneNorm{\tilde{\x}}.$ Thus,
\begin{eqnarray}
\nonumber \sum_{k=1}^n \matA_{kk}^2 \tilde{\x}_{k}^4
\E\left(1-\delta_{k}\right)^4 &\leq& \sum_{k \in I^{<1}} \tilde{\x}_{k}^4
\left(\frac{6\VOneNorm{\tilde{\x}}}{s\abs{\tilde{\x}_k}}+\frac{\VOneNorm{\tilde{\x}}^3}{s^3\abs{\tilde{\x}_k}^3}\right)\\
\nonumber & \leq& \sum_{k=1}^{n} \tilde{\x}_{k}^4
\left(\frac{6\VOneNorm{\tilde{\x}}}{s\abs{\tilde{\x}_k}}+\frac{\VOneNorm{\tilde{\x}}^3}{s^3\abs{\tilde{\x}_k}^3}\right)\\
\nonumber &=& \frac{6\VOneNorm{\tilde{\x}}}{s}\sum_{k=1}^{n} \abs{\tilde{\x}_{k}}^3
+\frac{\VOneNorm{\tilde{\x}}^3}{s^3}\sum_{k=1}^{n} \abs{\tilde{\x}_{k}}\\
\nonumber & =& \frac{6\VOneNorm{\tilde{\x}}\VThreeNorm{\tilde{\x}}^3}{s}
+\frac{\VOneNormQ{\tilde{\x}}}{s^3}.
\end{eqnarray}
Recall that $\VQNorm{\tilde{\x}} \leq \VPNorm{\tilde{\x}}$ for any $1\leq p\leq q\leq \infty$ and use $\TNorm{\tilde{\x}} \leq 1$ and $\VOneNorm{\tilde{\x}} \leq \sqrt{k}$ to get

\begin{eqnarray}
\nonumber\sum_{k=1}^n \matA_{kk}^2 \tilde{\x}_{k}^4
\E\left(1-\delta_{k}\right)^4 &\leq& \label{eqn:caseD}  \frac{6\VOneNorm{\tilde{\x}}\TNorm{\tilde{\x}}^3}{s}
+\frac{\VOneNormQ{\tilde{\x}}}{s^3}\\ &\leq& \frac{6\sqrt{k}}{s}
+\frac{k^2}{s^3}.
\end{eqnarray}
Combining all four cases (i.e., eqns.~(\ref{eqn:caseA}),~(\ref{eqn:caseB}),~(\ref{eqn:caseC}), and~(\ref{eqn:caseD})), we get
$$\E\left(\left(\tilde{\x}-\hat{\x}\right)^T\matA\left(\tilde{\x}-\hat{\x}\right)\right)^2 \leq 3k^2/s^2+k^2/s^3+6\sqrt{k}/s.$$
Using Markov's inequality and taking square roots concludes the proof of the lemma.
\end{proof}

We can now complete the proof of the lower bound for $\hat{\x}_{opt}^T\matA\hat{\x}_{opt}$.
First, combine Lemmas~\ref{lem:mainbound},~\ref{lem:firstbound}, and~\ref{lem:secondbound} to get
$$\abs{\tilde{\x}_{opt}^T\matA\tilde{\x}_{opt}-\hat{\x}_{opt} \matA\hat{\x}_{opt}} \leq 2\sqrt{8k/s} +
\left(24k^2/s^2+6k^2/s^3+54\sqrt{k}/s\right)^{1/2}.$$
Since each of Lemmas~\ref{lem:firstbound} and~\ref{lem:secondbound} fail with probability at most 1/8, it follows from the union bound that the above inequality holds with probability at least $1-2(1/8) = 3/4$. Therefore, setting $s = 200k/\epsilon^2$ guarantees (after some algebra) that with probability at least $3/4$,
$$\abs{\tilde{\x}_{opt}^T\matA\tilde{\x}_{opt}-\hat{\x}_{opt}^T\matA\hat{\x}_{opt}} \leq \epsilon.$$
Using the triangle inequality and Lemma~\ref{lem:relaxation} we get that with probability at least 3/4
$$\hat{\x}_{opt}^T \matA\hat{\x}_{opt} \geq \tilde{\x}_{opt}^T\matA\tilde{\x}_{opt} - \epsilon \geq \x_{opt}^T\matA\x_{opt} -\epsilon,$$
which concludes the proof of eqn.~(\ref{eqn:thm:main}).  
\section{Experiments}\label{sec:experiments}

We perform experiments on both real and synthetic datasets. We chose to compare our algorithm with the solution returned by the state-of-the-art \texttt{Spasm} toolbox of~\cite{KCLE12}, which implements the approach proposed in~\cite{ZHT12}. We also compare our solution with the simple \texttt{MaxComp} heuristic \cite{CJ95}: this method computes the top singular vector of matrix $\matA$ and returns a sparse vector by keeping the top $k$ largest components (in absolute value) and setting the remaining ones to zero.


\subsection{Experimental setup}\label{sec:setup}

Let $\matX \in \mathbb{R}^{m \times n}$ with $n \gg m$ denote the object-feature data matrix, where every column has zero mean, and recall that $\matA=\matX^T \matX$ in eqn.~\eqref{eqn:sparsePCArelaxed}. We use the following function to measure the quality of an approximate solution $\x\in\mathbb{R}^n$ to the sparse PCA problem:
\begin{eqnarray}\label{eqn:f_obj}
f(\x) = \x^T\matA \x/\TNorm{\matA}^2.
\end{eqnarray}
Notice that $ 0\leq f(\x) \leq 1$ for all $\x$ which satisfy $\TNorm{\x}\le 1$. The closer $f(\x)$ is to one
the more the vector $\x$ is parallel to the top singular vector of $\matA$, or, equivalently, the closer $f(\x)$ is to one the more $\x$
captures the variance of the matrix $\matA$ that corresponds to its top singular value. Our goal is to calculate sparse vectors $\x$ with $f(\x) \approx 1$.

Our approach first finds a stationary point of the optimization problem of eqn.~\eqref{eqn:sparsePCArelaxed} and then uses Algorithm~\ref{alg:sparsify} (with $s=k$) to obtain a sparse solution vector $\hat{\x}_{opt} \in \mathbb{R}^n$. We note that the chosen value of $s$ is much smaller than the $O\left(k/\epsilon^2\right)$ value stipulated by Theorem~\ref{thm:main}. Also, in practice, our choice of $s$ works very well and results in solutions that are comparable or better than competing methods in our data.

We note that in our use of \texttt{Spasm}, we used soft thresholding by setting the \texttt{STOP} parameter to $-\VZeroNorm{\hat{\x}_{opt}}$
and $\delta=-\infty$ (both \texttt{STOP} and $\delta$ are parameters of $\texttt{Spasm}$ toolbox). This implies that the solutions obtained by $\texttt{Spasm}$ and our approach will have the same number of non-zeros, thus making the comparison fair.  Similarly, for \texttt{MaxComp}, after computing the top singular vector of $\matA$, we select the $\VZeroNorm{\hat{\x}_{opt}}$ largest (in absolute value) coordinates to form the sparse solution. A final technical note is that the solutions obtained using either our method or \texttt{MaxComp} may result in vectors with non-unit Euclidean norms. In order to achieve a fair comparison in terms of eqn.~\eqref{eqn:f_obj}, there are two options. The first one (\textit{naive approach}) is to simply normalize the resulting vectors. However, a better approach (\textit{SVD-based}) is possible: given a sparse solution vector $\hat{\x}_{opt}$, we could keep the rows and columns of $\matA$ that correspond to the non-zero entries in $\hat{\x}_{opt}$ and compute the top singular vector of the induced matrix. Note that the induced matrix would be a $\VZeroNorm{\hat{\x}_{opt}} \times \VZeroNorm{\hat{\x}_{opt}}$ matrix and its top singular vector would be a $\VZeroNorm{\hat{\x}_{opt}}$-dimensional vector. Obviously, this latter vector would be a unit norm vector and it could be padded with zeros to derive a unit norm vector in $\mathbb{R}^n$ with the same sparsity pattern as $\hat{\x}_{opt}$. It is straight-forward to argue that this vector would improve the value of $f$ compared to the naive normalized vector $\hat{\x}_{opt}/\TNorm{\hat{\x}_{opt}}$. In our experimental evaluation, we will evaluate both the \textit{naive} and the \textit{SVD-based} normalization methods.

 We also compare the methods based on their wall-clock running times. All experiments were run on a Intel(R) Core(TM) i7-6700 machine running at 3.40GHz, with 16 GB of RAM.

In some of our experiments we will need to extract more than one sparse singular vectors. Towards that end, we can use a deflation approach (Algorithm~\ref{alg:deflation}) to construct more than one sparse singular vectors by first projecting the residual matrix into the space that is perpendicular to the top sparse singular vector and then computing the top sparse singular vector of the residual matrix. In Algorithm~\ref{alg:deflation}, \texttt{rspca} refers to the solution of the optimization problem of eqn.~\eqref{eqn:sparsePCArelaxed} followed by Algorithm~\ref{alg:sparsify}).
\begin{algorithm}[t]
\KwIn{ $\matX\in\mathbb{R}^{m\times n}$, integer $k$.}
\KwOut{${\matU}=\{\u_1,\dots, \u_k\}$, ${\matV}=\{\v_1,\dots, \v_k\}$.}
{${\matY}=\matX^\top$}
{$\v_1 = \texttt{rspca}(\matX)$ and $\u_1 = \texttt{rspca}({\matY})$}\\
\For{$i=2,\ldots, k$}{
$\matX = \matX - \matX \v_{i-1} \v_{i-1}^\top$ and ${\matY} = {\matY} - {\matY} \u_{i-1}\u_{i-1}^\top$\\
$\v_i = \texttt{rspca}(\matX)$ and $\u_i = \texttt{rspca}({\matY})$}
\caption{Computing $k$ sparse principal components}\label{alg:deflation}
\end{algorithm} 

\subsection{Data Sets}\label{subsec:datasets}

We used 22 matrices emerging from population genetics, namely the 22 matrices (one for each chromosome) that encode all autosomal genotypes that are available in both the Human Genome Diversity Panel~\cite{hapmap07} and the HAPMAP~\cite{Li08} project. Each of these matrices has approximately 2,500 samples (objects) described with respect to tens of thousands of features (Single Nucleotide Polymorphisms or SNPs); see~\cite{Paschou2010} for details. We also used a gene expression dataset (GSE10072, lung cancer gene expression data) from the NCBI Gene Expression Omnibus database. This dataset includes 107 samples (58 lung tumor cases and 49 normal lung controls) measured using 22,215 probes (features) from the GPL96 platform annotation table. Both the population genetics and the gene expression datasets are interesting in the context of sparse PCA beyond numerical evaluations, since the sparse components can be directly interpreted to identify small sets of SNPs or genes that capture the data variance.

Our synthetic dataset has been carefully designed in order to highlight a setting where the \texttt{MaxComp} heuristic will fail. More specifically, the absolute values of the entries of the largest singular vector of a matrix in this family of matrices is not a good indicator of the importance of the respective entry in a sparse PCA solution. Instead, the vector that emerges from the optimization problem of eqn.~(\ref{eqn:sparsePCArelaxed}) is a much better indicator. For a detailed description of the synthetic data generator see Section~\ref{sxn:appendix:synthetic} in the Appendix.

Our third data set comes from the field of text categorization and information retrieval. In such applications, documents are often represented as a ``bag of words'' and a vector space model is used. We use a subset of the Classic-3\footnote{\href{ftp://ftp.cs.cornell.edu/pub/smart/}{ftp://ftp.cs.cornell.edu/pub/smart/}} document collection, which we will call Classic-2. This subset consists of the CISI (Comit\'{e}s Interminist\'{e}riels pour la Soci\'{e}t\'{e} de l'Information) collection (1,460 information science abstracts) and the CRANFIELD collection (1,398 aeronautical systems abstracts). We created a document-by-term matrix using the Text-to-Matrix Generator (TMG)~\cite{TMG-incoll-06}\footnote{The matrix was created using the stoplist provided by TMG, the \texttt{tf-idf} weighting scheme, and document normalization. We removed numbers and alphanumerics and we didn't use stemming.}; the final matrix is a sparse $2,858 \times 12,427$ matrix with entries between zero and one, representing the weight of each term in the corresponding document.

\subsection{Results}

\begin{figure}[H]
\centering
	\subfloat[Synthetic data: $n=2^{12}$.]{%
			\label{fig:main:synthetic}%
		\includegraphics[width = 0.46\textwidth]{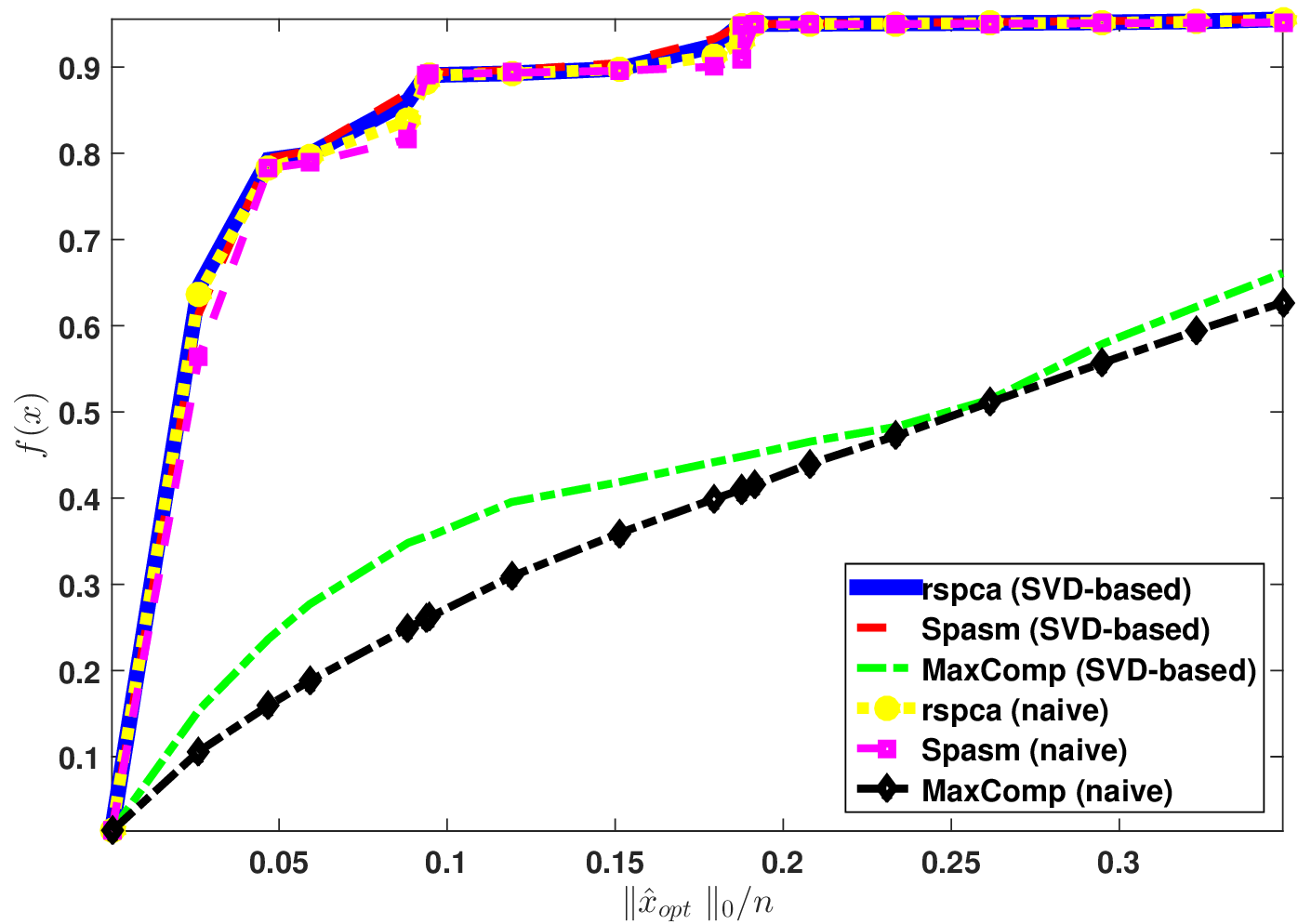}}		
         \quad
	\subfloat[HapMap+HGDP data (chromosome 1): $n=37493$.]{%
		\label{fig:main:SNP1}%
		\includegraphics[width = 0.46\textwidth]{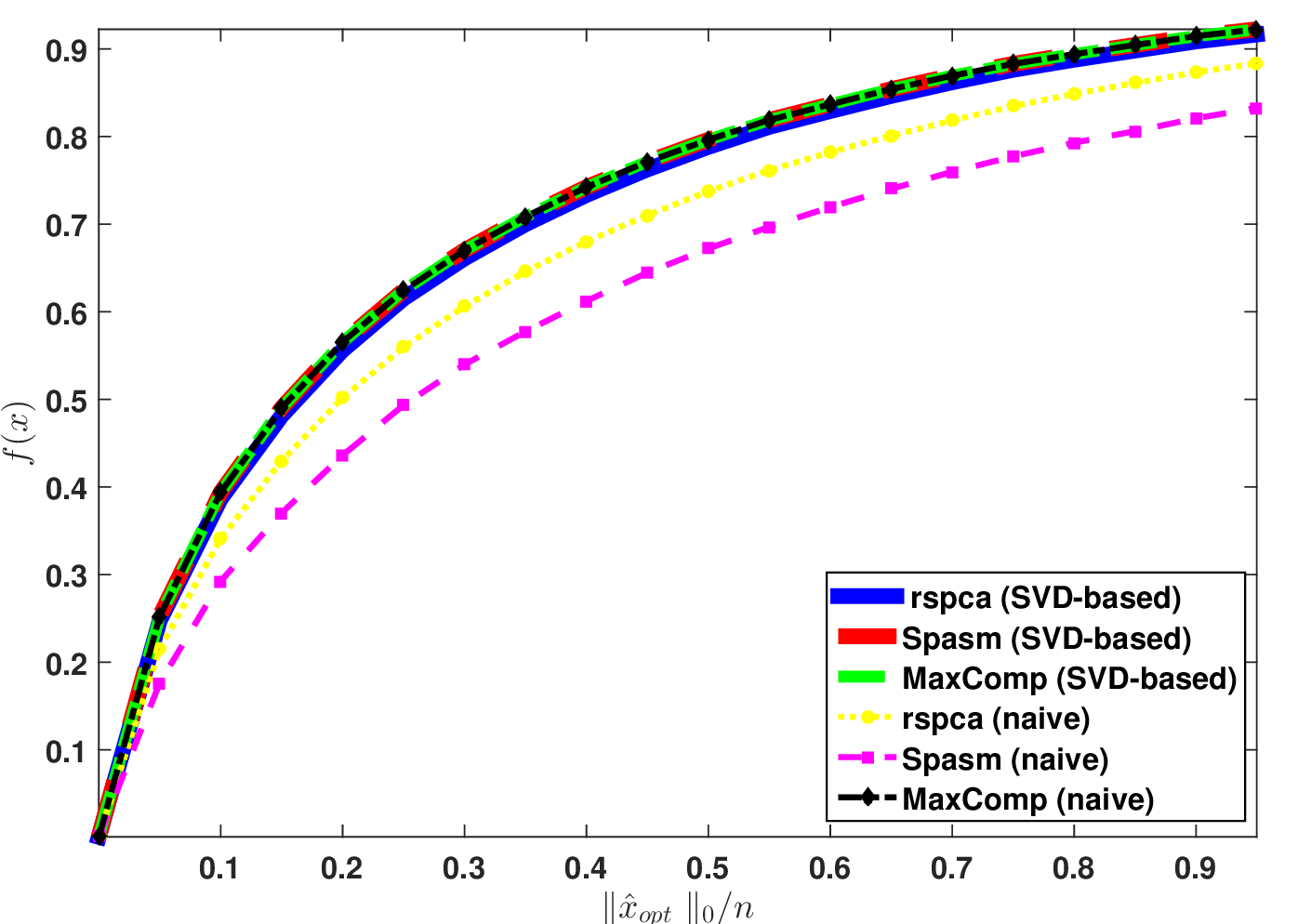}}
        \\
	\subfloat[HapMap+HGDP data (chromosome 2): $n=40844$.]{%
		\label{fig:main:SNP2}%
		\includegraphics[width = 0.46\textwidth]{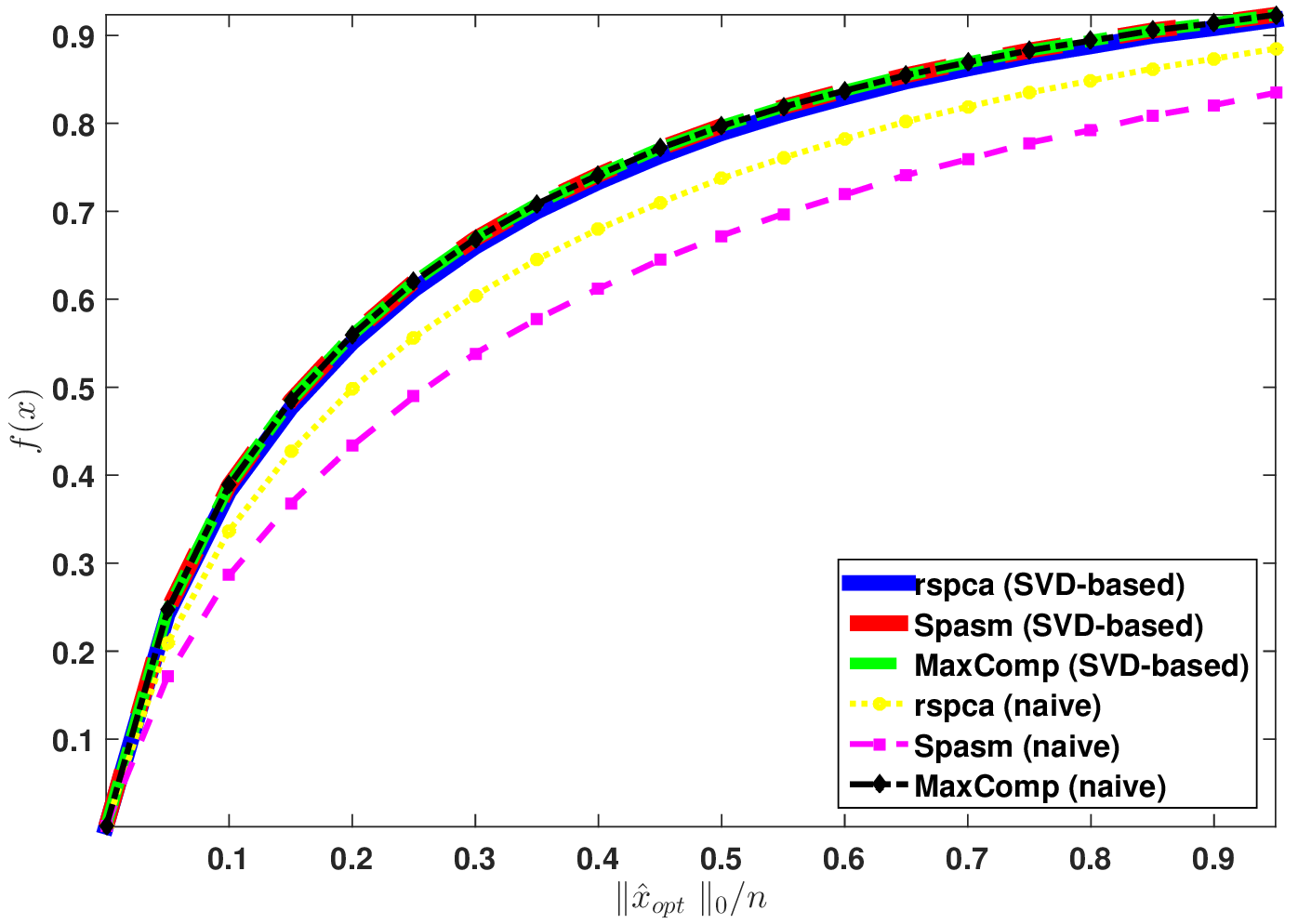}}
         \quad
	\subfloat[Gene Expression: $n=22,215$.]{%
		\label{fig:main:gene}%
		\includegraphics[width = 0.46\textwidth]{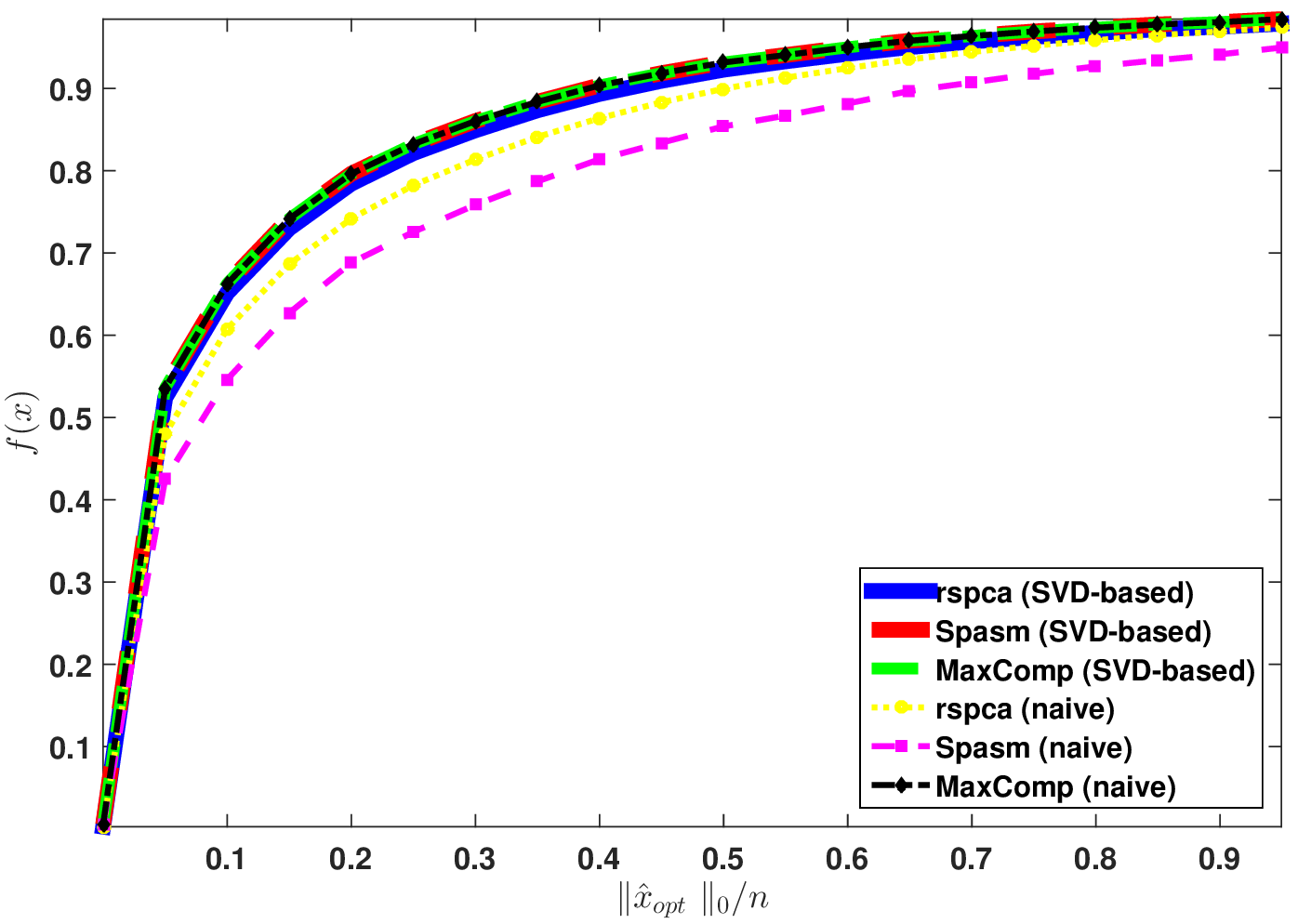}}
	\caption{$f(\x)$ vs. sparsity ratio $\VZeroNorm{\hat{\x}_{opt}}/n$ for various real and synthetic datasets.}
	\label{fig:main}%
\end{figure}

First we compare the performance of the different methods on a synthetic dataset, using the data generator which was described in Section~\ref{subsec:datasets} with
$m=2^7$, $n=2^{12}$, and $\theta\approx 0.27\pi$. In Figure~\ref{fig:main:synthetic} we plot (in the $y$-axis) the value of the performance ratio $f(\x)$ (as defined in eqn.~\eqref{eqn:f_obj}) for our method (\texttt{rspca}), the \texttt{Spasm} software, and the \texttt{MaxComp} heuristic. We also note that for each of the three methods, we use two different approaches to normalize the resulting sparse vector: the \textit{naive} and the \textit{SVD-based} ones (see the last paragraph in Section~\ref{sec:setup}). As a result, we have a total of six possible methods to create and normalize a sparse vector for sparse PCA. The $x$-axis shows the sparsity ratio of the resulting vector, namely $\VZeroNorm{\hat{\x}_{opt}}/n$. We remark that all six methods produce sparse vectors with exactly the same sparsity in order to have a fair comparison. Notice that in Figure~\ref{fig:main:synthetic}, the \texttt{MaxComp} heuristic has worse performance when compared to either our approach or to \texttt{Spasm}: this is expected, since we constructed this family of matrices in order to precisely guarantee that the largest components of the top singular vector would not be good elements to retain in the construction of a sparse vector. To further visualize this, we look at the
sparse vectors, returned by the different methods, in Figure~\ref{fig:syntheticSingVec} in the Appendix. In this figure, we present the resulting sparse vectors for each of the three methods (normalization, obviously, is not relevant in Figure~\ref{fig:syntheticSingVec}) for a particular choice of the sparsity ratio ($\VZeroNorm{\hat{\x}_{opt}}/n \approx 0.1$). Notice that \texttt{MaxComp} fails to capture the right sparsity pattern, whereas our method and \texttt{Spasm} succeed.

In the second experiment, we present the performance of the different methods on the real datasets described in Section~\ref{subsec:datasets}.
The results are shown in Figures~\ref{fig:main:SNP1},~\ref{fig:main:SNP2}, and~\ref{fig:main:gene}. (We only show results for the first two chromosomes of the joint HapMap and HGDP datasets; the other 20 chromosomes behave very similarly and are shown in Figures~\ref{fig:SNPs_1},~\ref{fig:SNPs_2},~\ref{fig:SNPs_3}, and~\ref{fig:SNPs_4} in the Appendix). We note that in the population genetics data, our method has approximately the same or better performance compared to both
\texttt{MaxComp} and \texttt{Spasm}. Not surprisingly, the \textit{naive} normalization approach is consistently worse than the \textit{SVD-based} one. It is worth noting that our \textit{SVD-based} normalization approach easily improves the output of \texttt{Spasm}. This is because \texttt{Spasm} does detect the correct sparsity pattern but fails to compute the appropriate coefficients of the resulting sparse vectors.

\begin{figure}[H]
	\centering
	\subfloat[HapMap+HGDP data (chromosome 1): $n=37493$.]{%
		\label{fig:main2:SNP1}%
		\includegraphics[width = 0.46\textwidth]{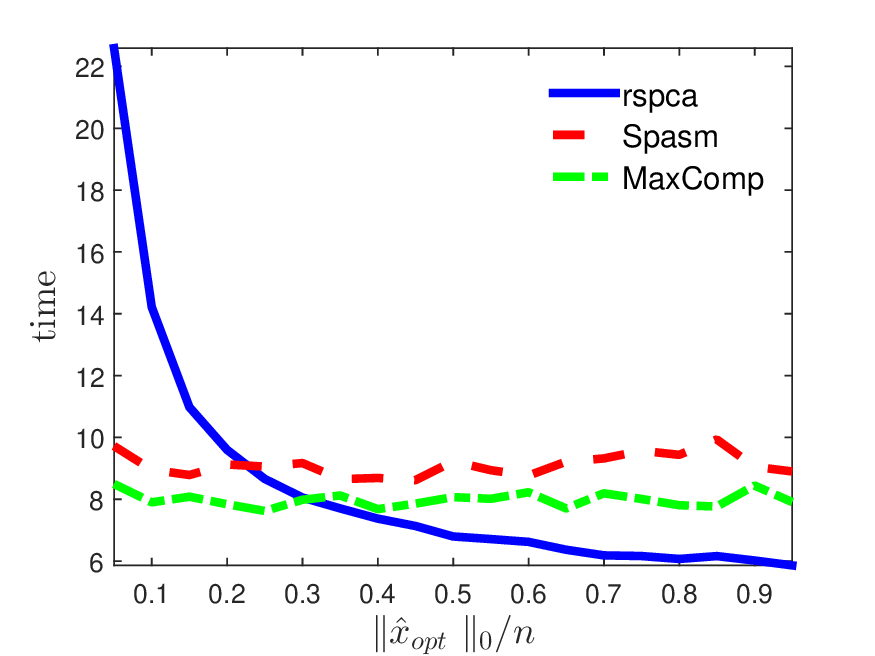}}		
	\quad
	\subfloat[HapMap+HGDP data (chromosome 2): $n=40844$.]{%
		\label{fig:main2:SNP2}%
		\includegraphics[width = 0.46\textwidth]{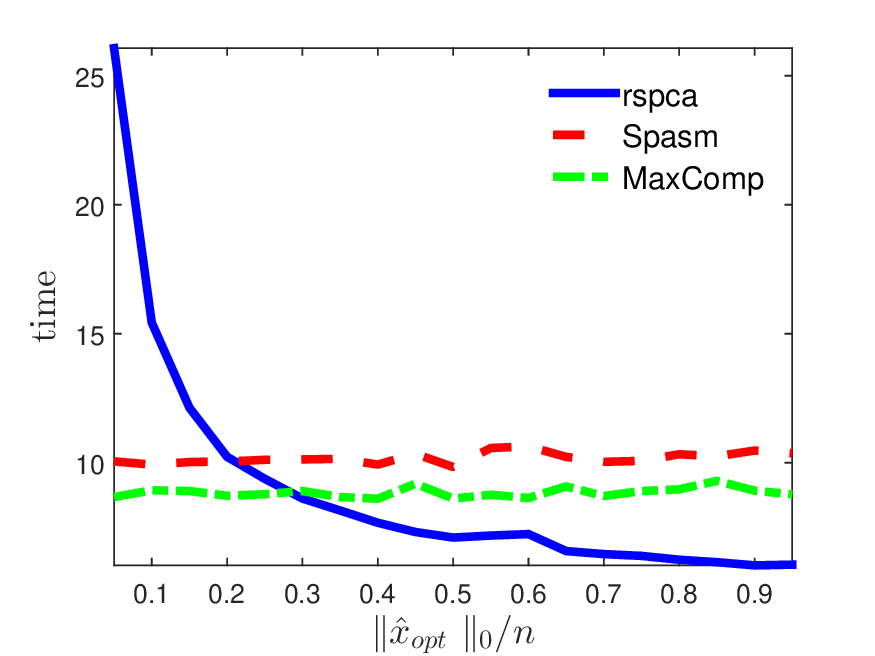}}
	\\
	\subfloat[HapMap+HGDP data (chromosome 3): $n=34258$.]{%
		\label{fig:main2:SNP3}%
		\includegraphics[width = 0.46\textwidth]{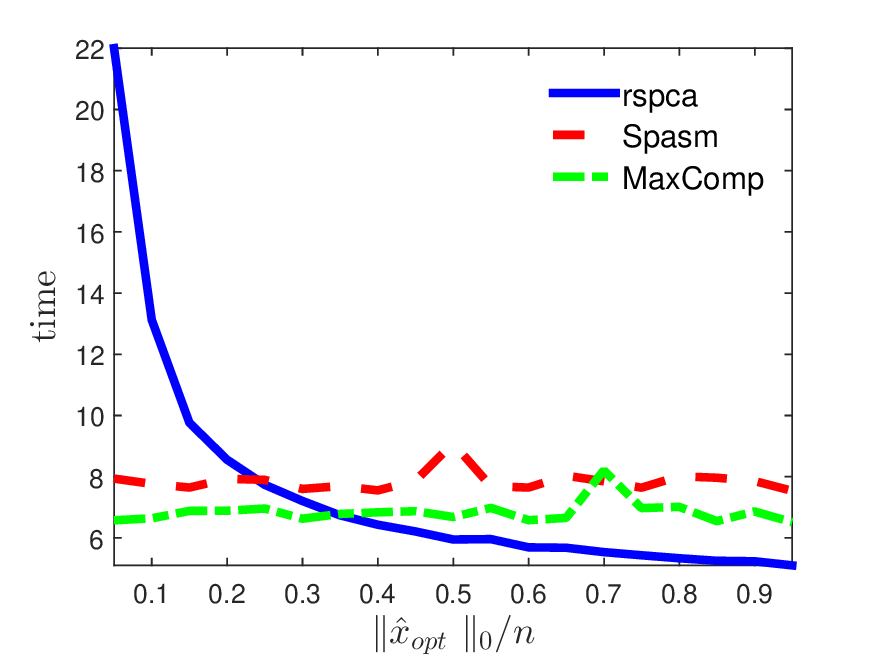}}
	\quad
	\subfloat[HapMap+HGDP data (chromosome 4): $n=30328$.]{%
		\label{fig:main2:SNP4}%
		\includegraphics[width = 0.46\textwidth]{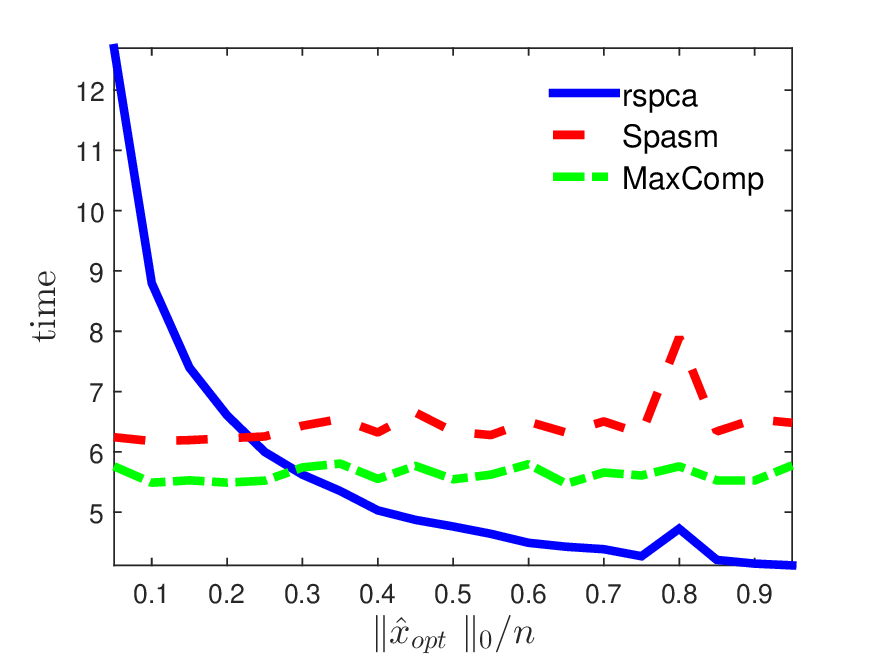}}
	\caption{Running time vs. sparsity ratio $\VZeroNorm{\hat{\x}_{opt}}/n$ for real data. }
	\label{fig:main2}%
\end{figure}

In Figure~\ref{fig:main2} (and Figures~\ref{fig:SNPs_time_1}, \ref{fig:SNPs_time_2}, \ref{fig:SNPs_time_3} at the Appendix) we plot (in the $y$-axis) the running time for our method (\texttt{rspca}), the \texttt{Spasm} software, and the \texttt{MaxComp} heuristic. We also note that for each of the methods we use the \textit{naive} approach (see the last paragraph in Section~\ref{sec:setup}). The $x$-axis shows the sparsity ratio of the resulting vector.
Our algorithm presents a variable time performance that tends to decrease as the sparsity ratio increases.
This is an artifact of the projected gradient ascent algorithm~\cite{JNRS2010} that we used to find stationary points for the problem of eqn. \eqref{eqn:sparsePCArelaxed}.
The smaller $k$ is in~\eqref{eqn:sparsePCArelaxed}, the harder the problem of \eqref{eqn:sparsePCArelaxed} becomes for projected gradient ascent.
In practice we observed that the smaller $k$ is the more iterations were needed for projected gradient ascent to converge since the algorithm balances between
the standard PCA objective and the sparsity constraint. 
Projected gradient ascent, which solves problem \eqref{eqn:sparsePCArelaxed} tends to be slower for small $k$ than the algorithm implemented in \texttt{Spasm} for solving the problem in~\cite{ZHT12}. 
We anticipate that this issue can be solved by using more sophisticated optimization algorithms, we leave this for future work.
Notice that \texttt{MaxComp} is the fastest method, but based on Figure~\ref{fig:main:synthetic} there are cases where it fails to find a good solution.
The running time of \texttt{MaxComp} and \texttt{Spasm} appears to be constant. This is because both methods require calculation of the first principal component which dominates the computational cost.
In particular, \texttt{MaxComp} is computing the first principal component which is then sparsified, while \texttt{Spasm} is computing the first principal component to initialize the algorithm that solves the problem in~\cite{ZHT12}.
All algorithms take into account the sparsity of the data. 

Finally, we evaluate the performance of our algorithm in a text mining application, using the Classic-2 document collection. First, we run Algorithm~\ref{alg:deflation} to obtain two sparse singular vectors and we use the number of their non-zero entries to compute two sparse singular vectors for \texttt{MaxComp} and \texttt{Spasm}\footnote{\texttt{Spasm} does not support sparse input matrices.}. This way we can guarantee the same sparsity levels for all three pairs of singular vectors. We repeat this procedure for 8 different values of $k$ (sparsity parameter in eqn.~\eqref{eqn:sparsePCA}). Table~\ref{tab:variance} summarizes, for each $k$, the variance and the sparsity (in parenthesis) captured by the top two principal components using PCA, randomized sPCA (\texttt{rspca}), \texttt{MaxComp} heuristic and \texttt{Spasm} or solving eqn.~\eqref{eqn:sparsePCArelaxed} (cvx).

\begin{table}[H]
	\centering
	\caption{Variance and sparsity captured by the principal components. PCA results in dense principal components, while \texttt{Spasm} and \texttt{MaxComp} share the same sparsity with \texttt{rspca}.  }{\vspace{0.2cm}
		\begin{tabular}{|l|c||c||c||c||c||c|}
			\cline{2-7}
			\multicolumn{1}{c|}{}          & \multicolumn{1}{c||}{\textbf{$k$}} &  \multicolumn{1}{c||}{\textbf{pca}} & \multicolumn{1}{c||}{\textbf{cvx}} & \multicolumn{1}{c||}{\textbf{\texttt{rspca}}} &\multicolumn{1}{c||}{\textbf{\texttt{MaxComp}}} & \multicolumn{1}{c|}{\textbf{\texttt{Spasm}}} \\\hline
			{\bf Top Principal Comp.}     & \multirow{2}{*}{100}  & 0.4351   & 0.3077 (99\%)  & 0.2942 (99\%)  & 0.1955   & 0.2768 \\\cline{1-1}\cline{3-7}
			{\bf Top two Principal Comp.} &                       & 0.6802   & 0.4897 (99\%)  & 0.4680 (99\%)  & 0.3391   & 0.4227 \\\hline
			{\bf Top Principal Comp.}     & \multirow{2}{*}{500}  & 0.4351   & 0.3880 (95\%)  & 0.3728 (98\%)  & 0.3353   & 0.3601	\\\cline{1-1}\cline{3-7}
			{\bf Top two Principal Comp.} &                       & 0.6802   & 0.6073 (95\%)  & 0.5864 (98\%)  & 0.5399   & 0.5701 \\\hline
			{\bf Top Principal Comp.}     & \multirow{2}{*}{1000} & 0.4351   & 0.4136 (90\%)  & 0.4005 (95\%)  & 0.3825   & 0.3912	\\\cline{1-1}\cline{3-7}
			{\bf Top two Principal Comp.} &                       & 0.6802   & 0.6486 (90\%)  & 0.6294 (95\%)  & 0.6074   & 0.6163 \\\hline
			{\bf Top Principal Comp.}     & \multirow{2}{*}{1500} & 0.4351   & 0.4242 (84\%)  & 0.4120 (93\%)  & 0.4013   & 0.4039	\\\cline{1-1}\cline{3-7}
			{\bf Top two Principal Comp.} &                       & 0.6802   & 0.6649 (82\%)  & 0.6470 (93\%)  & 0.6342   & 0.6361 \\\hline
			{\bf Top Principal Comp.}     & \multirow{2}{*}{2000} & 0.4351   & 0.4295 (75\%)  & 0.4190 (91\%)  & 0.4133   & 0.4131	\\\cline{1-1}\cline{3-7}
			{\bf Top two Principal Comp.} &                       & 0.6802   & 0.6730 (70\%)  & 0.6572 (91\%)  & 0.6503   & 0.6491 \\\hline
			{\bf Top Principal Comp.}     & \multirow{2}{*}{4000} & 0.4351   & 0.4350 (6\%)   & 0.4278 (81\%)  & 0.4284   & 0.4271	\\\cline{1-1}\cline{3-7}
			{\bf Top two Principal Comp.} &                       & 0.6802   & 0.6801 (3\%)   & 0.6700 (81\%)  & 0.6710   & 0.6690 \\\hline
			{\bf Top Principal Comp.}     & \multirow{2}{*}{8000} & 0.4351   & 0.4351 (0\%)   & 0.4324 (68\%)  & 0.4326   & 0.4316	\\\cline{1-1}\cline{3-7}
			{\bf Top two Principal Comp.} &                       & 0.6802   & 0.6802 (0\%)   & 0.6764 (69\%)  & 0.6768   & 0.6752 \\\hline
			{\bf Top Principal Comp.}     & \multirow{2}{*}{10500}& 0.4351   & 0.4351 (0\%)   & 0.4332 (63\%)  & 0.4333   & 0.4324 \\\cline{1-1}\cline{3-7}
			{\bf Top two Principal Comp.} &                       & 0.6802   & 0.6802 (0\%)   & 0.6776 (64\%)  & 0.6778   & 0.6764 \\\hline
		\end{tabular}}
		\label{tab:variance}
	\end{table}
	
	It seems that \texttt{Spasm} and \texttt{MaxComp} capture less variance than the randomized sPCA. Furthermore, the variance captured by randomized sPCA is constantly close to the one captured by the solution of eqn.~\eqref{eqn:sparsePCArelaxed}, but with a sparser component as Table~\ref{tab:variance} indicates.
	
	
	Table~\ref{tab:terms} summarizes the terms with non-zero weights in randomized sPCA principal components with sparsity parameter $k = 100$. The terms are ranked in descending order with respect to their weights. Notice that the first principal component reveals terms that appear more often in the CRANFIELD collection while the second principal component reveals terms that appear mostly in the CISI collection. CRANFIELD's terms are more singular than these of CISI's and they tend to dominate the singular vectors since they tend to appear more in the documents associated with the CRANFIELD collection than in the entire CLASSIC-2 collection (e.g., the word ``boundary'' has one appearance in CISI and 459 in CRANFIELD). The exact opposite happens for terms in CISI: a significant amount of these terms appear with high weights in both collections (e.g., the word ``information'' has 664 appearances in CISI and 44 in CRANFIELD). This indicates that these terms will appear in singular vectors that do not separate the two collections.
	\newpage
	
	\begin{table}[H]
		\scriptsize
		\centering
		\caption{Non-zero terms of the randomized sPCA when $k=100$}{\vspace{0.2cm}
			\begin{tabular}{||c||c||}
				\hline
				{\bf 1st Principal Component} & {\bf 2nd Principal Component}\\
				\hline
				boundary   		&                 information  \\\hline
				layer      		&                 library       \\\hline
				heat       		&                 retrieval    \\\hline
				flow       		&                 systems      \\\hline
				transfer   		&                 system       \\\hline
				laminar    		&                 scientific   \\\hline
				plate      		&                 science      \\\hline
				shock      		&                 libraries    \\\hline
				turbulent  		&                 research      \\\hline
				hypersonic 		&                 layer        \\\hline
				pressure   		&                 services      \\\hline
				temperature		&                 boundary       \\\hline
				wall       		&                 indexing     \\\hline
				flat        	&                 search        \\\hline
				surface      	&                 literature    \\\hline
				mach          	&                 heat           \\\hline
				friction      	&                 university    \\\hline
				compressible  	&                 documents     \\\hline
				velocity      	&                 users         \\\hline
				reynolds      	&                 classification \\\hline
				stagnation    	&                 technology  \\\hline
				skin          	&               \multicolumn{1}{|c}{}\\\cline{1-1}
				number        	&               \multicolumn{1}{|c}{}\\\cline{1-1}
				stream        	&               \multicolumn{1}{|c}{}\\\cline{1-1}
				equations     	&               \multicolumn{1}{|c}{}\\\cline{1-1}
				transition    	&               \multicolumn{1}{|c}{}\\\cline{1-1}
				solution      	&               \multicolumn{1}{|c}{}\\\cline{1-1}
				viscous         &               \multicolumn{1}{|c}{}\\\cline{1-1}
				layers          &               \multicolumn{1}{|c}{}\\\cline{1-1}
				separation      &               \multicolumn{1}{|c}{}\\\cline{1-1}
				solutions       &               \multicolumn{1}{|c}{}\\\cline{1-1}
				gradient        &              \multicolumn{1}{|c}{}\\\cline{1-1}
				injection       &              \multicolumn{1}{|c}{}\\\cline{1-1}
				dimensional     &              \multicolumn{1}{|c}{}\\\cline{1-1}
				incompressible  &              \multicolumn{1}{|c}{}\\\cline{1-1}
				fluid           &              \multicolumn{1}{|c}{}\\\cline{1-1}
				edge            &              \multicolumn{1}{|c}{}\\\cline{1-1}
				cylinder        &               \multicolumn{1}{|c}{}\\\cline{1-1}
				shear           &               \multicolumn{1}{|c}{}\\\cline{1-1}
				point           &              \multicolumn{1}{|c}{}\\\cline{1-1}
				interaction     &              \multicolumn{1}{|c}{}\\\cline{1-1}
				numbers         &              \multicolumn{1}{|c}{}\\\cline{1-1}
				body            &             \multicolumn{1}{|c}{}\\\cline{1-1}
				wave            &             \multicolumn{1}{|c}{}\\\cline{1-1}
				method          &             \multicolumn{1}{|c}{}\\\cline{1-1}
				leading         &             \multicolumn{1}{|c}{}\\\cline{1-1}
				region          &             \multicolumn{1}{|c}{}\\\cline{1-1}
				bodies          &             \multicolumn{1}{|c}{}\\\cline{1-1}
				supersonic      &             \multicolumn{1}{|c}{}\\\cline{1-1}
				gas             &             \multicolumn{1}{|c}{}\\\cline{1-1}
				measurements    &             \multicolumn{1}{|c}{}\\\cline{1-1}
				approximate     &             \multicolumn{1}{|c}{}\\\cline{1-1}
				local           &             \multicolumn{1}{|c}{}\\\cline{1-1}
				equation        &             \multicolumn{1}{|c}{}\\\cline{1-1}
				case            &             \multicolumn{1}{|c}{}\\\cline{1-1}
				constant        &            \multicolumn{1}{|c}{}\\\cline{1-1}			
			\end{tabular}}
			\label{tab:terms}
		\end{table}

\section{Open problems}

From a theoretical perspective, it would be interesting to explore whether other relaxations of the sparse PCA problem of eqn.~(\ref{eqn:sparsePCA}), combined with randomized rounding procedures, could improve our error bounds in Theorem~\ref{thm:main}.
It would also be interesting to formally analyze the deflation algorithm that computes more than one sparse singular vectors in a randomized manner (Algorithm~\ref{alg:deflation}). Finally, from a complexity theory perspective, we are not aware of any inapproximability results for the sparse PCA problem; to the best of our knowledge, it is not known whether relative error approximations are possible (without any assumptions on the input matrix). 
\newpage
\appendix
\section*{APPENDIX}
\setcounter{section}{1}
\subsection{Synthetic data generator}\label{sxn:appendix:synthetic}

In order to generate our synthetic dataset, we generated the following matrix:
$$
\matX = \matU\mat\Sigma \matV^\intercal + \matE_\sigma,
$$
where $\matU\in\mathbb{R}^{m\times m}$ and $\matV\in\mathbb{R}^{n\times n}$ are orthonormal. The matrix $\mat\Sigma\in\mathbb{R}^{m\times n}$ has
$m$ distinct singular values $\sigma_i$ in its diagonal and the matrix $\matE_{\sigma}\in\mathbb{R}^{m\times n}$ is a noise matrix
parameterized by $\sigma>0$.

\begin{figure}[H]
\centering
\includegraphics[scale=0.37]{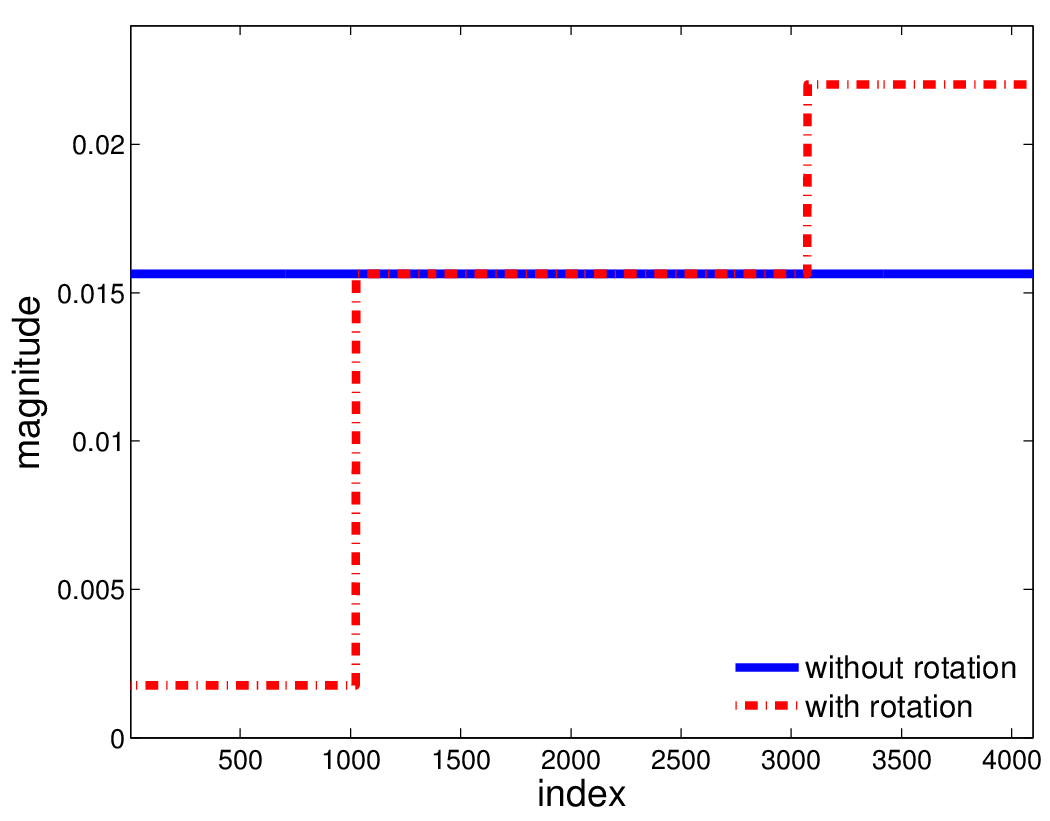}
\caption
{
The red dashed line corresponds to the sorted absolute values of the components
of the first column of matrix $\matV$. Similarly, the blue line corresponds to the first
column of $\tilde{\matV}$.
}
\label{fig:rotationComparison}
\end{figure}

We set $\matU$ to be a Hadamard matrix with normalized columns; we set $\mat\Sigma$ to have entries $\sigma_1=100$ and $\sigma_i=1/e^i$ for all $i=2,\ldots,m$. The entries of the matrix $E_\sigma$ follow a zero-mean normal distribution with standard deviation $\sigma=10^{-3}$.
We now describe how to construct the matrix $\matV$:
we set $\matV=\matG_d(\theta)\tilde{\matV}$, where $\tilde{\matV}$ is also a Hadamard matrix with normalized columns. Here $\matG_n(\theta)$ is a composition of Givens rotations. In particular, $\matG_n(\theta)$ is a composition of $n/4$ Givens rotations with the same angle $\theta$ for every rotation. More precisely,
let $\matG(i,j,\theta)\in\mathbb{R}^{n \times n}$ be a Givens rotation matrix, which rotates the plane $i$-$j$ by an angle $\theta$:
\begin{equation*}
\matG(i,j,\theta) =
\begin{bmatrix}   1   & \cdots &    0       & \cdots &    0      & \cdots &    0   \\
                      \vdots & \ddots & \vdots  &           & \vdots &            & \vdots \\
                         0      & \cdots &    c_1      & \cdots &    -c_2     & \cdots &    0   \\
                      \vdots &            & \vdots & \ddots & \vdots &            & \vdots \\
                         0      & \cdots &   c_2       & \cdots &    c_1      & \cdots &    0   \\
                      \vdots &            & \vdots &            & \vdots & \ddots & \vdots \\
                         0     & \cdots &    0       & \cdots &    0      & \cdots &    1
       \end{bmatrix},
\end{equation*}
where $i,j\in\{1,\ldots,n\}$, $c_1 = \cos \theta$ and $c_2=\sin \theta$. We define the composition as follows:
$$
\matG_n(\theta) = \matG(i_1,j_1,\theta)\matG(i_2,j_2,\theta)\cdots\cdots ,\matG(i_k,j_k,\theta),\cdots,\matG(i_{n/4},j_{n/4},\theta)
$$
with
$$
i_k = \frac{n}{2} + 2k-1, \quad j_k = \frac{n}{2} + 2k \quad \mbox{for } \ k=1,\ldots,\frac{n}{4}.
$$
The matrix $\matG_n(\theta)$ rotates (in a pairwise manner) the bottom $n/2$ components of the columns of $\tilde{\matV}$. Since the Hadamard matrix has entries equal to +1 or -1 (up to normalization), we will pick a value of $\theta$ that guarantees that, after rotation, $n/4$ components of the columns of $\tilde{\matV}$ will be almost zero. Thus, the resulting matrices will have about a quarter of components set at a large value, a quarter of their components set at roughly zero, and the rest set at a moderate value. For example, let $n=2^{12}$ and $\theta\approx 0.27\pi$; then, the difference between the first column of matrix $\matV$ and $\tilde{\matV}$ is presented in Figure~\ref{fig:rotationComparison}, where we plotted the (sorted) absolute values of the components of the first column of the matrices $\matV$ and $\tilde{\matV}$.

%

\begin{figure}[H]
\centering
	\subfloat[Actual eigenvector]{%
			\label{synthetic_cs_real}%
		\includegraphics[width = 0.46\textwidth]{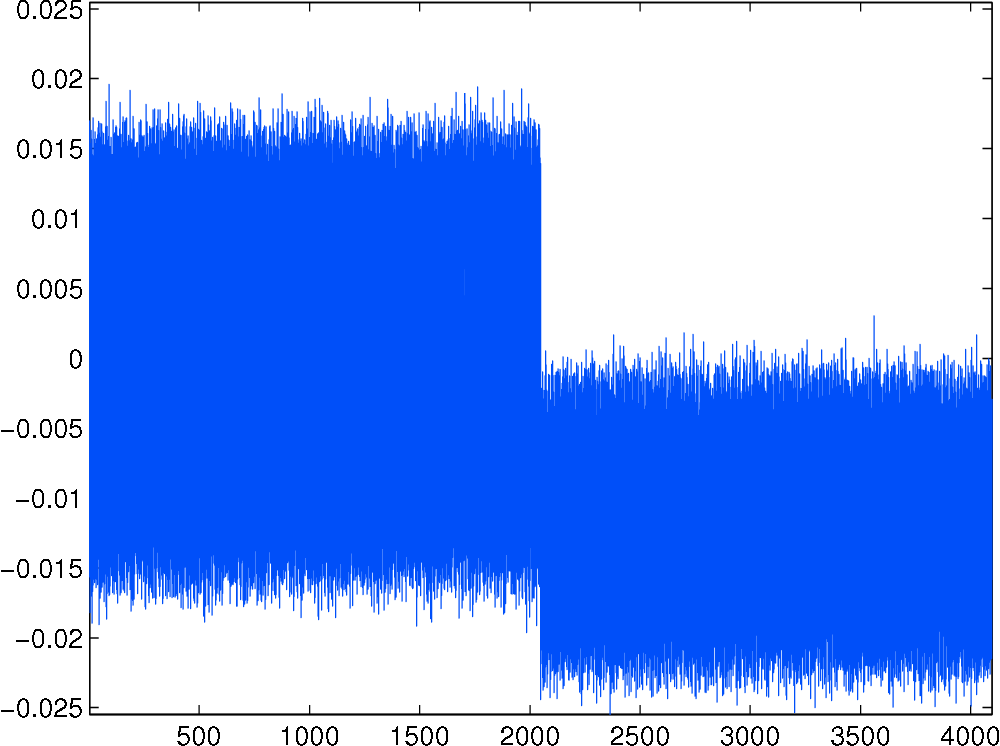}}		
         \quad
	\subfloat[\texttt{rspca}]{%
		\label{synthetic_cs_our}%
		\includegraphics[width = 0.46\textwidth]{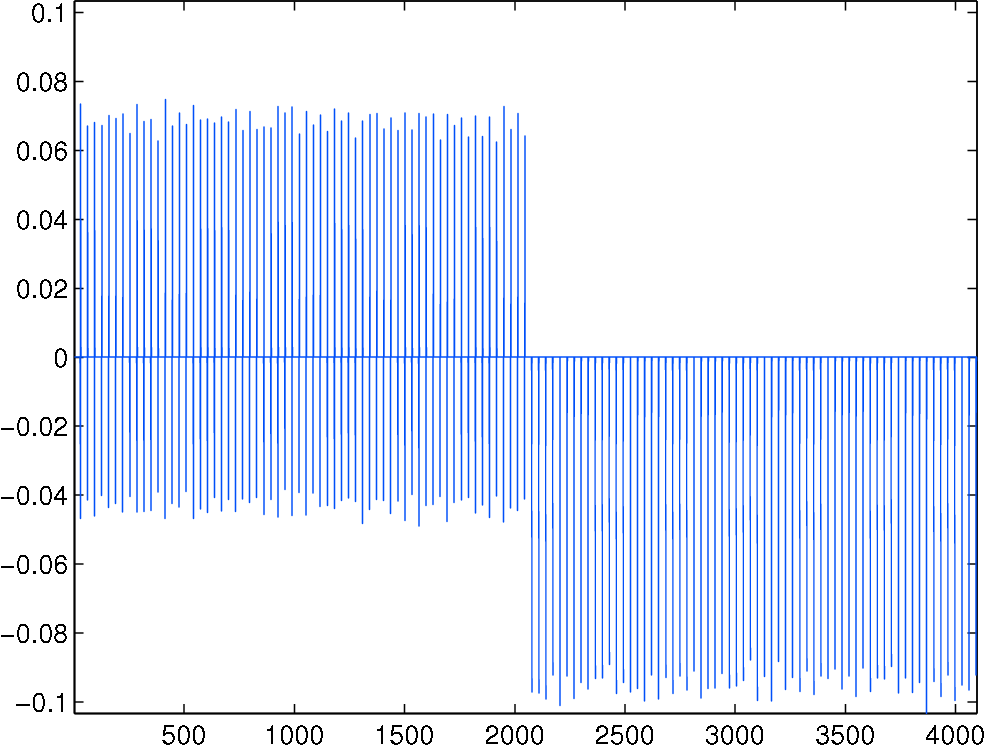}}
  	\\
  	\subfloat[\texttt{Spasm}]{%
		\label{synthetic_cs_spasm}%
		\includegraphics[width = 0.46\textwidth]{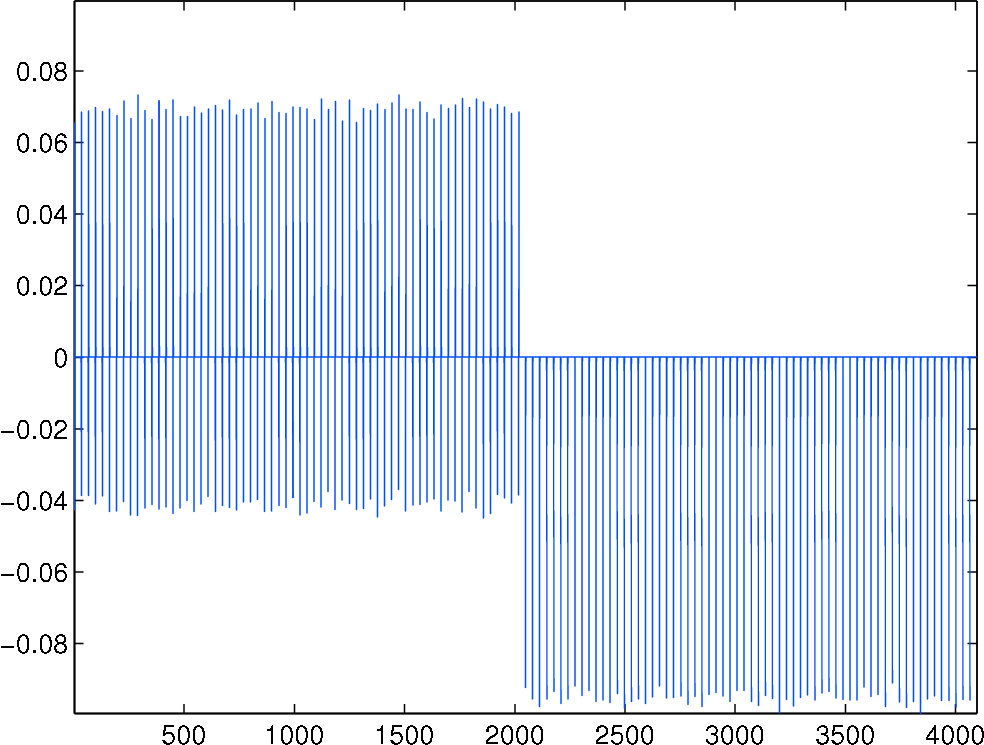}}
    \quad
	\subfloat[\texttt{MaxComp}]{%
		\label{synthetic_cs_mxreal}%
		\includegraphics[width = 0.46\textwidth]{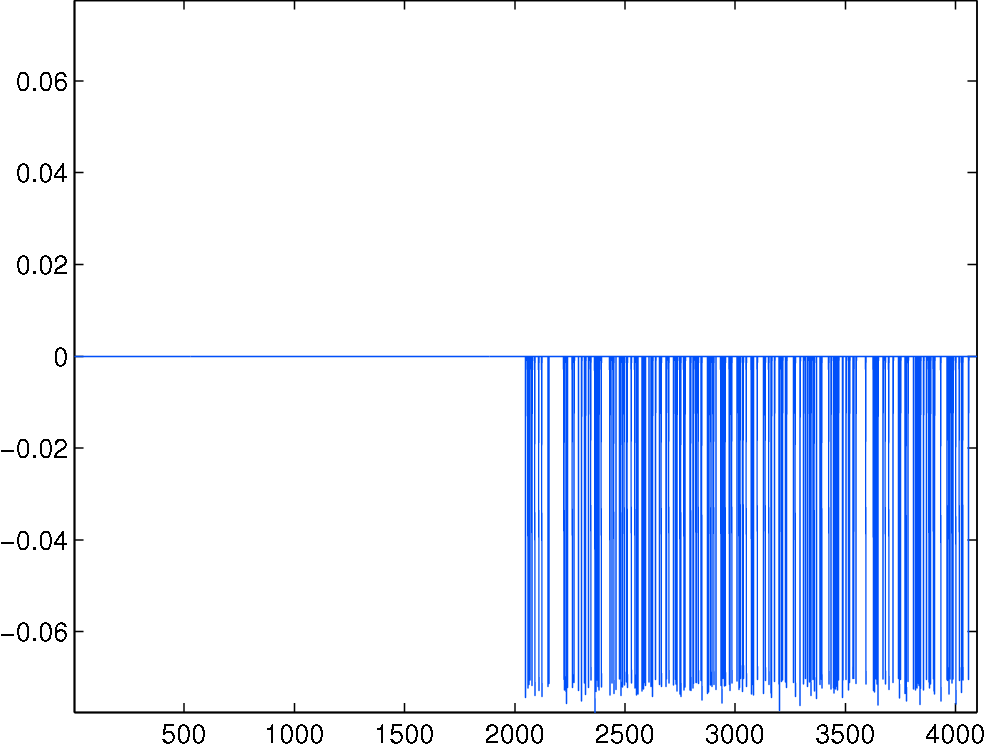}}
	\caption{Sparse PCA solutions for a synthetic experiment.}
	\label{fig:syntheticSingVec}
\end{figure}


\begin{figure}[H]
\centering
	\subfloat[HapMap+HGDP data (chromosome 3): $n=34258$.]{%
			\label{fig:snp3}%
		\includegraphics[width = 0.46\textwidth]{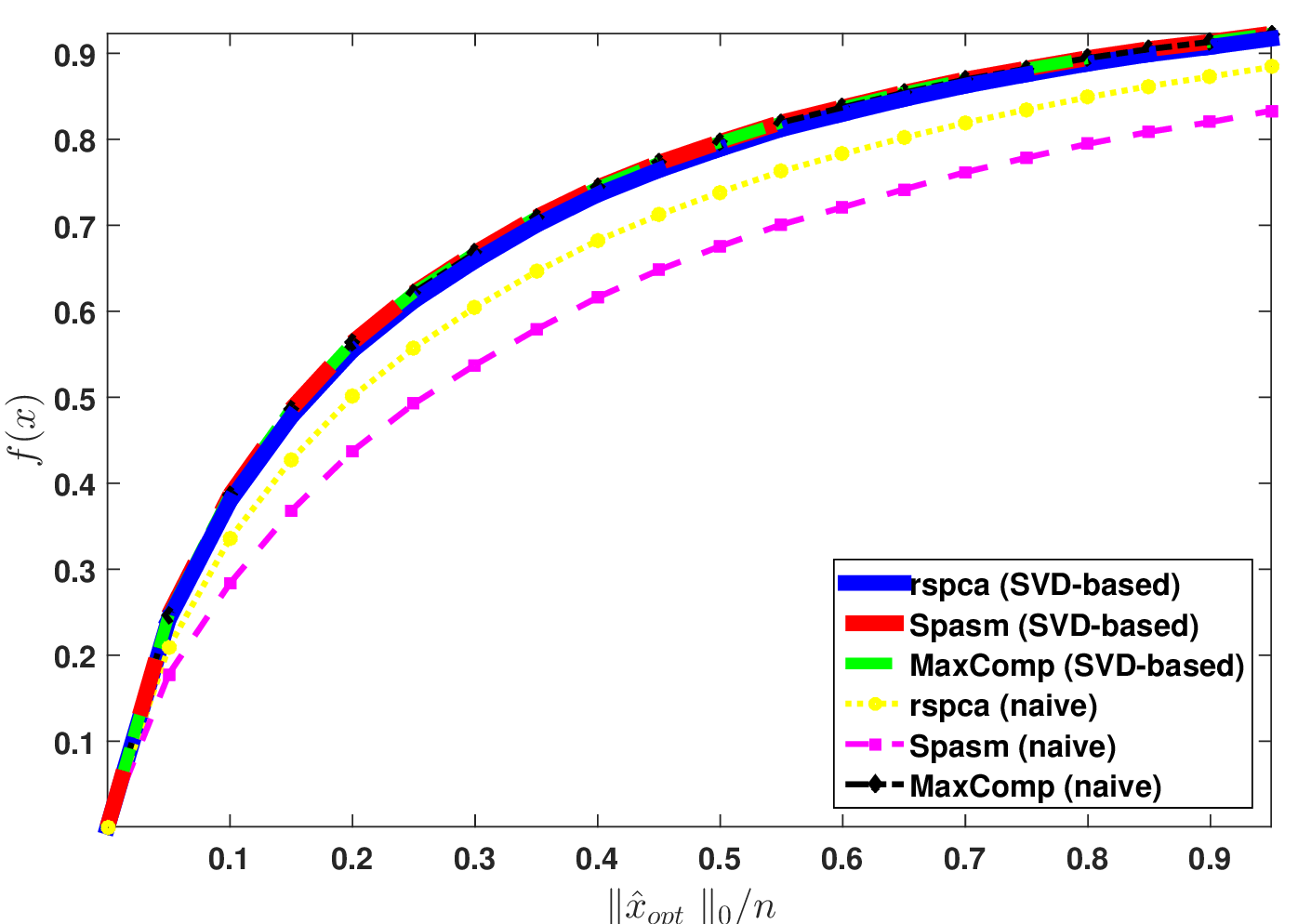}}		
        \quad
	\subfloat[HapMap+HGDP data (chromosome 4): $n=30328$.]{%
			\label{fig:snp4}%
		\includegraphics[width = 0.46\textwidth]{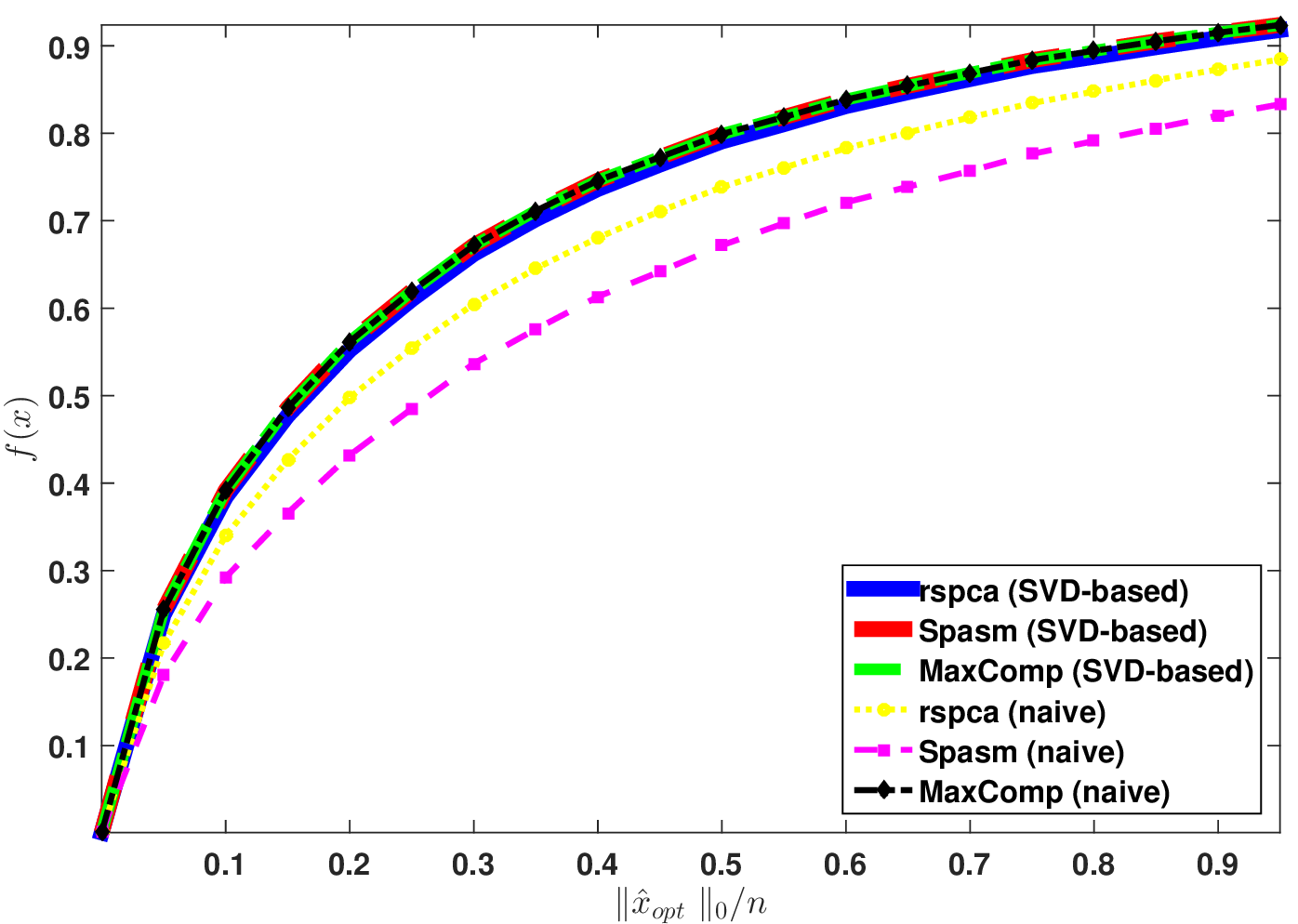}}		
  \\
	\subfloat[HapMap+HGDP data (chromosome 5): $n=31479$.]{%
			\label{fig:snp5}%
		\includegraphics[width = 0.46\textwidth]{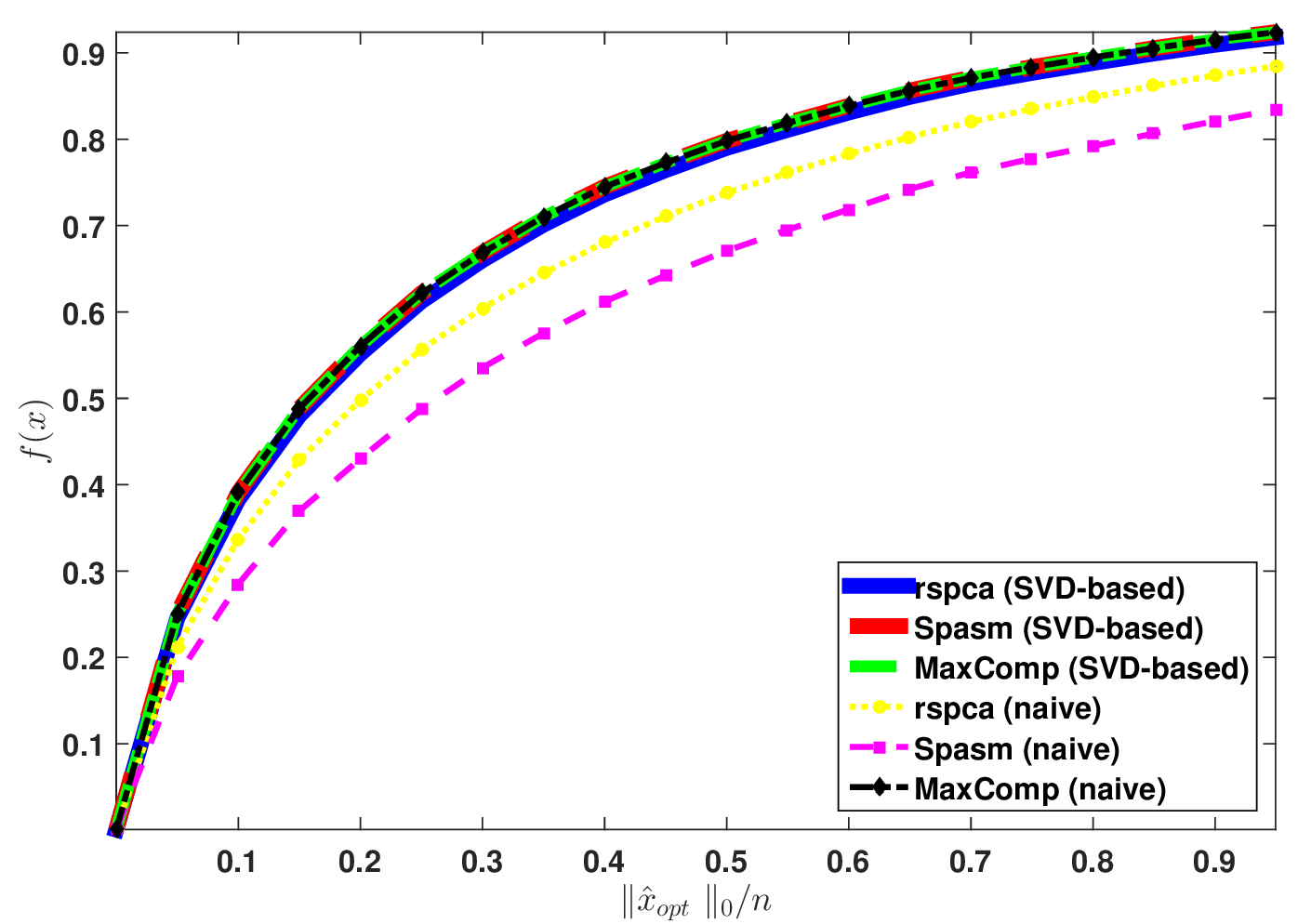}}		
       \quad
	\subfloat[HapMap+HGDP data (chromosome 6): $n=32800$.]{%
			\label{fig:snp6}%
		\includegraphics[width = 0.46\textwidth]{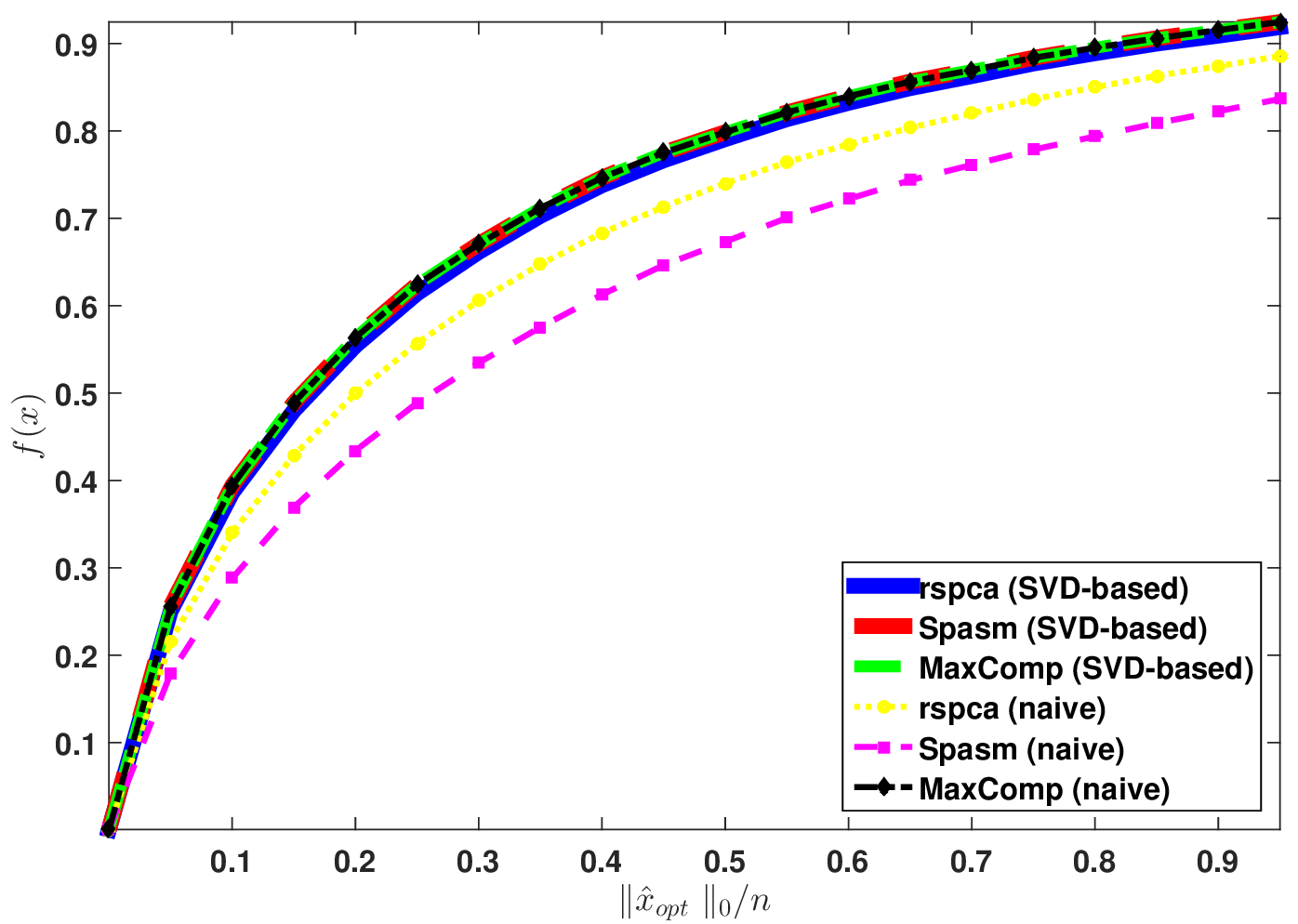}}	
   \\
   	\subfloat[HapMap+HGDP data (chromosome 7): $n=27130$.]{%
			\label{fig:snp7}%
		\includegraphics[width = 0.46\textwidth]{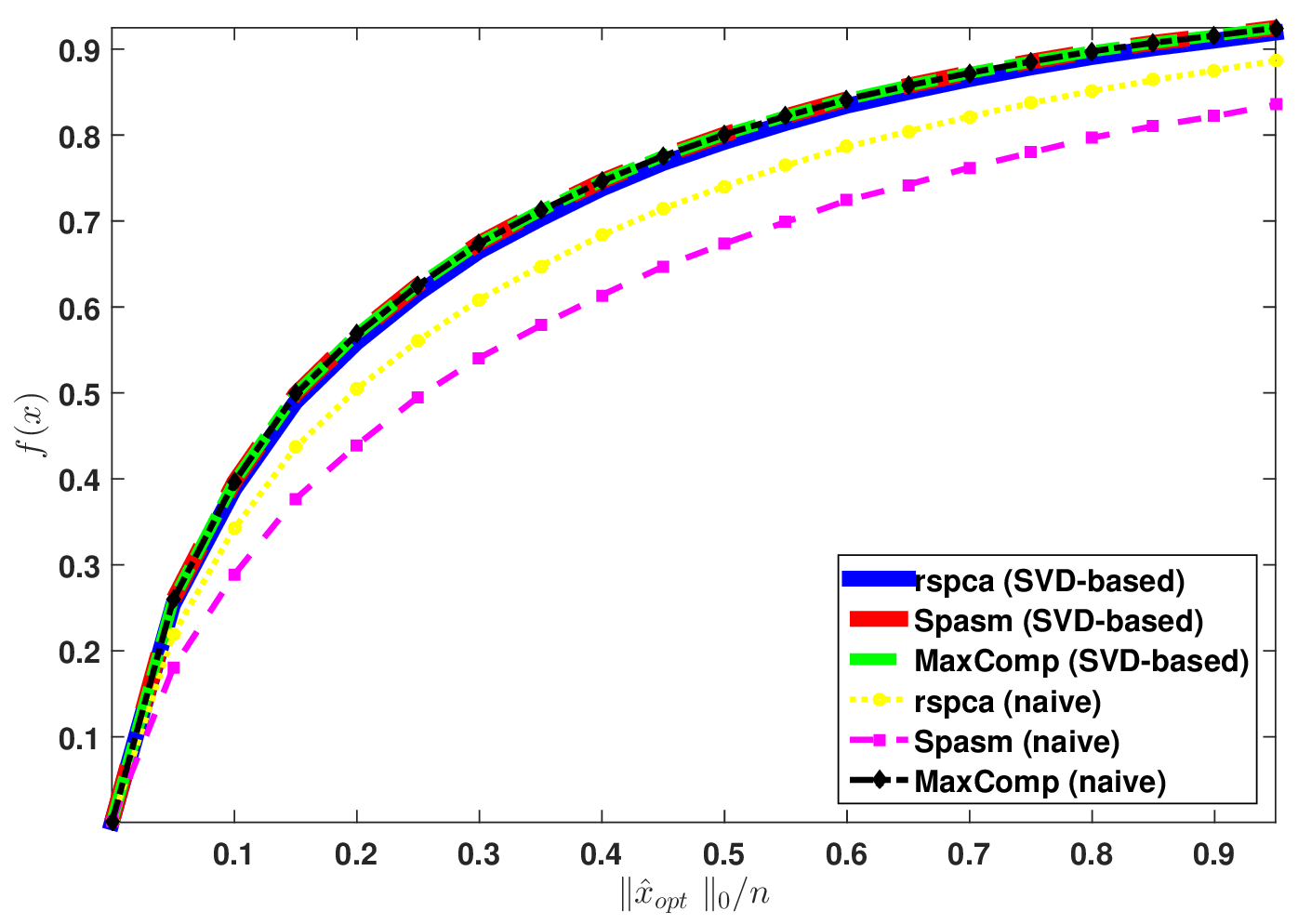}}		
         \quad
	\subfloat[HapMap+HGDP data (chromosome 8): $n=28508$.]{%
			\label{fig:snp8}%
		\includegraphics[width = 0.46\textwidth]{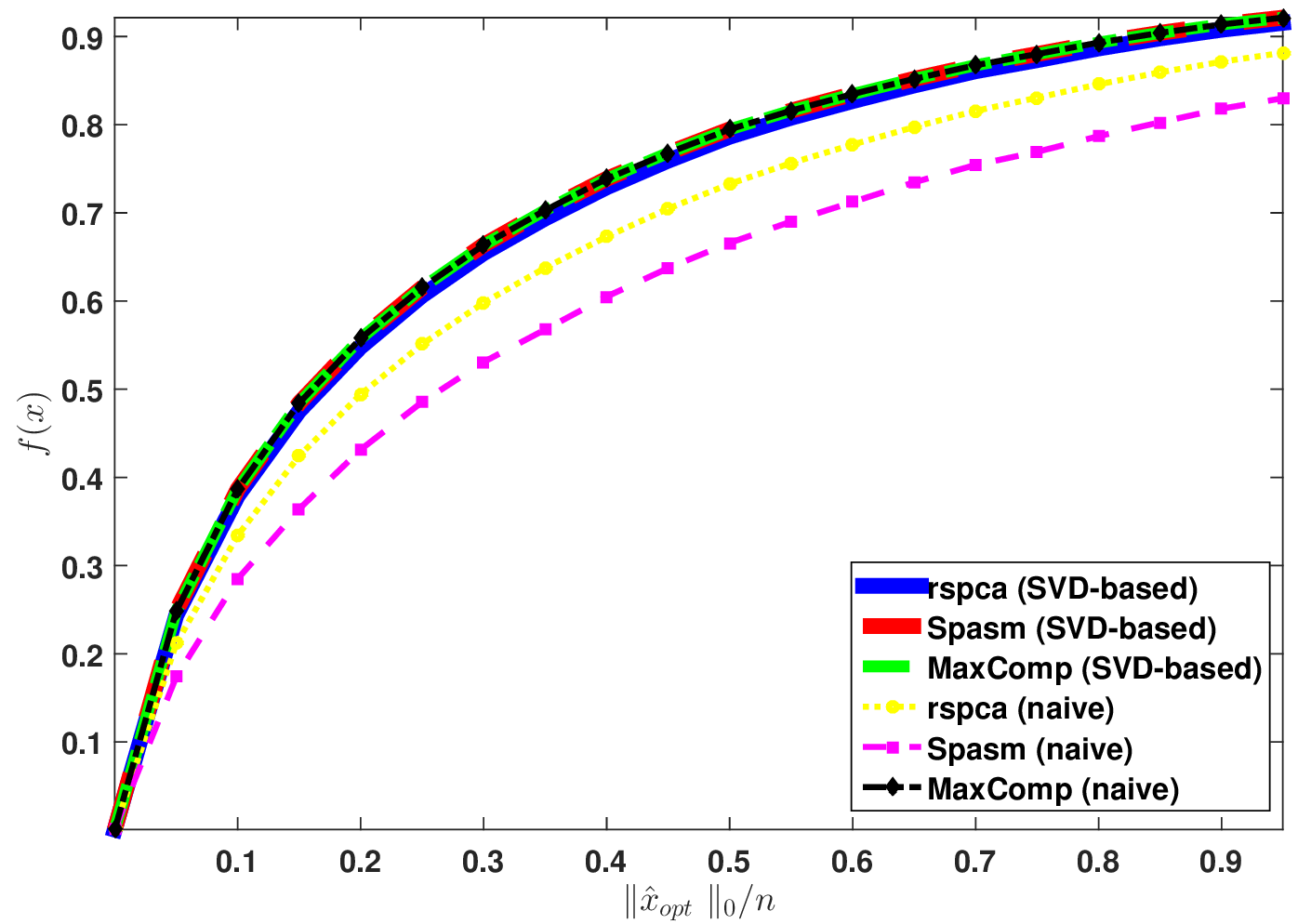}}
\caption{Performance of sparse PCA algorithms on additional population genetics data (chromosomes 3-8).}
\label{fig:SNPs_1}%
\end{figure}

 \begin{figure}[H]
\centering
	\subfloat[HapMap+HGDP data (chromosome 9): $n=24070$.]{%
			\label{fig:snp9}%
		\includegraphics[width = 0.46\textwidth]{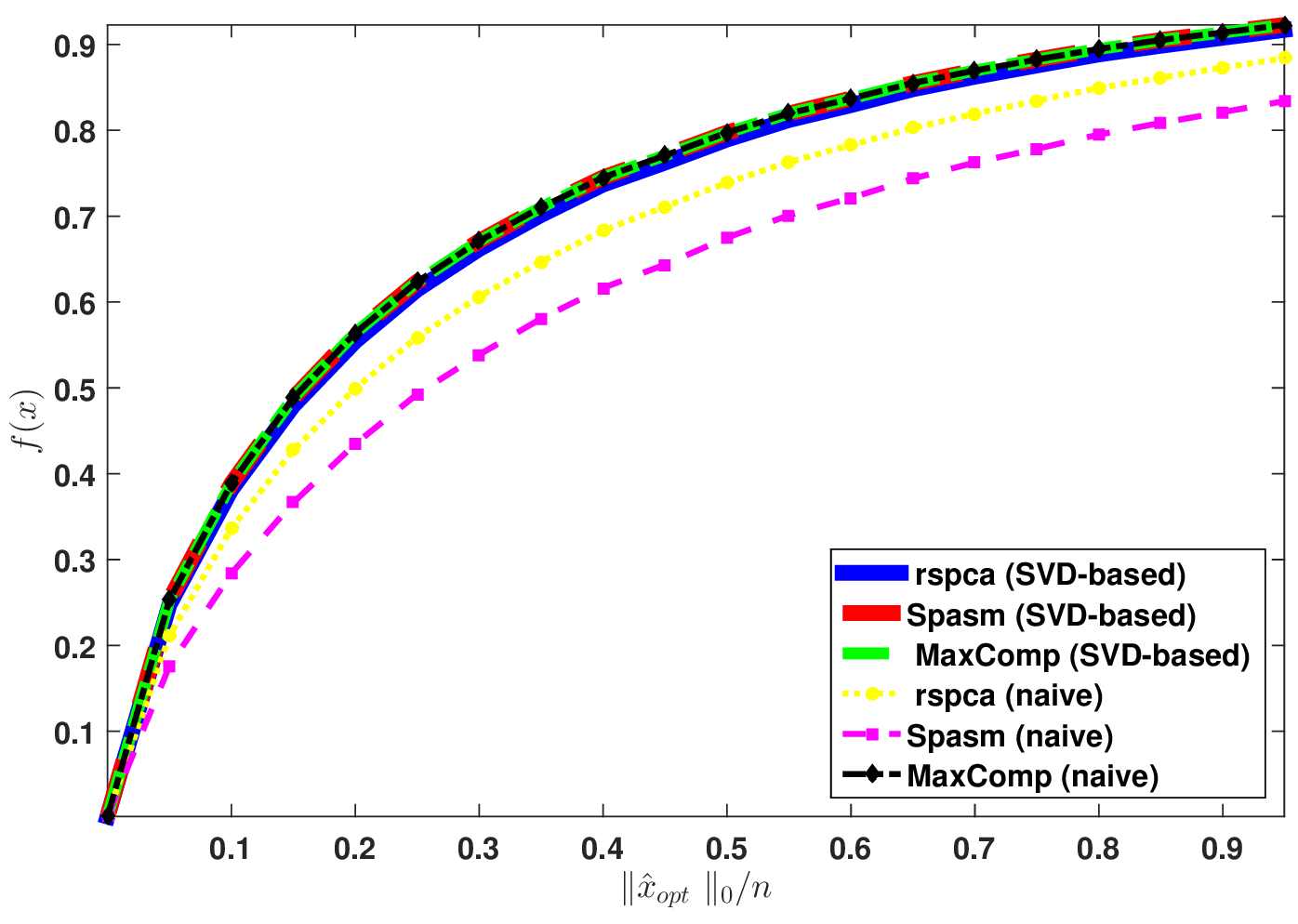}}		
         \quad
	\subfloat[HapMap+HGDP data (chromosome 10): $n=26327$.]{%
			\label{fig:snp10}%
		\includegraphics[width = 0.46\textwidth]{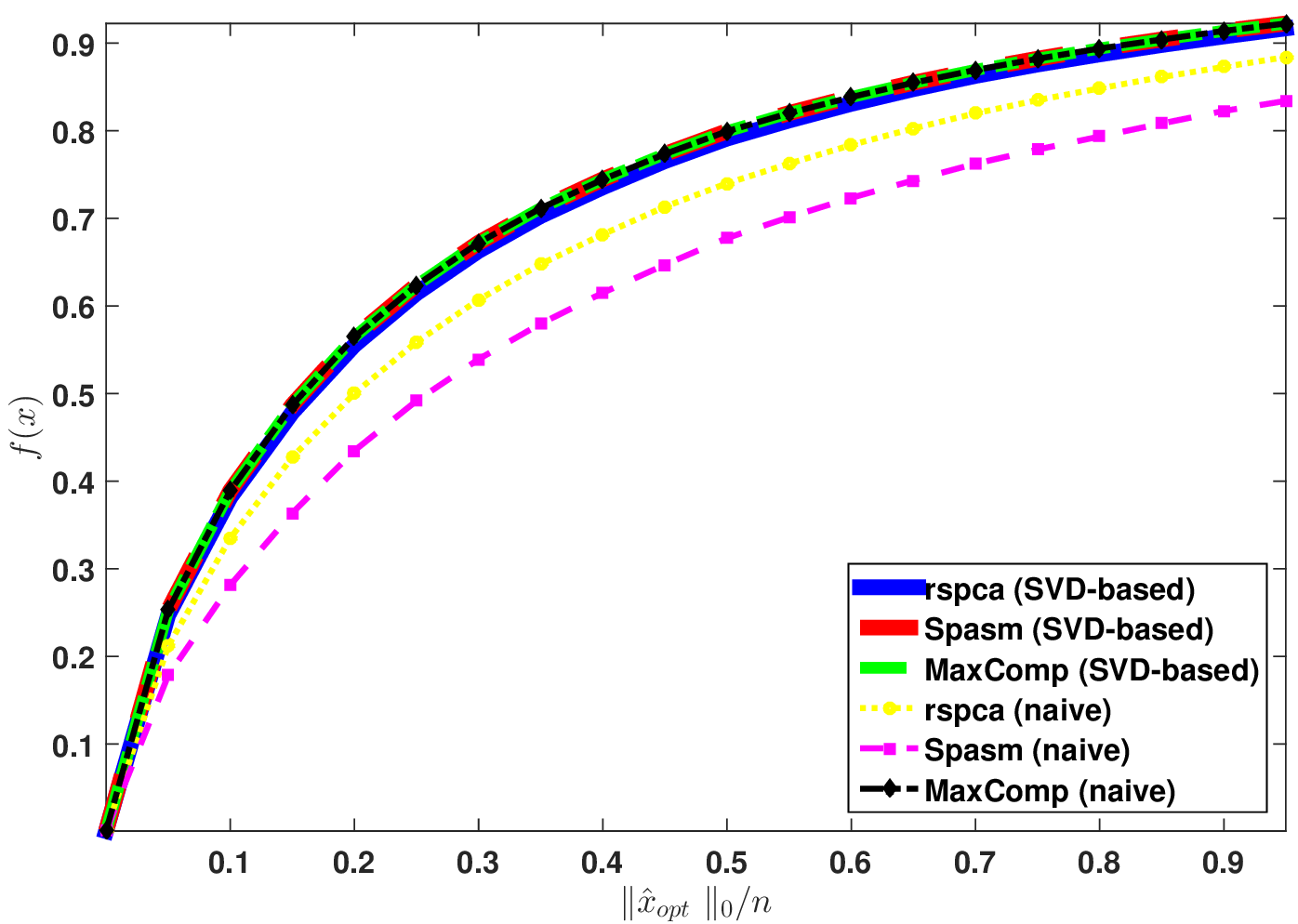}}	
	\\
	\subfloat[HapMap+HGDP data (chromosome 11): $n=24440$.]{%
			\label{fig:snp11}%
		\includegraphics[width = 0.46\textwidth]{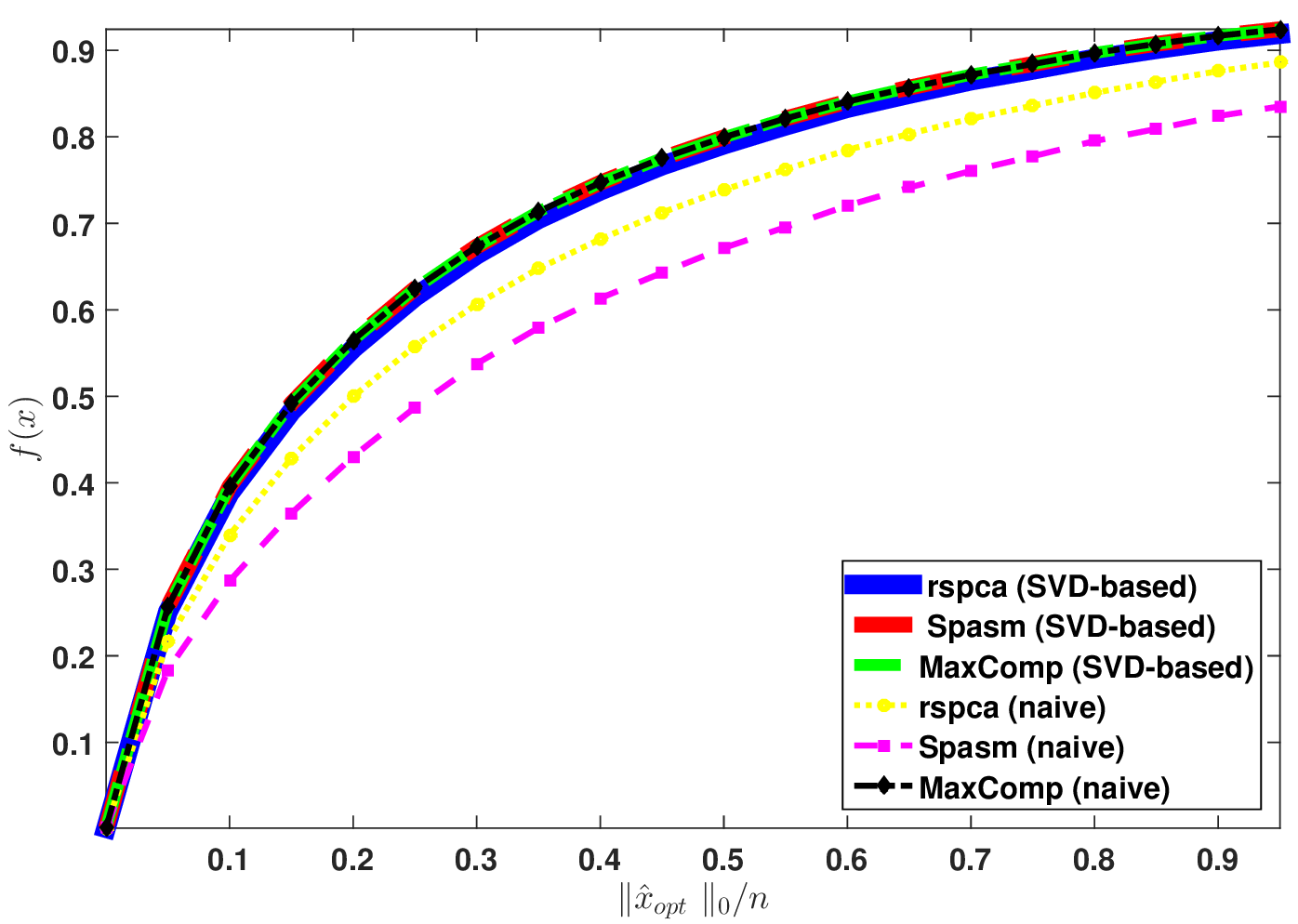}}		
     \quad
	\subfloat[HapMap+HGDP data (chromosome 12): $n=24084$.]{%
			\label{fig:snp12}%
		\includegraphics[width = 0.46\textwidth]{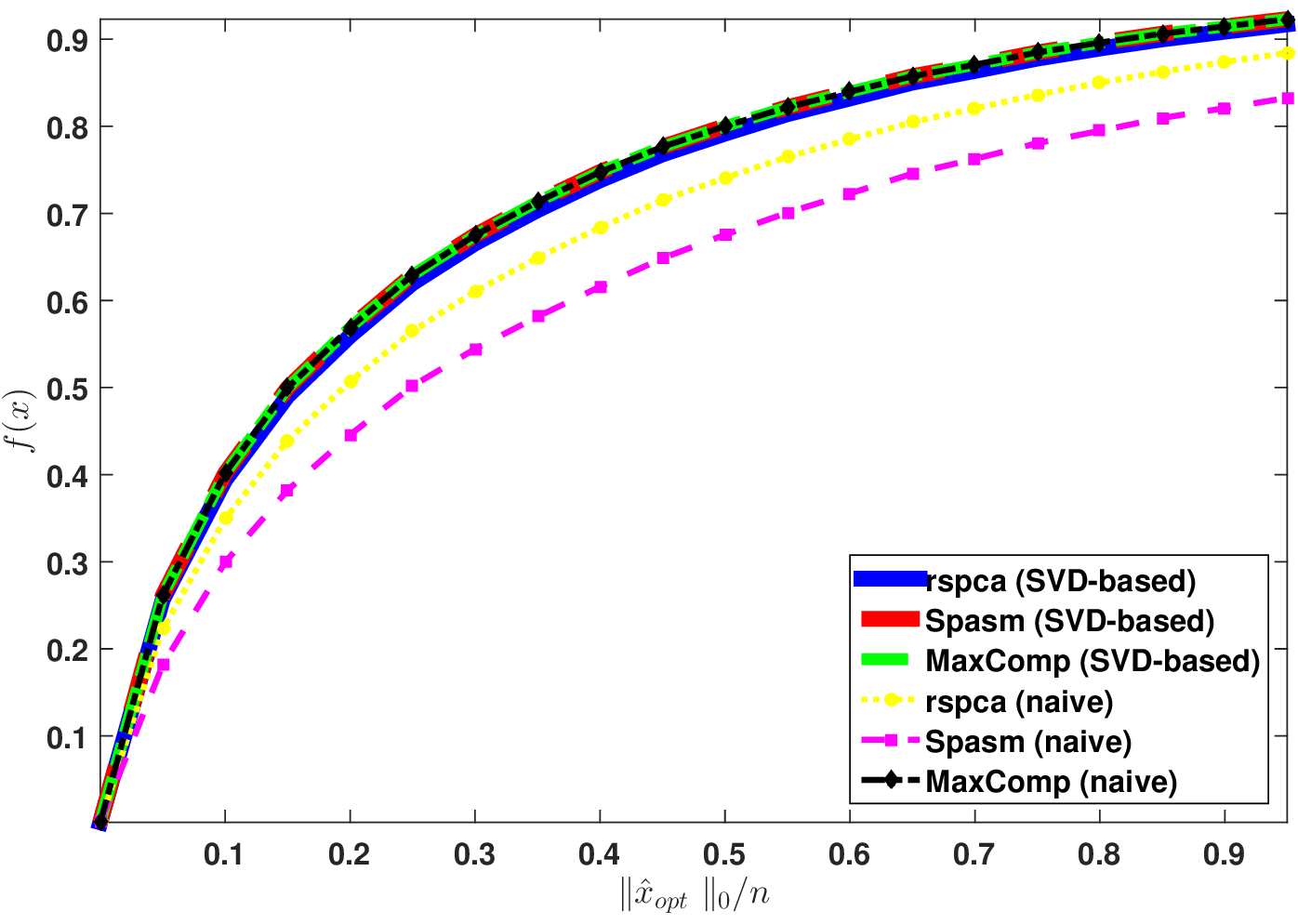}}		
        \\
	\subfloat[HapMap+HGDP data (chromosome 13): $n=18958$.]{%
			\label{fig:snp13}%
		\includegraphics[width = 0.46\textwidth]{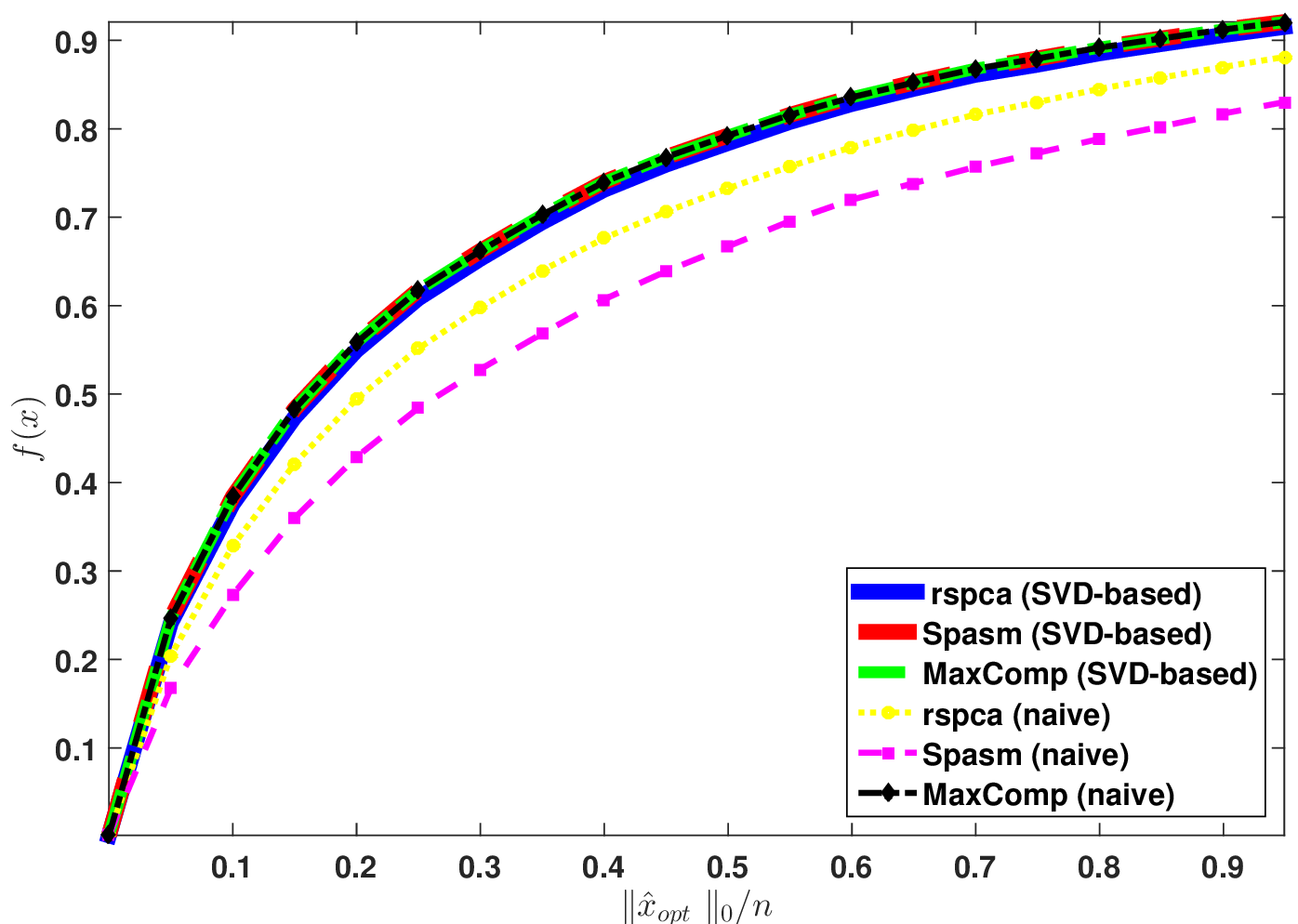}}		
         \quad
	\subfloat[HapMap+HGDP data (chromosome 14): $n=16469$.]{%
			\label{fig:snp14}%
		\includegraphics[width = 0.46\textwidth]{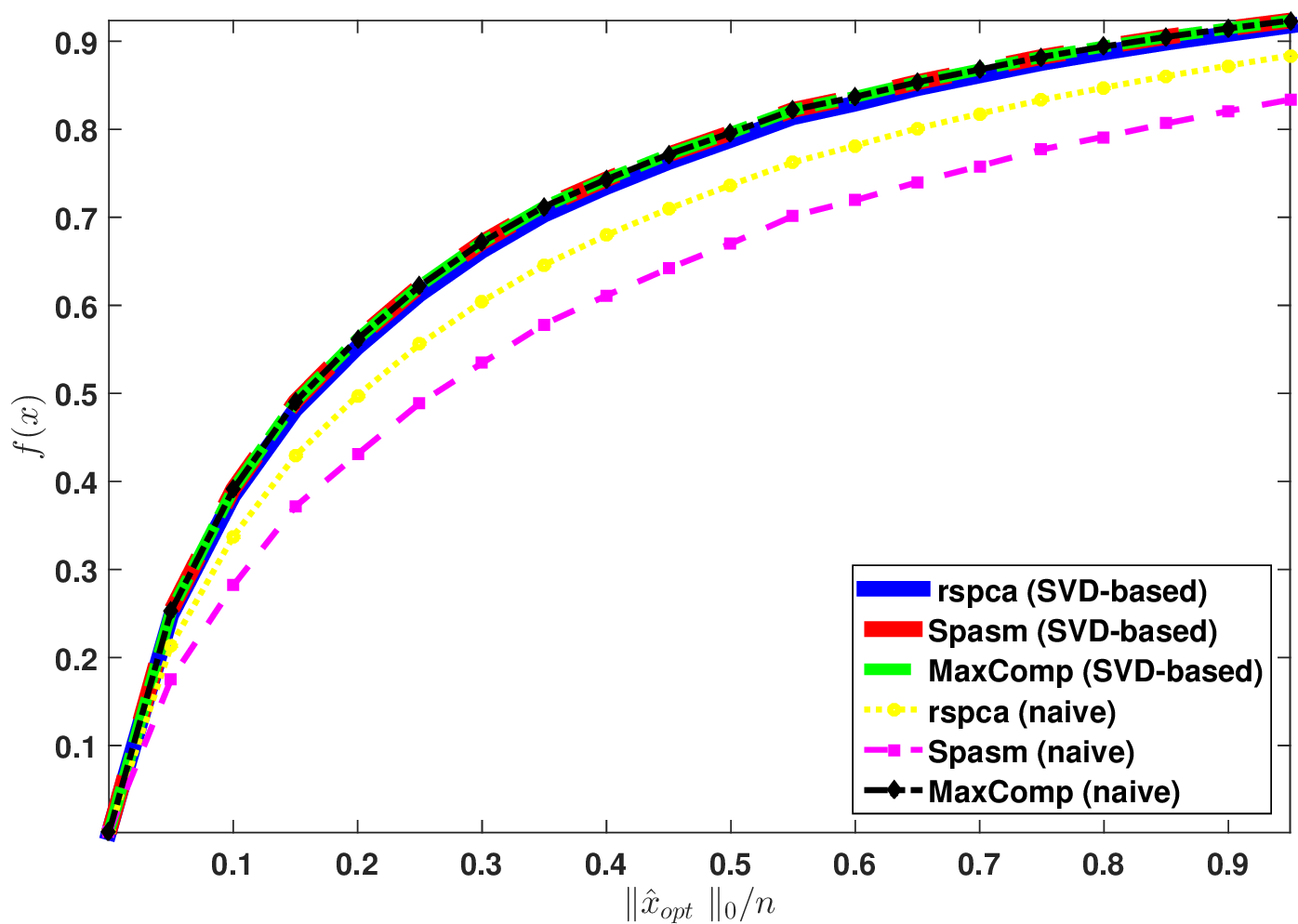}}	
		\caption{Performance of sparse PCA algorithms on additional population genetics data (chromosomes 9-14).}
		\label{fig:SNPs_2}%
\end{figure}

\begin{figure}[H]
\centering
	\subfloat[HapMap+HGDP data (chromosome 15): $n=15351$.]{%
			\label{fig:snp15}%
		\includegraphics[width = 0.46\textwidth]{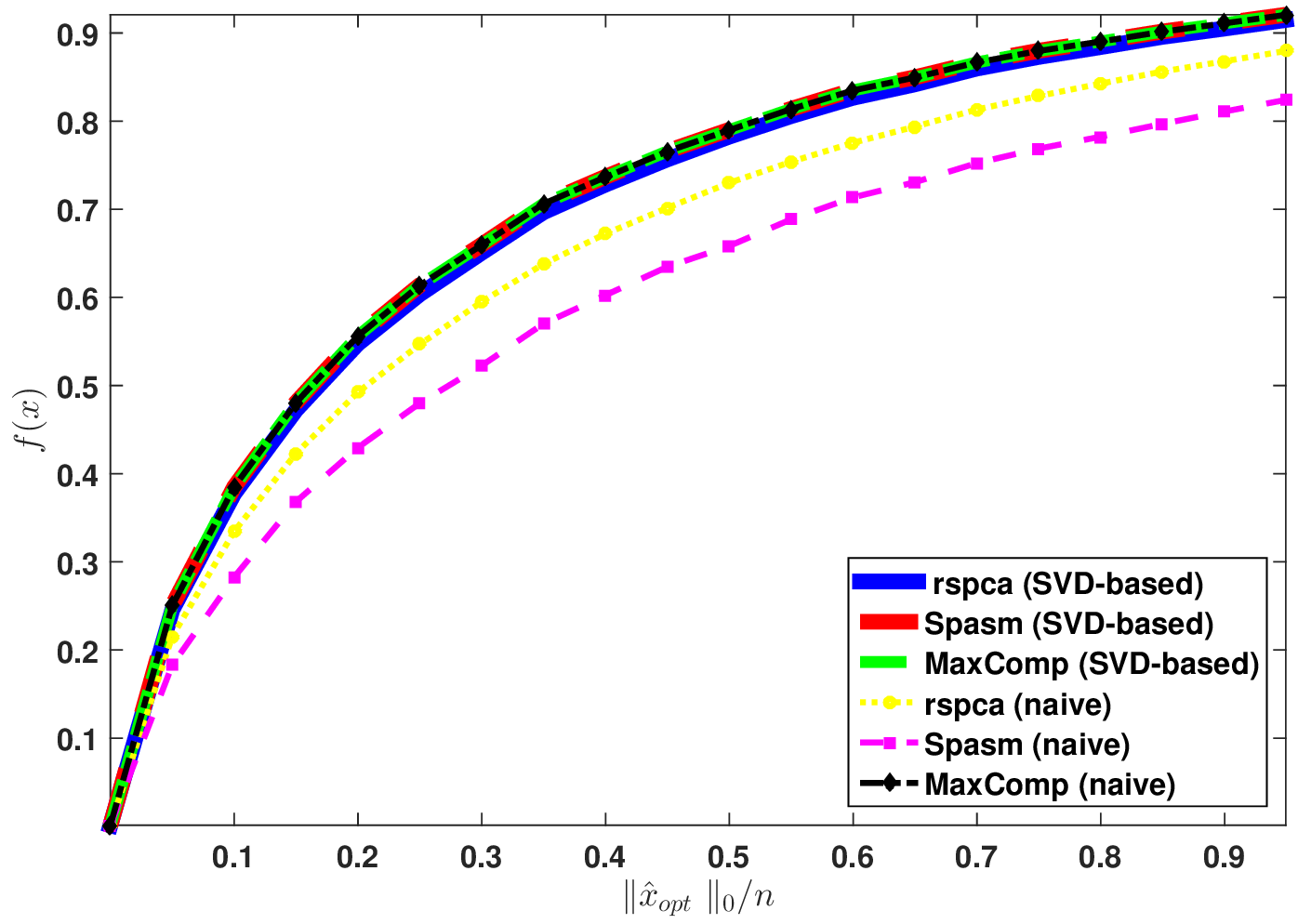}}		
         \quad
	\subfloat[HapMap+HGDP data (chromosome 16): $n=15289$.]{%
			\label{fig:snp16}%
		\includegraphics[width = 0.46\textwidth]{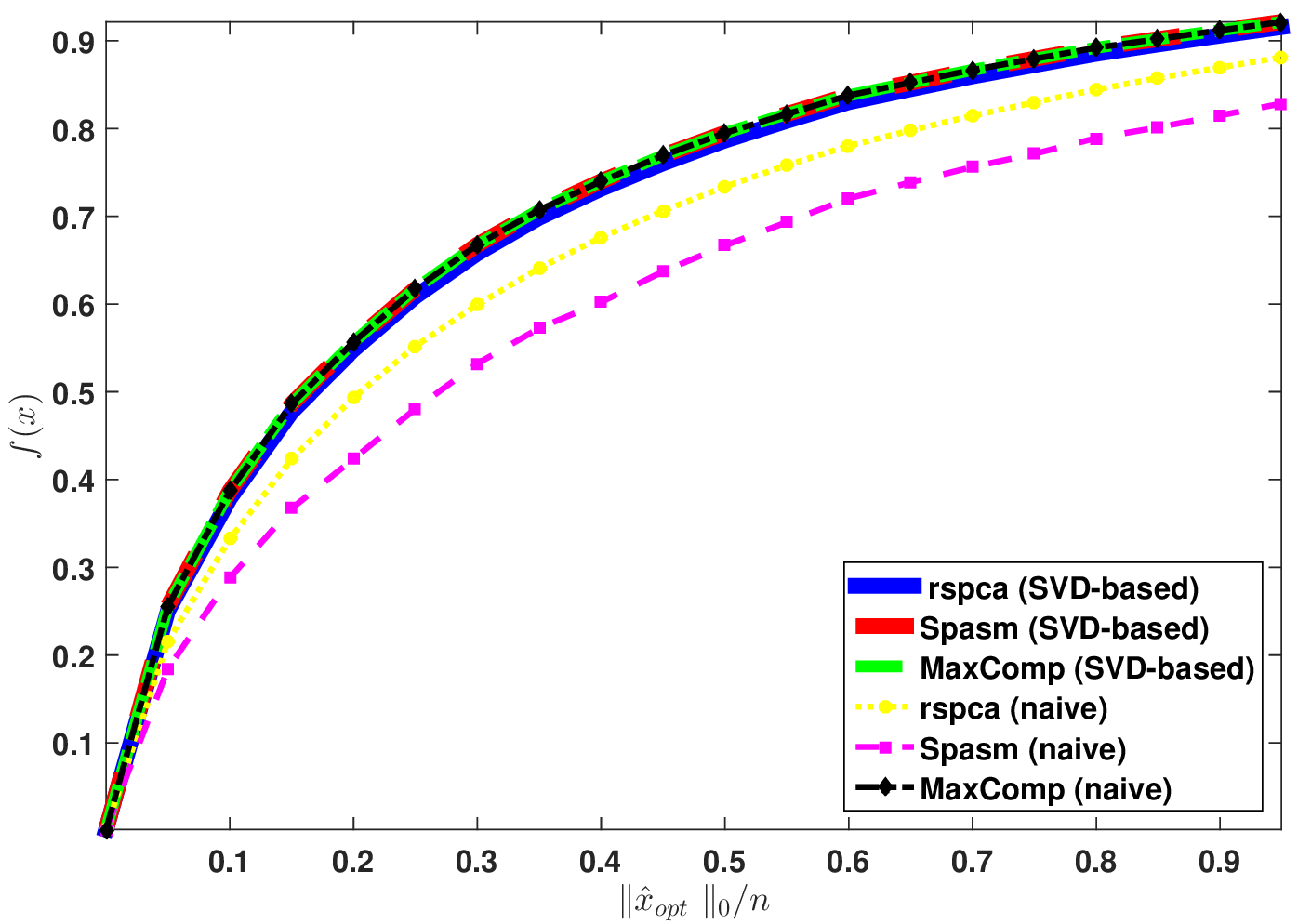}}
        \\
	\subfloat[HapMap+HGDP data (chromosome 17): $n=12945$.]{%
			\label{fig:snp17}%
		\includegraphics[width = 0.46\textwidth]{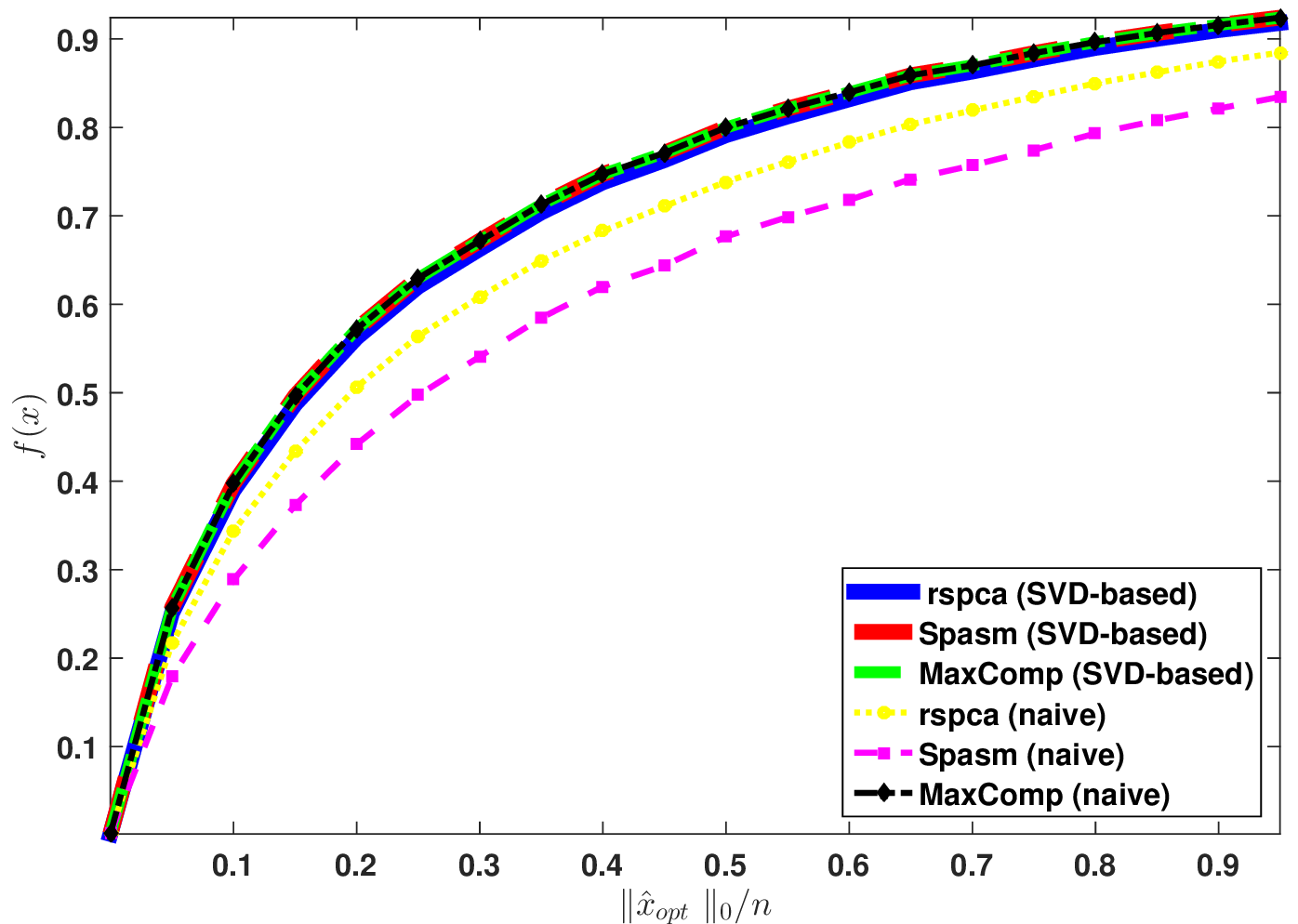}}		
         \quad
	\subfloat[HapMap+HGDP data (chromosome 18): $n=15373$.]{%
			\label{fig:snp18}%
		\includegraphics[width = 0.46\textwidth]{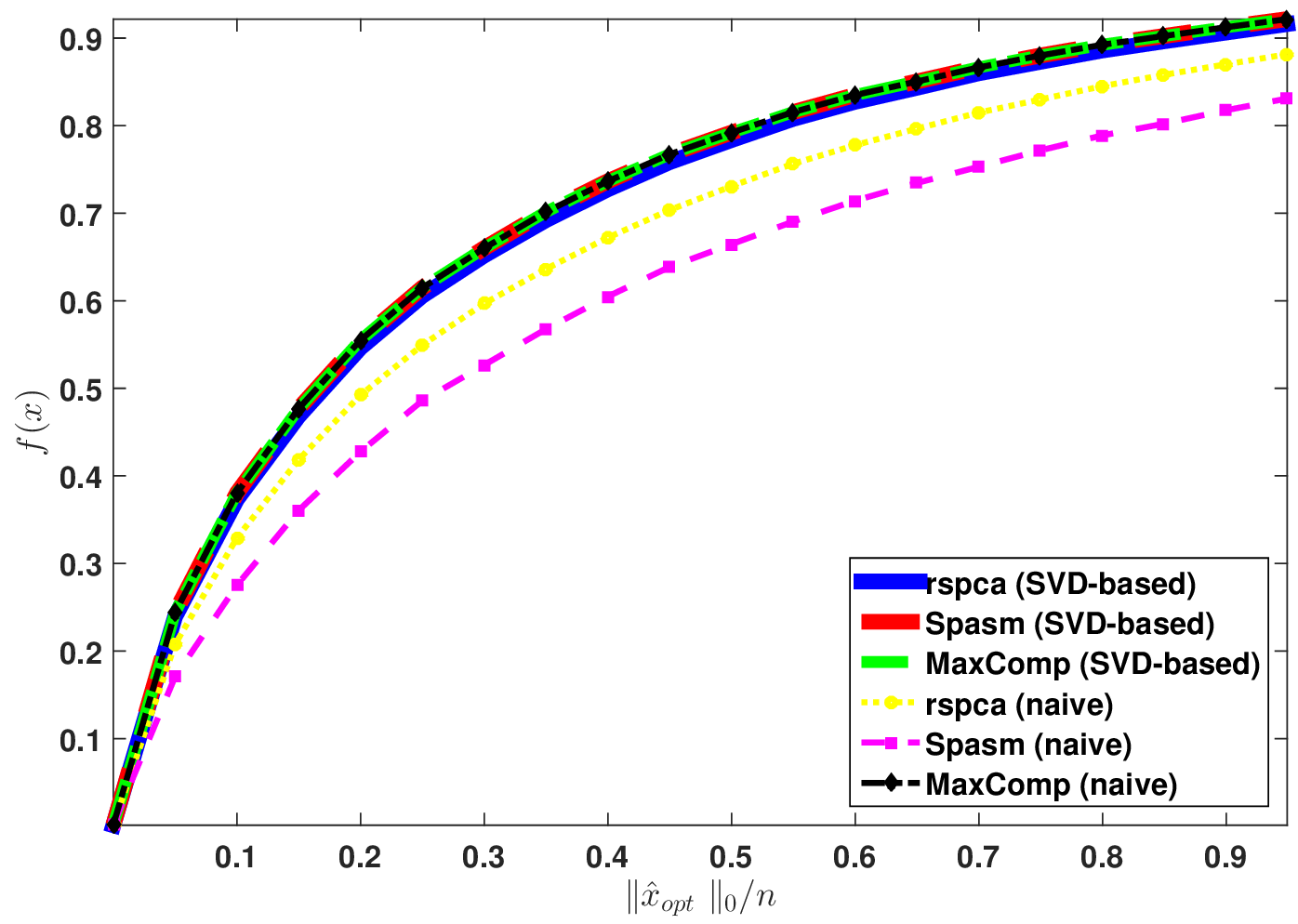}}	
	\\
	\subfloat[HapMap+HGDP data (chromosome 19): $n=8465$.]{%
			\label{fig:snp19}%
		\includegraphics[width = 0.46\textwidth]{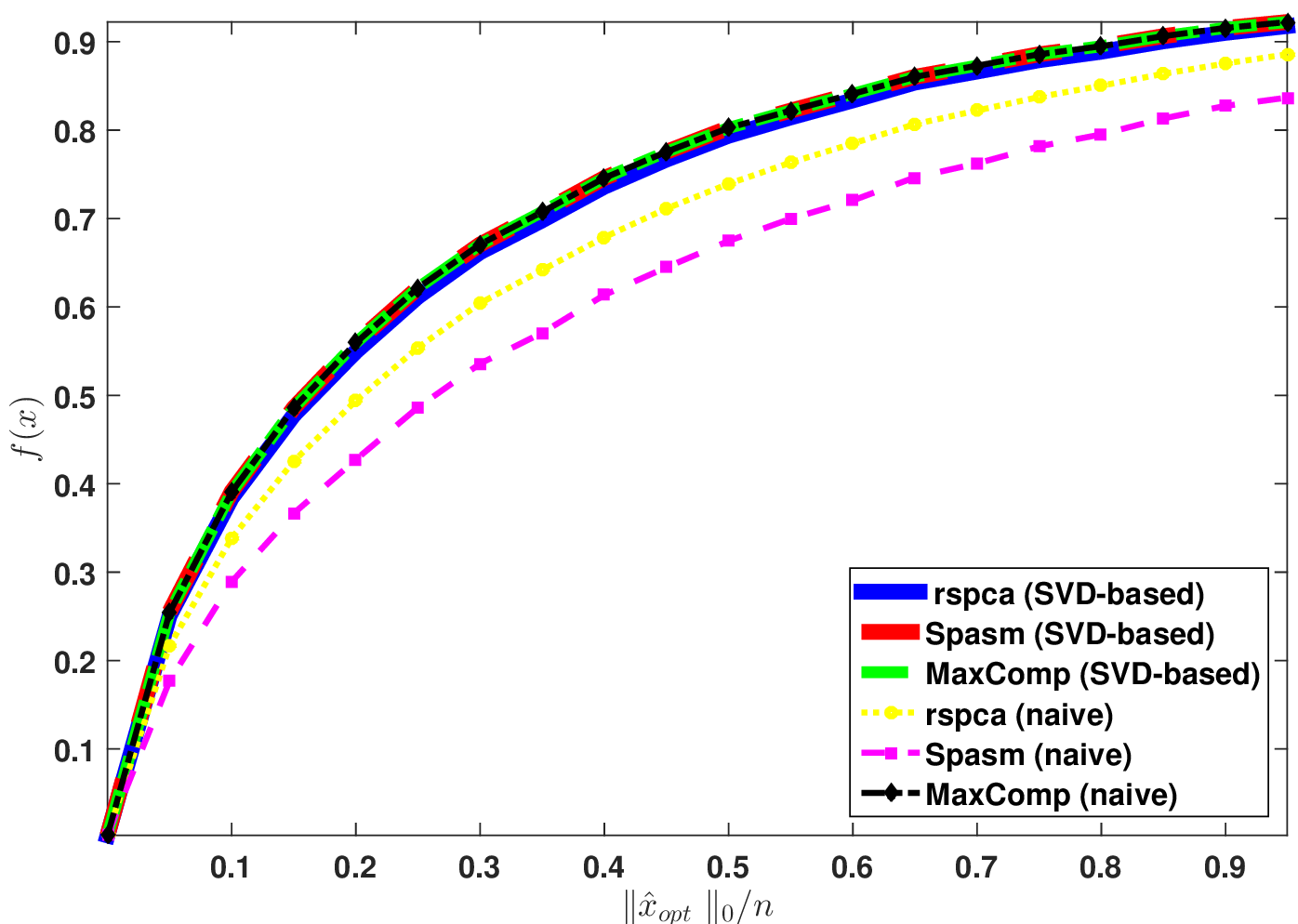}}		
         \quad
	\subfloat[HapMap+HGDP data (chromosome 20): $n=13015$.]{%
			\label{fig:snp20}%
		\includegraphics[width = 0.45\textwidth]{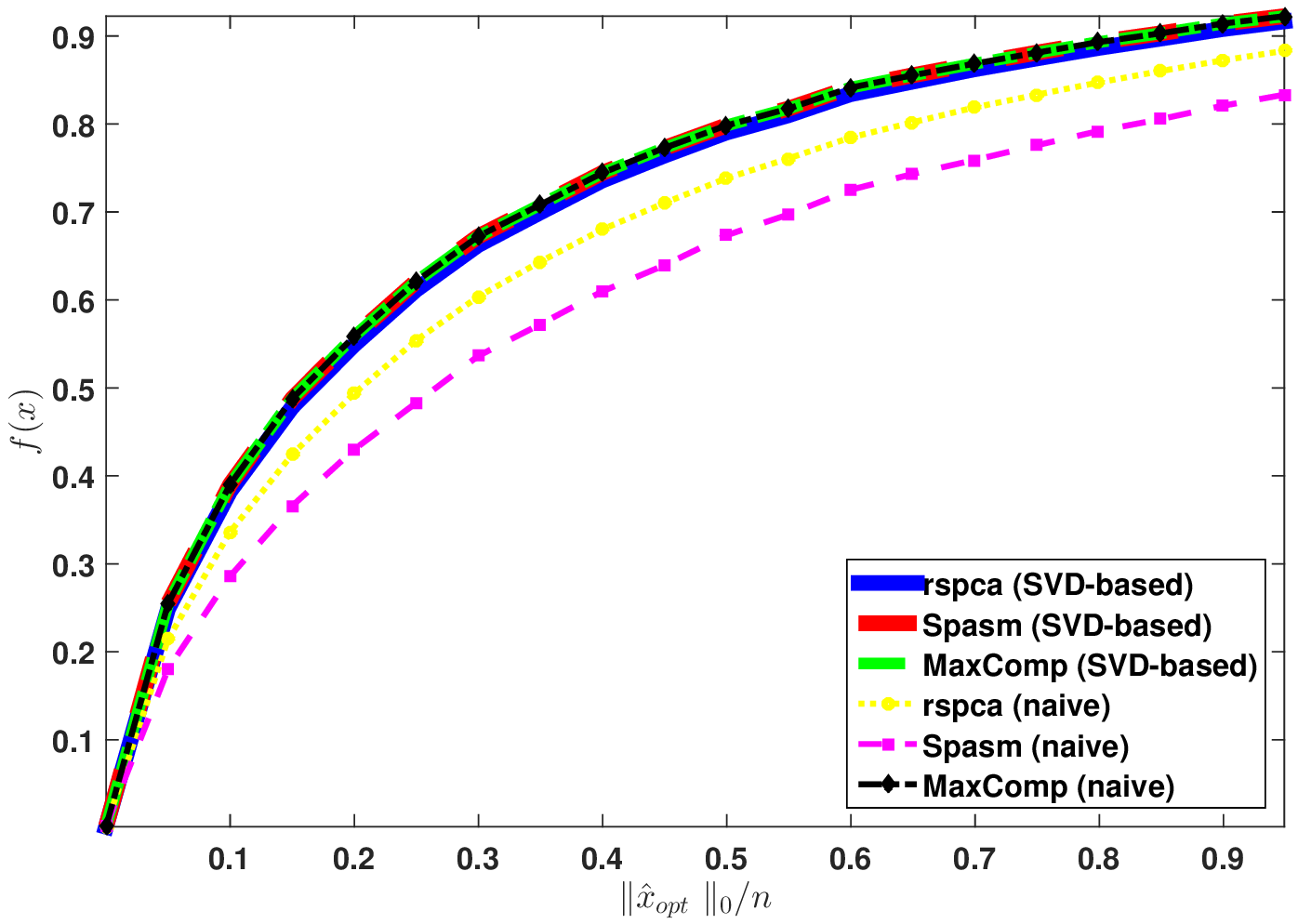}}		
	\caption{Performance of sparse PCA algorithms on additional population genetics data (chromosomes 15-20).}
	\label{fig:SNPs_3}%
\end{figure}

\begin{figure}[H]
\centering
	\subfloat[HapMap+HGDP data (chromosome 21): $n=7556$.]{%
			\label{fig:snp21}%
		\includegraphics[width = 0.46\textwidth]{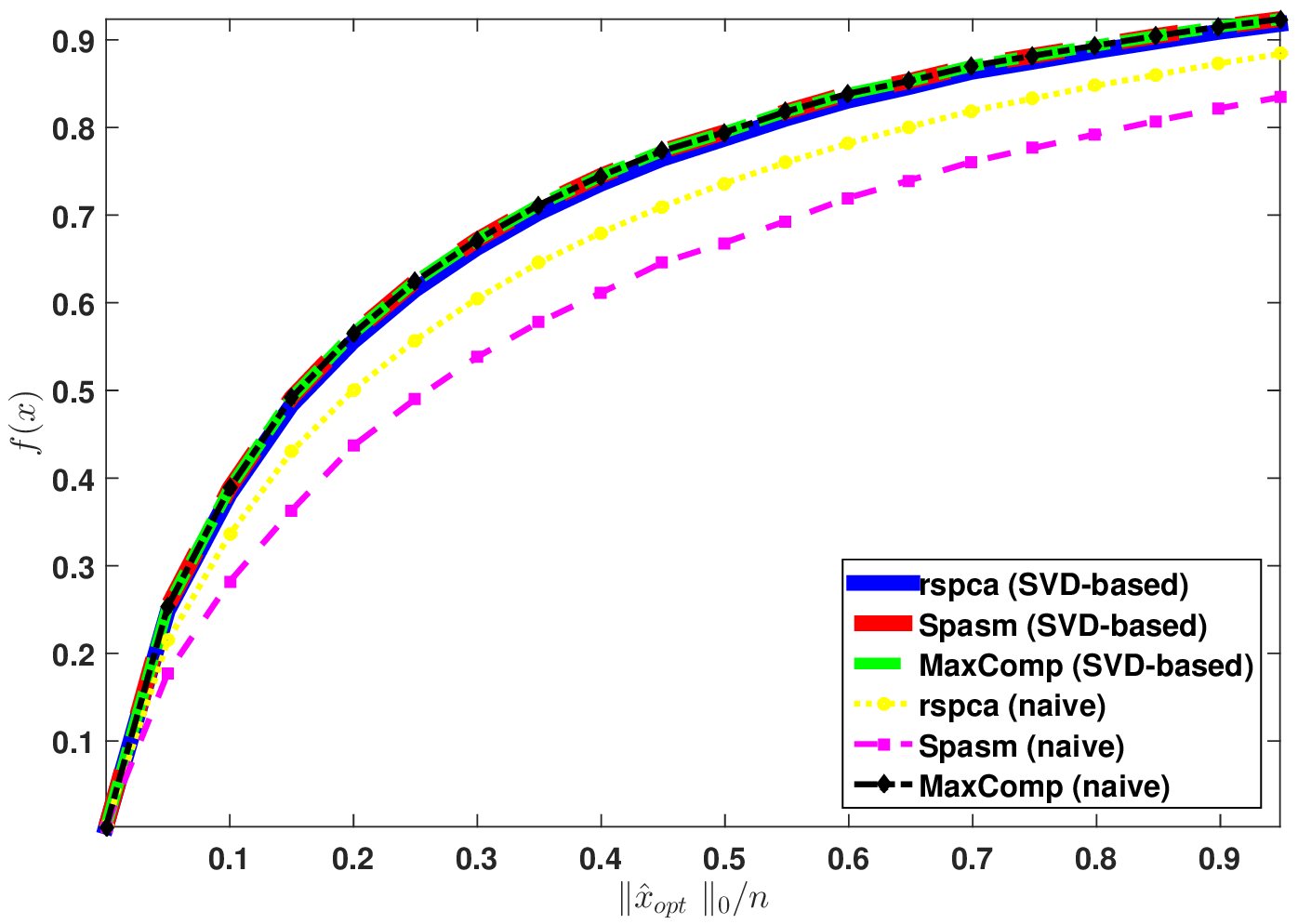}}		
         \quad
	\subfloat[HapMap+HGDP data (chromosome 22): $n=7334$.]{%
			\label{fig:snp22}%
		\includegraphics[width = 0.46\textwidth]{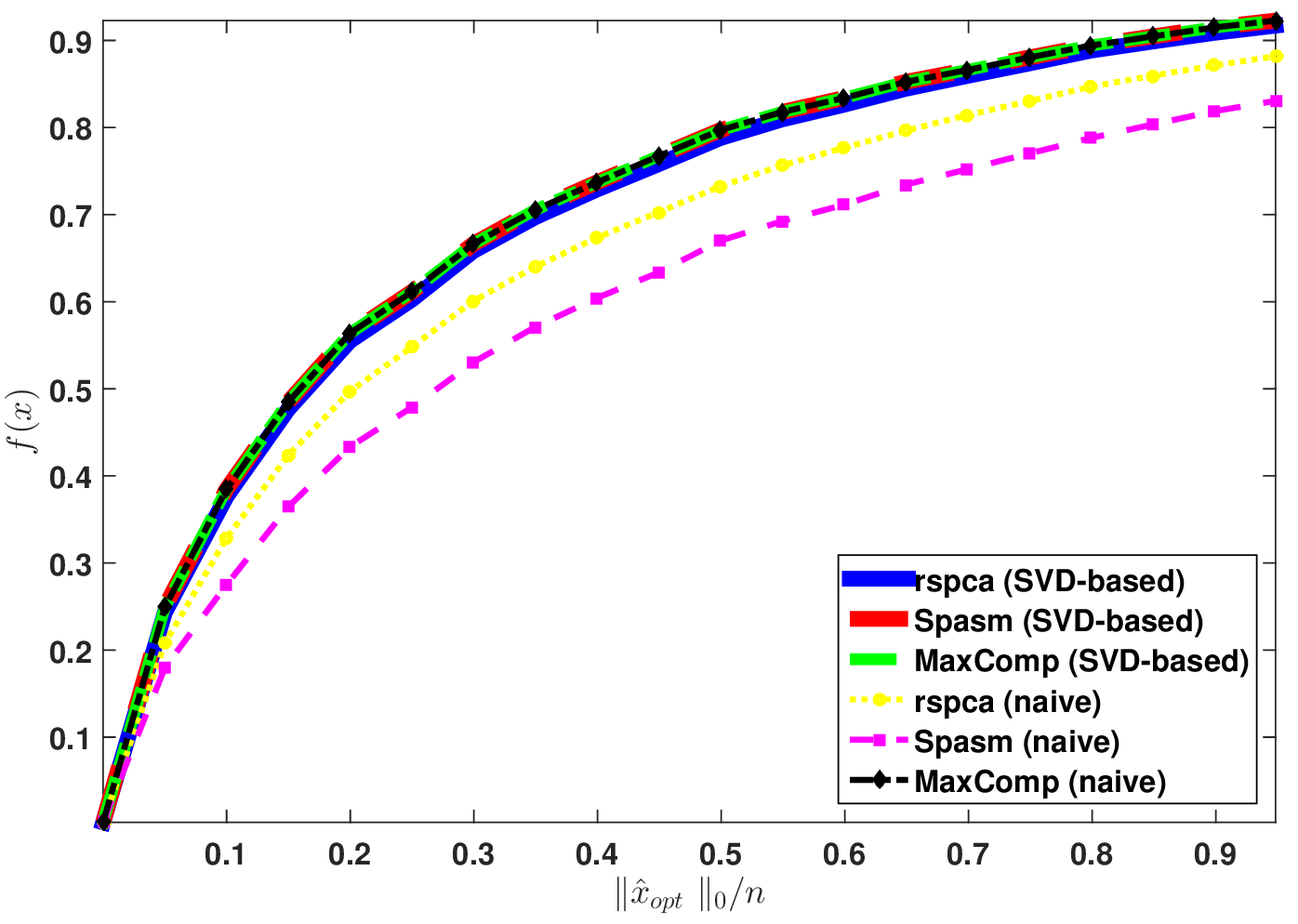}}
 	\caption{Performance of sparse PCA algorithms on additional population genetics data (chromosomes 21-22).}
	\label{fig:SNPs_4}%
	
\end{figure}

\newpage 

\begin{figure}[H]
	\centering
	\subfloat[HapMap+HGDP data (chromosome 5): $n=31479$.]{%
		\label{fig:snp5_time}%
		\includegraphics[width = 0.4\textwidth]{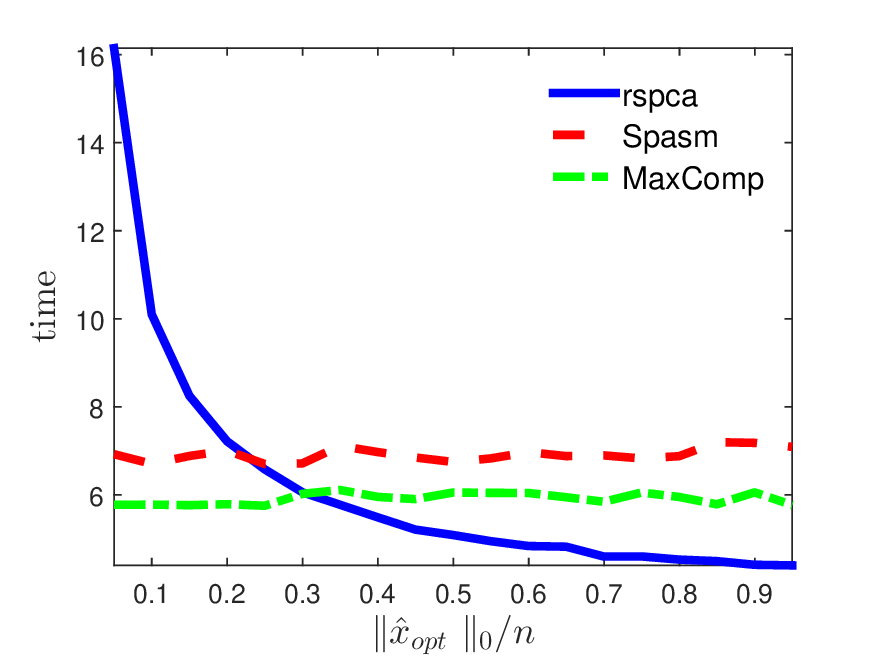}}		
	\quad
	\subfloat[HapMap+HGDP data (chromosome 6): $n=32800$.]{%
		\label{fig:snp6_time}%
		\includegraphics[width = 0.4\textwidth]{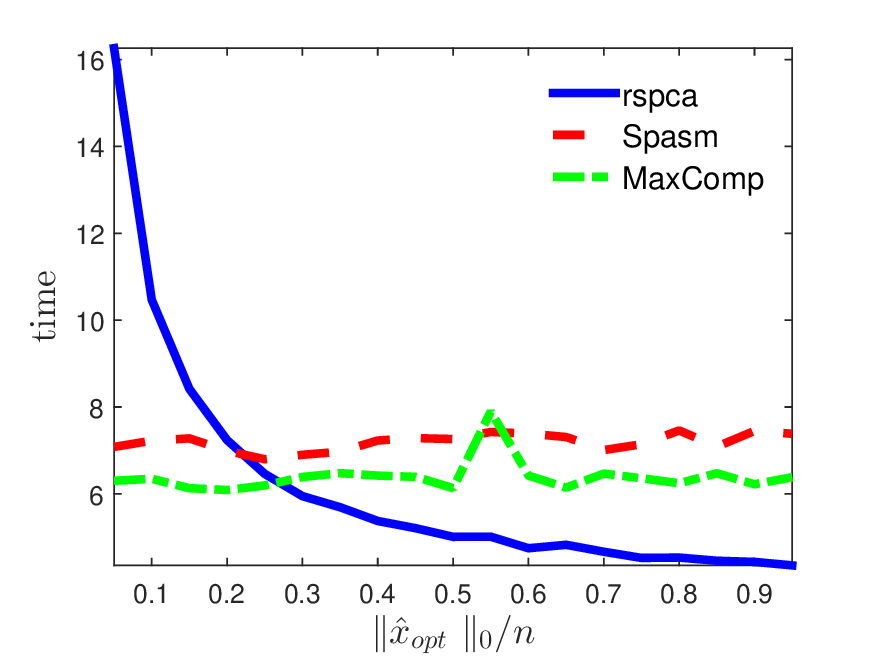}}	
	\\
	\subfloat[HapMap+HGDP data (chromosome 7): $n=27130$.]{%
		\label{fig:snp7_time}%
		\includegraphics[width = 0.4\textwidth]{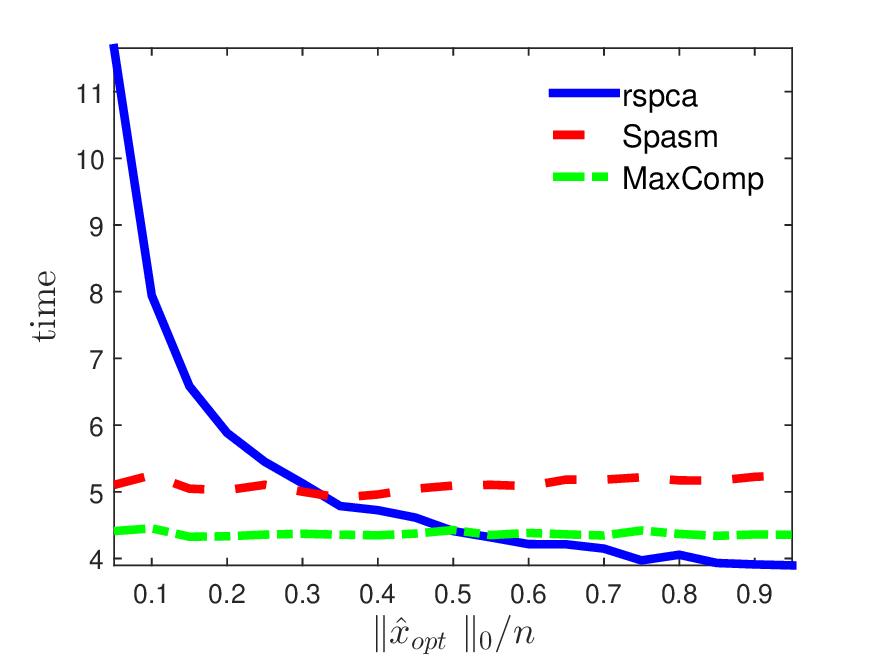}}		
	\quad
	\subfloat[HapMap+HGDP data (chromosome 8): $n=28508$.]{%
		\label{fig:snp8_time}%
		\includegraphics[width = 0.4\textwidth]{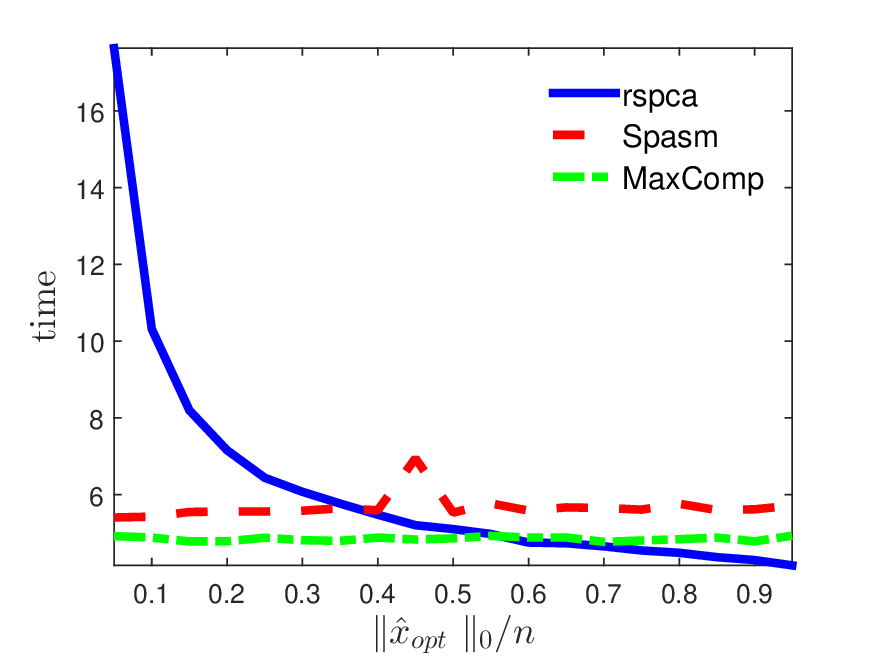}}
	\\
	\centering
	\subfloat[HapMap+HGDP data (chromosome 9): $n=24070$.]{%
		\label{fig:snp9_time}%
		\includegraphics[width = 0.4\textwidth]{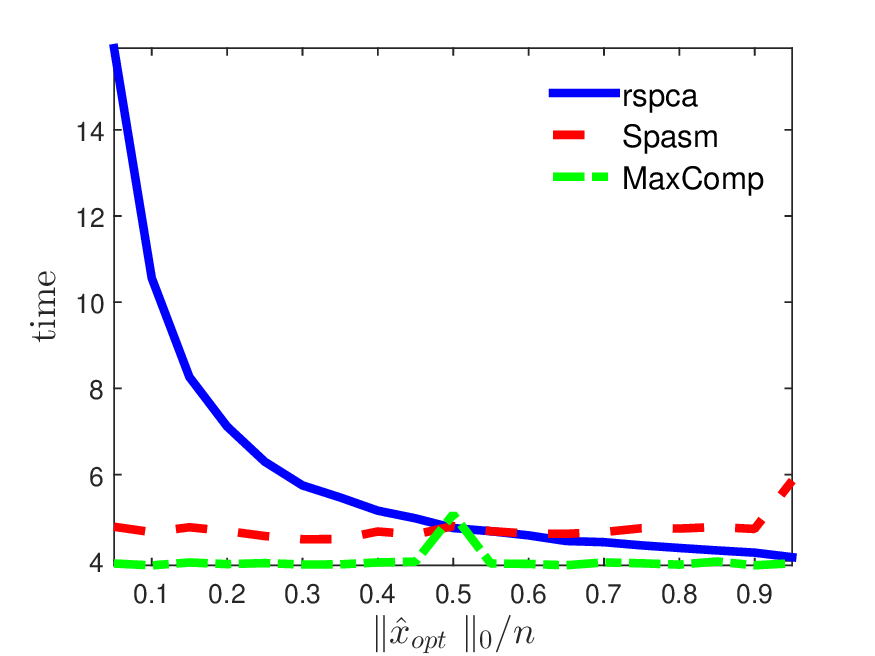}}		
	\quad
	\subfloat[HapMap+HGDP data (chromosome 10): $n=26327$.]{%
		\label{fig:snp10_time}%
		\includegraphics[width = 0.4\textwidth]{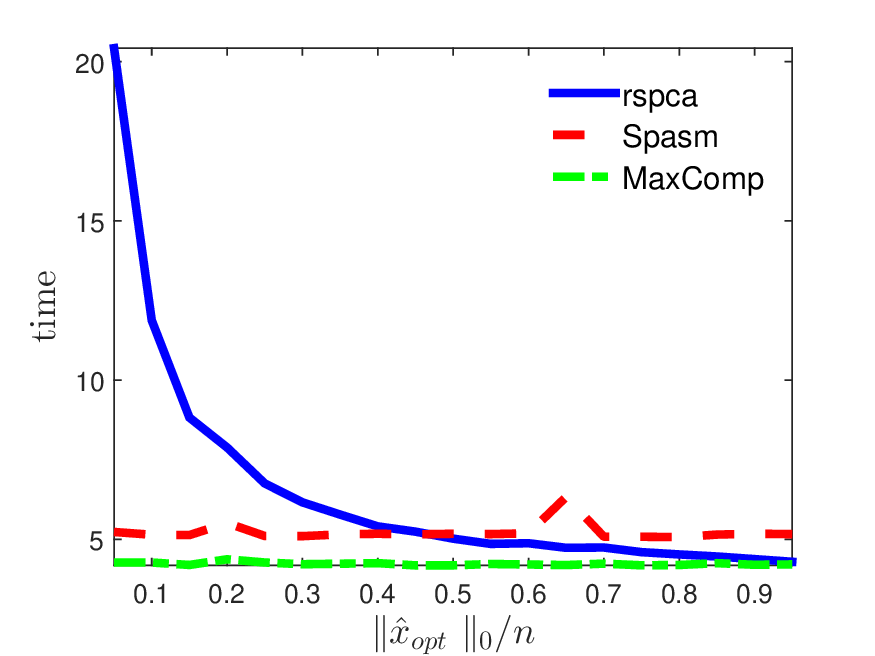}}	
		\caption{Running time of sparse PCA algorithms on additional population genetics data (chromosomes 5-10).}
		\label{fig:SNPs_time_1}
	\end{figure}
	\begin{figure}[H]
	\centering
	\subfloat[HapMap+HGDP data (chromosome 11): $n=24440$.]{%
			\label{fig:snp11_time}%
			\includegraphics[width = 0.4\textwidth]{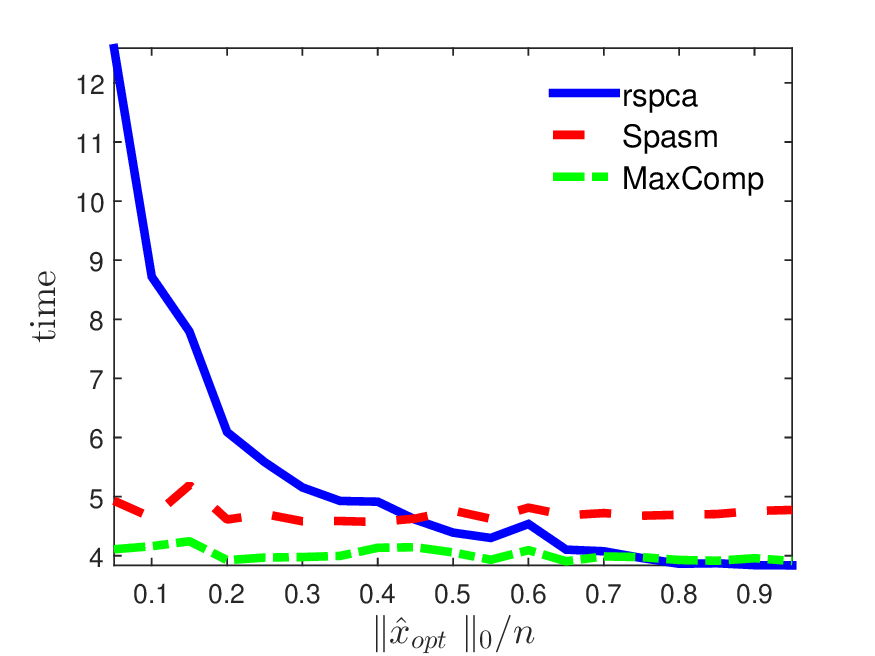}}		
		\quad
		\subfloat[HapMap+HGDP data (chromosome 12): $n=24084$.]{%
			\label{fig:snp12_time}%
			\includegraphics[width = 0.4\textwidth]{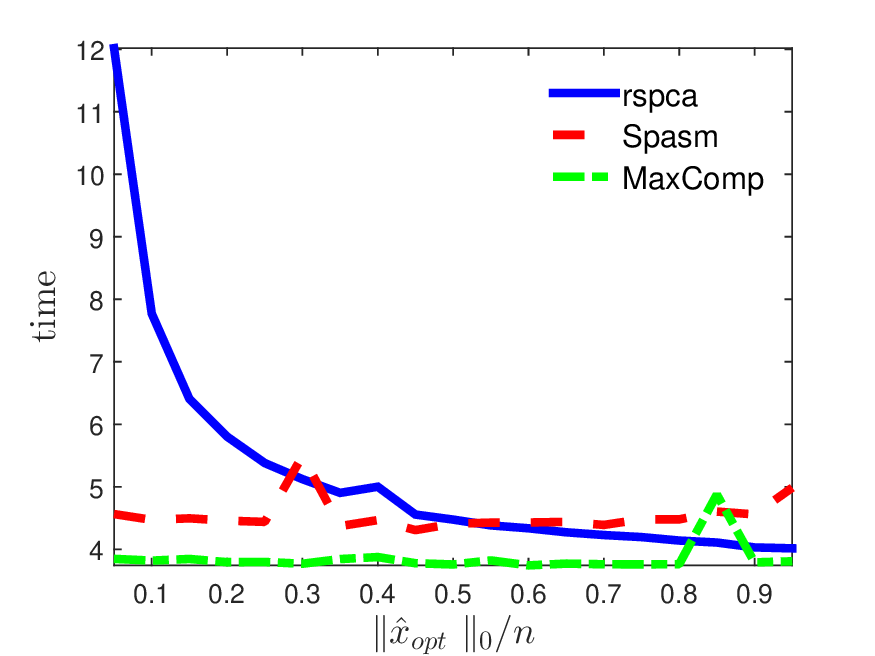}}
\\
 	\subfloat[HapMap+HGDP data (chromosome 13): $n=18958$.]{%
 		\label{fig:snp13_time}%
 		\includegraphics[width = 0.4\textwidth]{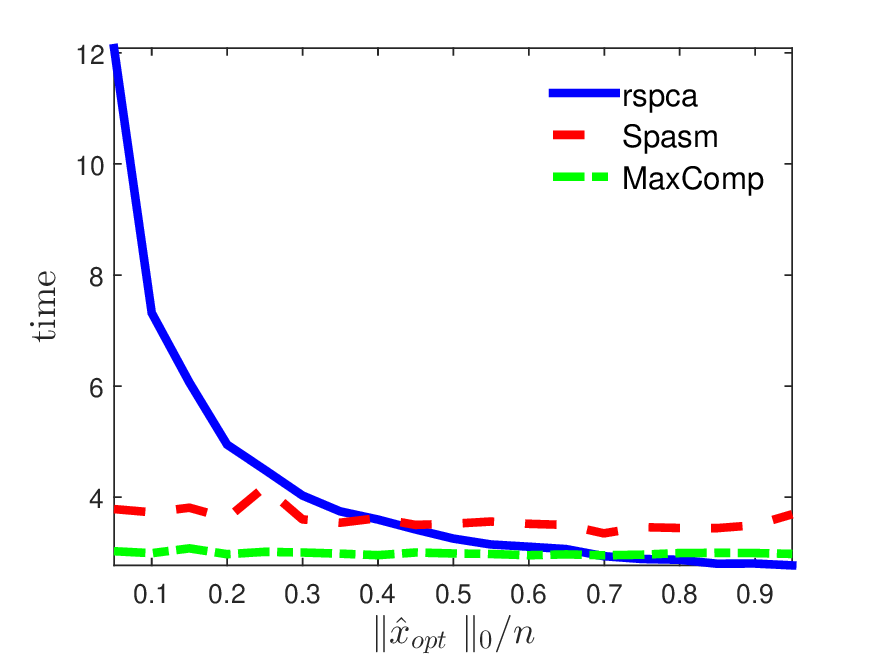}}		
 	\quad
 	\subfloat[HapMap+HGDP data (chromosome 14): $n=16469$.]{%
 		\label{fig:snp14_time}%
 		\includegraphics[width = 0.4\textwidth]{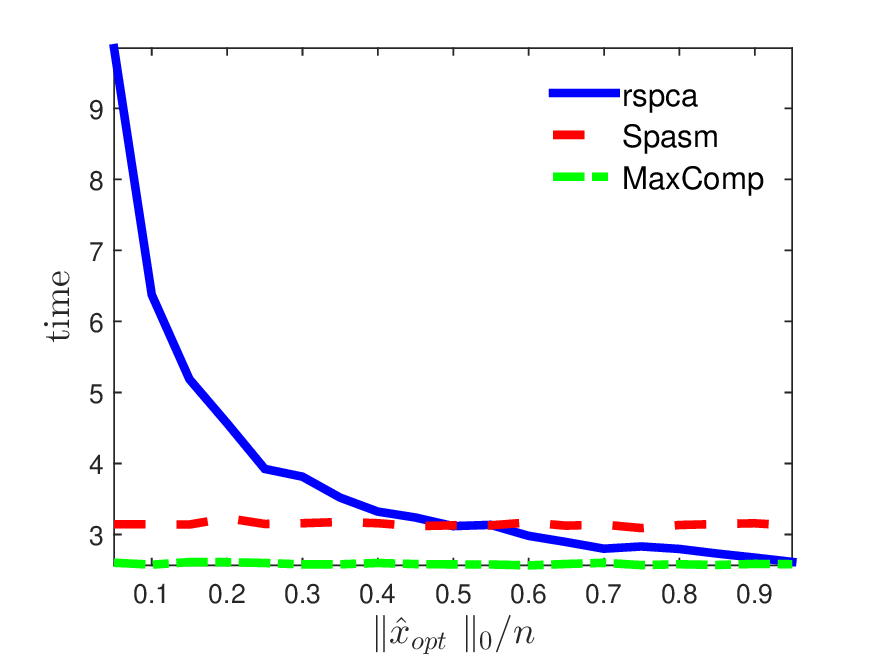}}
 	\\
	\subfloat[HapMap+HGDP data (chromosome 15): $n=15351$.]{%
		\label{fig:snp15_time}%
		\includegraphics[width = 0.4\textwidth]{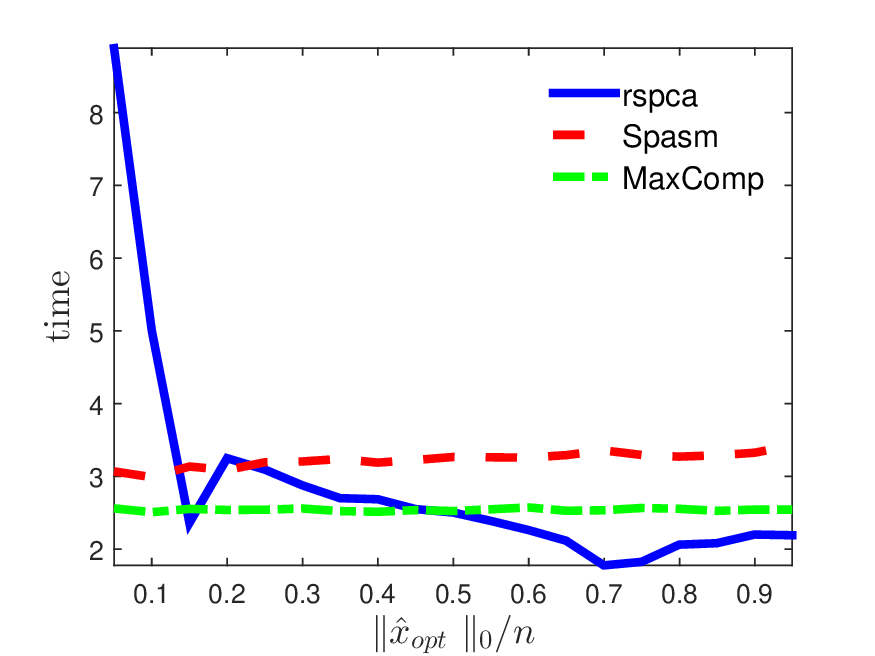}}		
	\quad
	\subfloat[HapMap+HGDP data (chromosome 16): $n=15289$.]{%
		\label{fig:snp16_time}%
		\includegraphics[width = 0.4\textwidth]{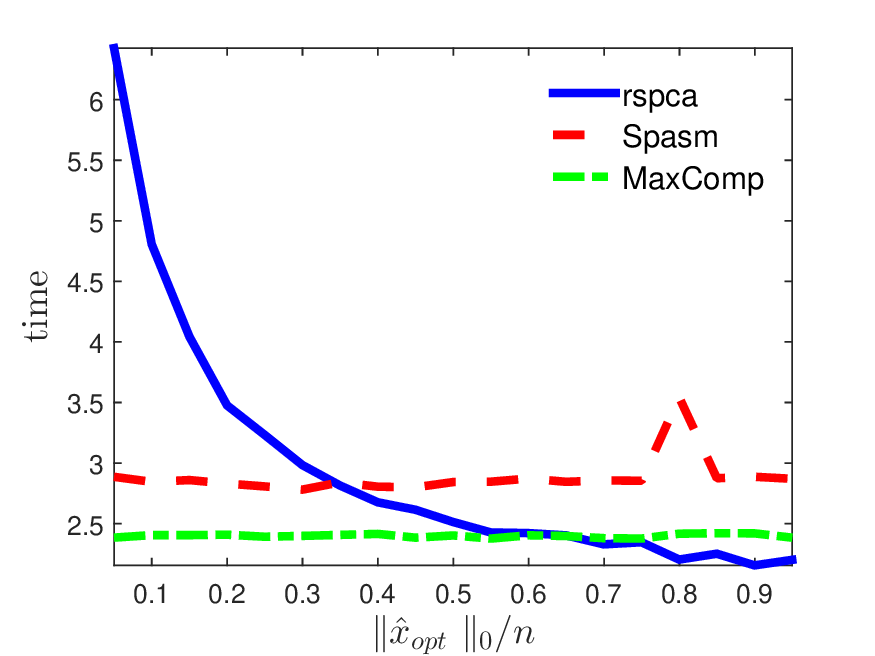}}	
	\caption{Running time of sparse PCA algorithms on additional population genetics data (chromosomes 11-16). }
	\label{fig:SNPs_time_2}%
\end{figure}		

\begin{figure}[H]
	\centering
    \subfloat[HapMap+HGDP data (chromosome 17): $n=12945$.]{%
		\label{fig:snp17_time}%
		\includegraphics[width = 0.4\textwidth]{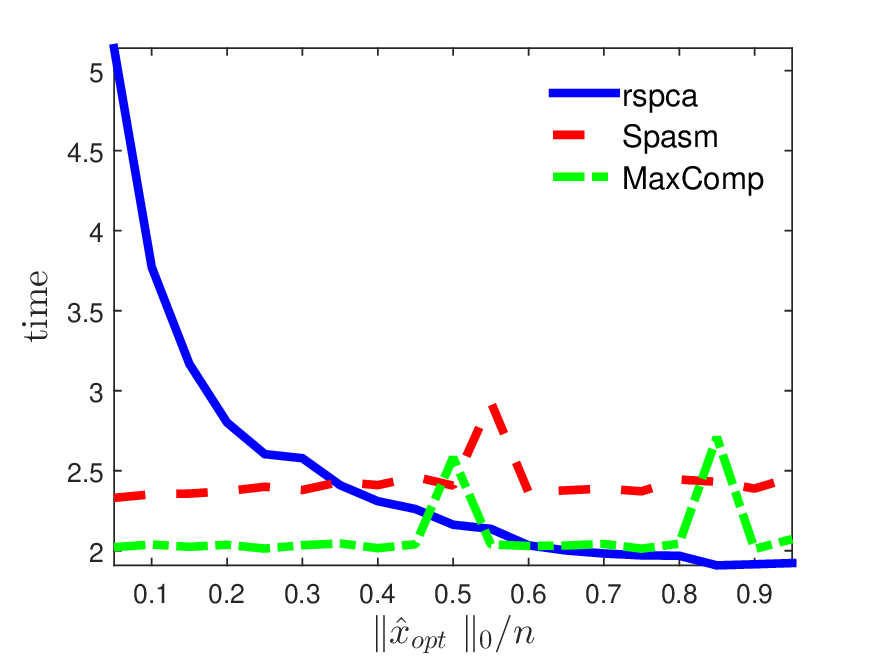}}		
	\quad
	\subfloat[HapMap+HGDP data (chromosome 18): $n=15373$.]{%
		\label{fig:snp18_time}%
		\includegraphics[width = 0.4\textwidth]{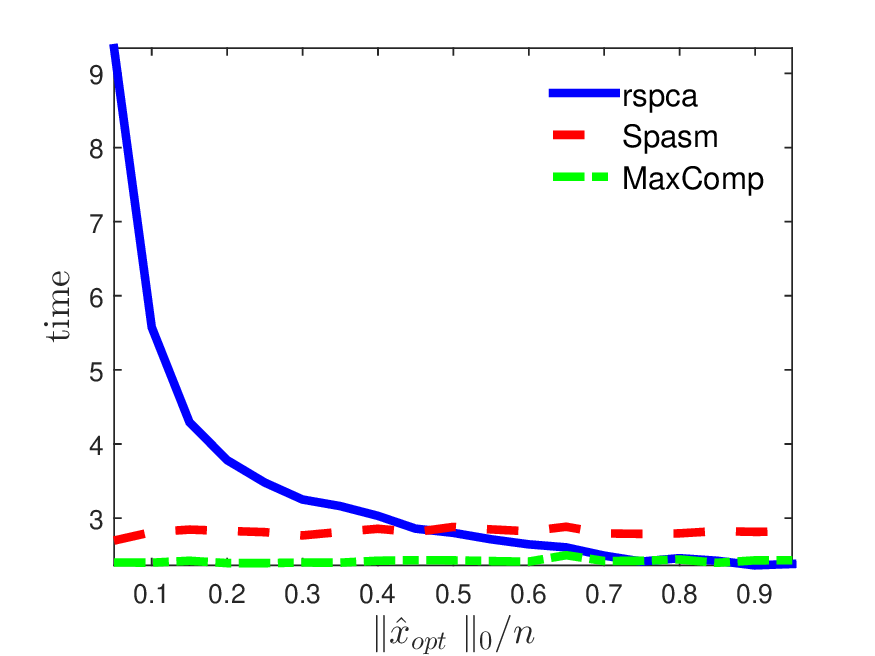}}	
    \\
 	\subfloat[HapMap+HGDP data (chromosome 19): $n=8465$.]{%
 		\label{fig:snp19_time}%
 		\includegraphics[width = 0.4\textwidth]{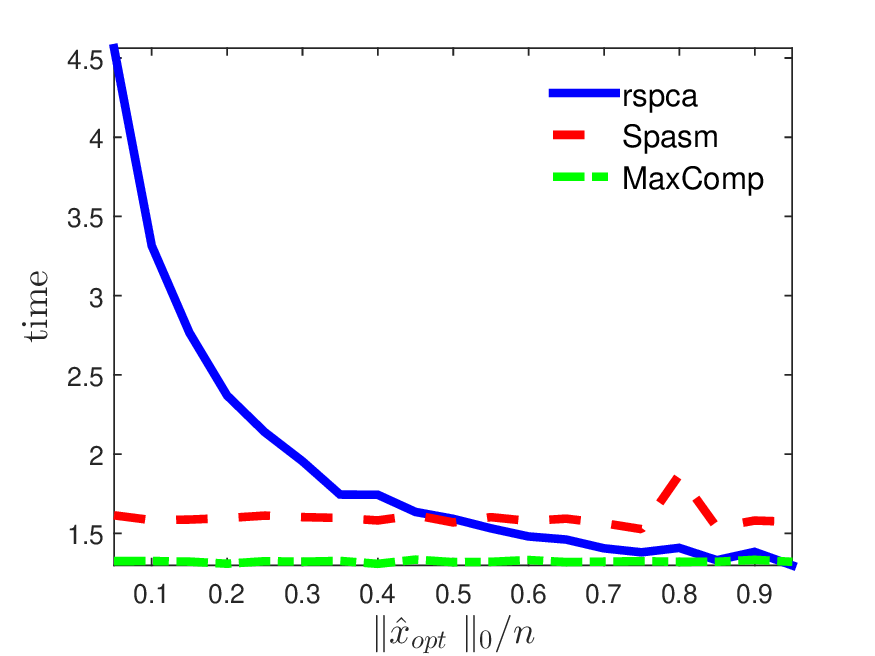}}		
 	\quad
 	\subfloat[HapMap+HGDP data (chromosome 20): $n=13015$.]{%
 		\label{fig:snp20_time}%
 		\includegraphics[width = 0.4\textwidth]{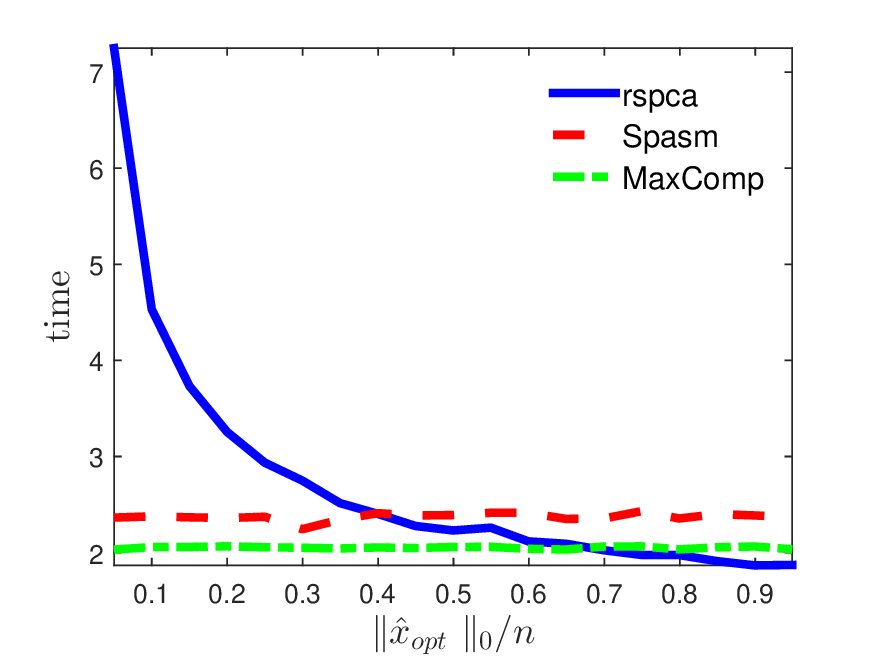}}		
 	\\
 	\subfloat[HapMap+HGDP data (chromosome 21): $n=7556$.]{%
 		\label{fig:snp21_time}%
 		\includegraphics[width = 0.4\textwidth]{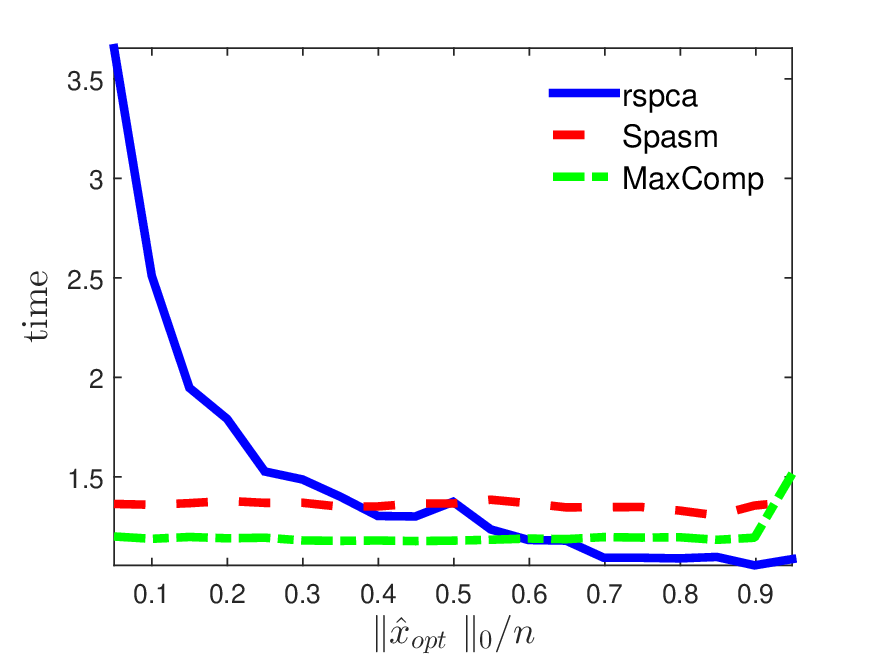}}		
 	\quad
 	\subfloat[HapMap+HGDP data (chromosome 22): $n=7334$.]{%
 		\label{fig:snp22_time}%
 		\includegraphics[width = 0.4\textwidth]{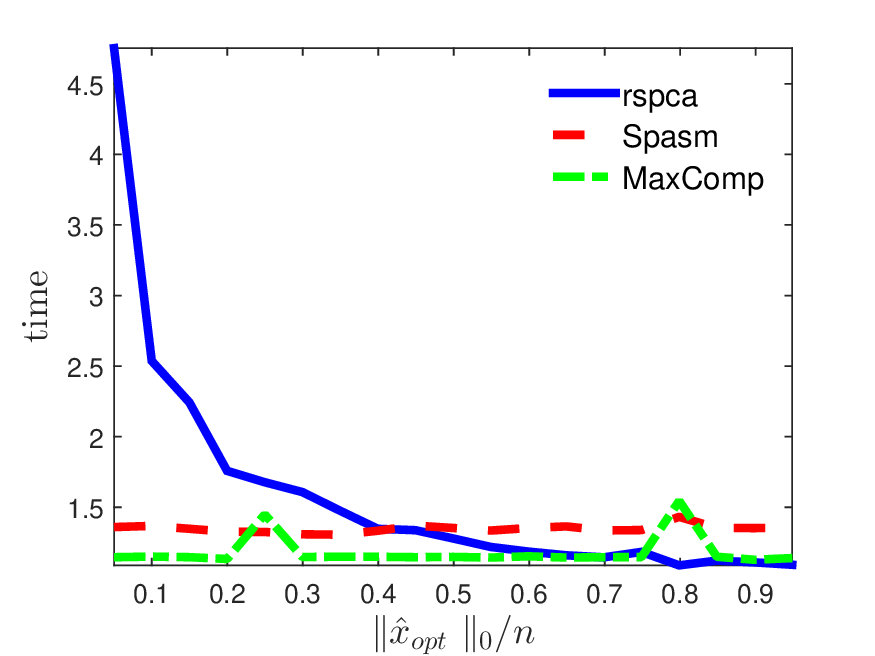}}
 \caption{Running time of sparse PCA algorithms on additional population genetics data (chromosomes 17-22). }
 \label{fig:SNPs_time_3}%
 \end{figure}

\newpage
\bibliographystyle{alpha}
\bibliography{sparsification}

\end{document}